\documentclass{aa}  

\bibpunct{(}{)}{;}{a}{}{,}

\usepackage{graphicx}
\usepackage{txfonts}
\usepackage{subfig}

\begin{document} 

   \title{The Perseus ALMA Chemical Survey (PEACHES).}
% Evidence of accretion shocks in a Class I protostar 

   \subtitle{III. Sulfur-bearing species tracing accretion and ejection processes in young protostars}

   \author{E. Artur de la Villarmois
        \inst{1,2,3} \and
        V. V. Guzm\'an 
        \inst{1,2} \and
        Y.-L. Yang
        \inst{4} \and
        Y. Zhang
        \inst{5,4} \and
        N. Sakai
        \inst{4}
        }

\institute{Instituto de Astrof\'isica, Pontificia Universidad Cat\'olica de Chile, Av. Vicu\~{n}a Mackenna 4860, 7820436 Macul, Santiago, Chile\\
        \email{Elizabeth.ArturdelaVillarmois@eso.org}
\and N\'ucleo Milenio de Formaci\'on Planetaria (NPF), Chile 
\and European Southern Observatory, Alonso de C\'ordova 3107, Casilla 19, Vitacura, Santiago, Chile
\and RIKEN Cluster for Pioneering Research, 2-1 Hirosawa, Wako-shi, Saitama 351-0198, Japan
\and Department of Astronomy, University of Virginia, Charlottesville, VA 22904-4235, USA
             }

   \abstract
    {Sulfur chemistry is poorly understood in the process of low-mass star and planet formation, where the main carriers of sulfur in both the gas and the dust phase are still unknown. Furthermore, the chemical evolution of sulfur-bearing species is not fully understood given that simple S-bearing molecules, such as SO and SO$_{2}$, are commonly seen in embedded Class 0/I sources but hardly detected in more evolved Class II disks. Despite the fact that simple S-bearing molecules are usually detected toward embedded sources, large surveys of S-bearing molecules with high angular resolution and sensitive observations are currently lacking. }   
     {The goal of this work is to present an unbiased survey of simple sulfur-bearing species in protostars and provide new statistics on detection rates, emitting regions, and molecular column densities. In addition, we investigate the role of S-bearing molecules in accretion processes and the connection between (non-)detection of complex organic molecules (COMs) and S-related species. }
    {We present the observations of sulfur-bearing species (CS, SO, $^{34}$SO, and SO$_{2}$) that are part of the Perseus ALMA Chemical Survey (PEACHES). We analyzed a total of 50 Class 0/I sources   with observations that have an average angular resolution of about 0$\farcs$6 ($\sim$180 au) in ALMA band 6.}
    {Class 0 sources show detection rates of 97$\%$ for CS, 86$\%$ for SO, 31$\%$ for $^{34}$SO, and 44$\%$ for SO$_{2}$, while Class I sources present detection rates of 71$\%$ for CS, 57$\%$ for SO, 36$\%$ for $^{34}$SO, and 43$\%$ for SO$_{2}$. When $^{34}$SO is detected, the SO/$^{34}$SO ratio is lower than the canonical value of 22, suggesting optically thick emission, and the lowest values are found for those sources that are rich in COMs. When SO$_{2}$ is detected, those sources that show CS and SO emission parallel to the outflow direction are usually very rich in COMs, while for sources where the CS and SO emission is perpendicular to the outflow direction,  only a few or no COMs are detected. When CH$_{3}$OH and SO$_{2}$ are detected, the comparison between CH$_{3}$OH and SO$_{2}$ abundances shows a positive trend and CH$_{3}$OH is between 10 and 100 times more abundant than SO$_{2}$. The SO$_{2}$ abundances toward the PEACHES sample are, on average, two orders of magnitude lower than values from the Ophiuchus star-forming region and comparable with sources in Taurus.}
    {The SO/$^{34}$SO ratio seems to be a good tracer of the inner high-density envelope and it could be used in the future to infer the presence of multiple COMs. The detection of multiple COMs seems to be related to the presence of collimated outflows (seen in CS and SO emission), where a high column density of warm material is expected close to the protostar, and SO$_{2}$ emission seems to trace the warm gas in those sources where CH$_{3}$OH is also detected. The difference in SO$_{2}$ abundances between different star-forming regions might indicate that the sulfur depletion in the gas-phase could depend on the external UV radiation toward the molecular cloud. Finally, the SO$_{2}$ emission detected in different evolutionary stages seems to arise from different physical mechanisms: high column density of warm material in Class 0 sources, shocks in Class I/II, and exposure to UV radiation from the protostar in more evolved Class II disks. }


\maketitle

\section{Introduction}
In the early stages of the low-mass star formation process the envelope, disk, and outflow components play a key role. Material infalls from the envelope to the disk, part of this material is accreted from the protoplanetary disk into the central protostar and part of the parental cloud material goes on to be ejected through energetic jets and outflows \citep{Reipurth2001, Pudritz2007}. \cite{Tychoniec2020} and \cite{Cridland2022} proposed that the first planetary embryos start to form in the embedded Class 0/I stages, therefore, the chemical content of the early disk will potentially influence the natal composition of emerging planets. In this sense, it is essential to identify and study the chemical composition of embedded Class 0/I sources.

One of the current enigmas is the sulfur chemistry, which is poorly understood in the process of low-mass star and planet formation. The sulfur content in dense molecular gas was found to be depleted by several orders of magnitude in comparison to diffuse clouds \citep{Ruffle1999, Laas2019, Riviere2019}, suggesting that most of the sulfur is locked into dust grains \citep{Ruffle1999, Laas2019, Kama2019, Shingledecker2020, Cazaux2022}. More recently, \cite{Fuente2023} proposed that this sulfur-depletion in the gas-phase might depend on the environment, finding a larger depletion factor for Taurus and Perseus, in comparison with Orion A. In the solid phase, it is also unclear what is the main sulfur carrier. \cite{Laas2019} predicted that most of the missing sulfur is in the form of organo-sulfur species, while \cite{Kama2019} proposed that the main sulfur carrier is much more refractory than water ice, consistent with sulfide minerals such as FeS. In addition, \cite{Shingledecker2020} and \cite{Cazaux2022} concluded that sulfur-chain molecules, such as S$_{8}$, should be abundant in dust grains. These results are not reflected in the detection of sulfur-bearing molecules in interstellar ices, where OCS is the only S-bearing species unambiguously detected \citep{Geballe1985, Palumbo1995, McClure2023}. SO$_{2}$ has been tentatively detected by \cite{Boogert1997}, \cite{Yang2022}, and \cite{McClure2023}, while the dominant S-bearing ice in comets \citep[H$_{2}$S; ][]{Calmonte2016} remains undetected \citep{Jimenez2011, McClure2023}. Therefore, the main carrier of sulfur is still unknown in both, the gas and the solid-phase. 

CS, SO, and SO$_{2}$ are simple S-bearing species commonly detected in the gas phase toward embedded sources \citep{Artur2019a, Tychoniec2021}. CS emission typically traces the envelope material and outflow cavity walls in young Class 0 sources \citep{Tychoniec2021}, SO emission has been detected in outflow cavity walls and the disk component of Class 0/I sources \citep{Sakai2014, Harsono2014, Tychoniec2021} and SO$_{2}$ is commonly seen in compact emission toward Class 0 sources and has been proposed to trace accretion shocks at the disk-envelope interface in more evolved Class I sources \citep{Oya2019, Artur2022}. In addition, \cite{Artur2019a} studied a sample of 10 Class I sources in Ophiuchus and found that SO$_{2}$ was only detected in sources with high bolometric luminosities (\textit{L$_\mathrm{bol}$}~$\geq$~1.4~L$_{\odot}$). In contrast to embedded sources, a handful of more evolved Class II disks show emission of CS, H$_{2}$CS \citep{LeGal2019}, SO \citep{Riviere2020, Booth2023, Huang2023}, and H$_{2}$S \citep{Phuong2018, Riviere2021, Riviere2022}, while SO$_{2}$ and CCS emission has only been detected in IRS 48 \citep{Booth2021} and GG Tau \citep{Phuong2021}, respectively. In addition, only one sulfur-bearing complex organic molecule (COM) has been detected so far \citep[CH$_{3}$SH toward the Class 0 source IRAS 16293-2422; ][]{Drozdovskaya2018}. The chemical evolution of S-bearing species from Class 0 to Class II disks is not very well understood and there is a lack of a statistically unbiased survey of sulfur-bearing molecules toward embedded Class 0/I sources, where they are more easily detected. 

In this work, we present the analysis of the simple sulfur-bearing molecules, CS, SO, $^{34}$SO, and SO$_{2}$ toward a sample of 50 Class 0/I sources in the Perseus star-forming region. A special focus in the analysis of the SO$_{2}$ emission is given, as this molecule has not been observed as much as the others and it usually shows centrally peaked emission (not commonly seen in outflow cavity walls or extended structures). The Perseus molecular cloud is one of the most active nearby star-forming regions, with distances between 234 and 331~pc \citep{Zucker2020}. The data analyzed in this paper are part of the PErseus ALMA CHEmical Survey \citep[PEACHES; ][]{Zhang2018, Yang2021, Zhang2023}, with an average angular resolution of 0$\farcs$6 in ALMA band 6. A more detailed study of COM detections toward the PEACHES sample is presented in \cite{Yang2021} and the connection between sulfur-bearing species and dust polarization of nine of the sources has been studied by \cite{Zhang2023}.  

Section 2 describes the observational procedure, calibration, and the parameters of the observed molecular transitions. Results are presented in Sect. 3, with moment 0 maps, spectra, and detection rates. Section 4 is dedicated to the analysis and discussion, where fluxes and column densities are analyzed as a function of the protostellar physical properties, SO$_{2}$ abundances are compared with other sources in the literature and the detectability of COMs is investigated in those sources where S-bearing molecules are also detected. Finally, we end with a summary in Sect. 5.

\section{Observations}
The data analyzed in this work is part of the PEACHES survey \citep{Yang2021} observed with ALMA (project codes: 2016.1.01501.S and 2017.1.01462; PI: N. Sakai). J0237+2848 was used as bandpass calibrator, while J0238+1636 and J0336+3218 were employed as flux and phase calibrators, respectively. The integration time corresponds to 10 minutes per source in ALMA band 6. The molecular transitions analyzed in this paper are listed in Table~\ref{table:molecules} and they are the only molecular lines detected in the PEACHES sample that contain sulfur. Self-calibration to the phase and amplitude was performed in most of the sources using the continuum data after the normal calibration and applied to the molecular line data. The exceptions were Per-emb 15, Per-emb 20, Per-emb 54, Per-emb 60, L1448 IRS2E, and L1455 IRS2, given that the continuum emission toward these sources was not bright enough to perform self-calibration. 

A robust parameter of 0.5 and an auto-mask option were employed in the cleaning process, and primary beam responses of 0.1 were used to maximize the field of view. The synthesized beam is about 0$\farcs$6~$\times$~0$\farcs$4 averaged across all spectral windows and, assuming a distance of 300~pc to the Perseus molecular cloud, this is equivalent to 180~$\times$~120~au. The final spectral resolution is 0.14~km~s$^{-1}$ (121~kHz) for both SO transitions, 0.15~km~s$^{-1}$ (122~kHz) for the CS and SO$_{2}$ transitions, and 1.2~km~s$^{-1}$ (987~kHz) for the $^{34}$SO transition. The source sample is listed in Table~\ref{table:observations} in the appendix, with other common names, coordinates, and physical properties. More detailed information on the PEACHES survey and other observed molecular transitions, as well as the calibration strategy is presented in \cite{Yang2021}.

% TABLE 1
\begin{table}[h!]
        \caption{Spectral setup and parameters of the observed molecular transitions.}
        \label{table:molecules}
        \centering
        \begin{tabular}{l l l l c c}
                \hline\hline
                Species 		& Transition                            	& Frequency  	& \textit{E$_\mathrm{up}$}      	& \textit{A$_{ij}$}                             	\\
                                         	&                                         	& (GHz)           	& (K)                                   	& ($\times$10$^{-5}$~s$^{-1}$)		\\
                \hline
                CS                  	& 5--4                                  	& 244.9356    	& 35                                        	& 30                                           		\\
                SO             	& 6$_{6}$--5$_{5}$                 	& 258.2558    	& 56                                       	& 21                                            	\\
                SO                    	& 7$_{6}$--6$_{5}$                  	& 261.8437    	& 48                                      	& 23                                          		\\
                $^{34}$SO     	& 5$_{6}$--4$_{5}$                  	& 246.6636   	& 50                                   	& 18                                             	\\
                SO$_{2}$       	& 14$_{0,14}$--13$_{1,13}$ 	& 244.2542    	& 94                                       	& 16                                        		\\
                
                \hline\hline
        \end{tabular}
        \tablefoot{Values from the CDMS database \citep{Muller2001}}
\end{table}

\section{Results}
In this section, we present moment 0 maps of CS, SO, $^{34}$SO, and SO$_{2}$, integrated spectra, detection rates, the physical components where the emission of the different S-bearing species are observed, continuum and molecular fluxes, and column density calculations. A 2D Gaussian was employed to fit the continuum emission, retrieving the coordinates and continuum fluxes listed in Table~\ref{table:observations} in the appendix. Continuum emission maps are not presented in this paper but can be found in Figure 1 of \cite{Yang2021}, as well as convolved sizes, deconvolved sizes, and positions angles (PA) of 2D Gaussian fits (Table 4). Comparing the beam size with deconvolved sizes, most of the sources are not resolved. The only resolved sources in the continuum are Per-emb 2, Per-emb 11 B/C, Per-emb 12 A/B, Per-emb 13, Per-emb 20, and Per-emb 44. 

\subsection{Moment 0 maps}

Moment 0 maps of CS, SO, $^{34}$SO, and SO$_{2}$ were created by integrating the data cubes over the velocity ranges shown in Table~\ref{table:integration} in the appendix. Figures~\ref{fig:Mom0_1} and \ref{fig:Mom0_1_ClassI} show moment 0 maps for Class 0 and Class I sources, respectively. Moment 0 maps of those sources where the value of the bolometric temperature (\textit{T$_\mathrm{bol}$}) is unknown, are presented in Fig.~\ref{fig:Mom0_1_nd}, given that some of them (SVS 13 A2 and RAC1999 VLA 20) show emission of CS and/or SO.

% FIGURE 1
\begin{figure*}[h!]
        \centering
        \includegraphics[width=.92\textwidth]{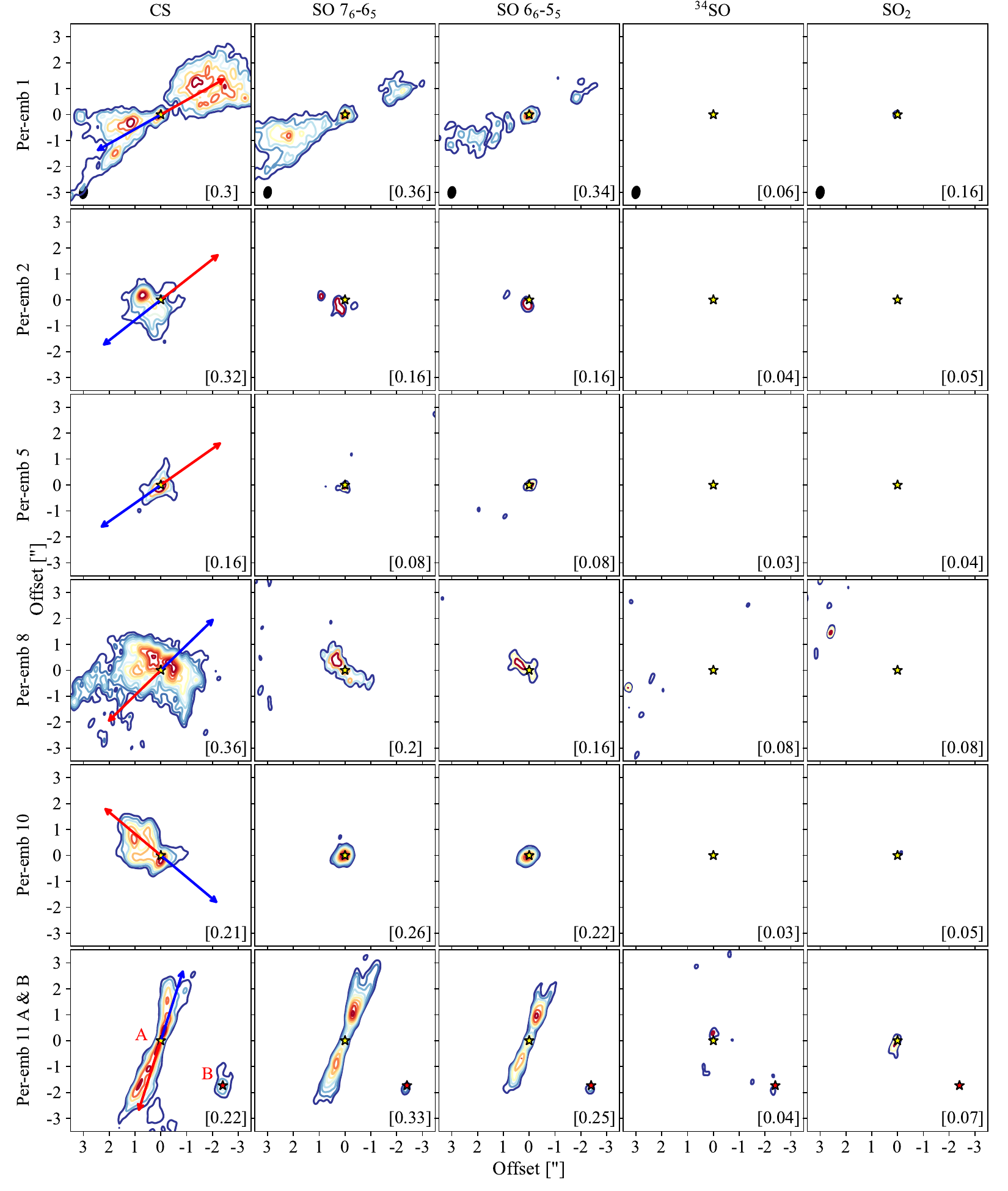}
        \caption[]{\label{fig:Mom0_1}
        Class 0 sources. Moment 0 maps of CS, SO, $^{34}$SO, and SO$_{2}$. The contours start at 3$\sigma$, follow steps of 1$\sigma$, and the emission has been integrated over the velocity ranges listed in Table~\ref{table:integration}. Maximum values (in units of Jy~beam$^{-1}$~km~s$^{-1}$) are indicated in the bottom-right corner of each panel. Red and blue arrows represent the outflow direction and the yellow star indicates the position of the continuum peak. Magenta stars represent binary components and the synthesized beam for each molecular transition is shown in the panels of the first row. Sources with no detections are not shown in this Figure. 
        }
\end{figure*}

\begin{figure*}[h!]
        \ContinuedFloat
        \centering
        \includegraphics[width=.92\textwidth]{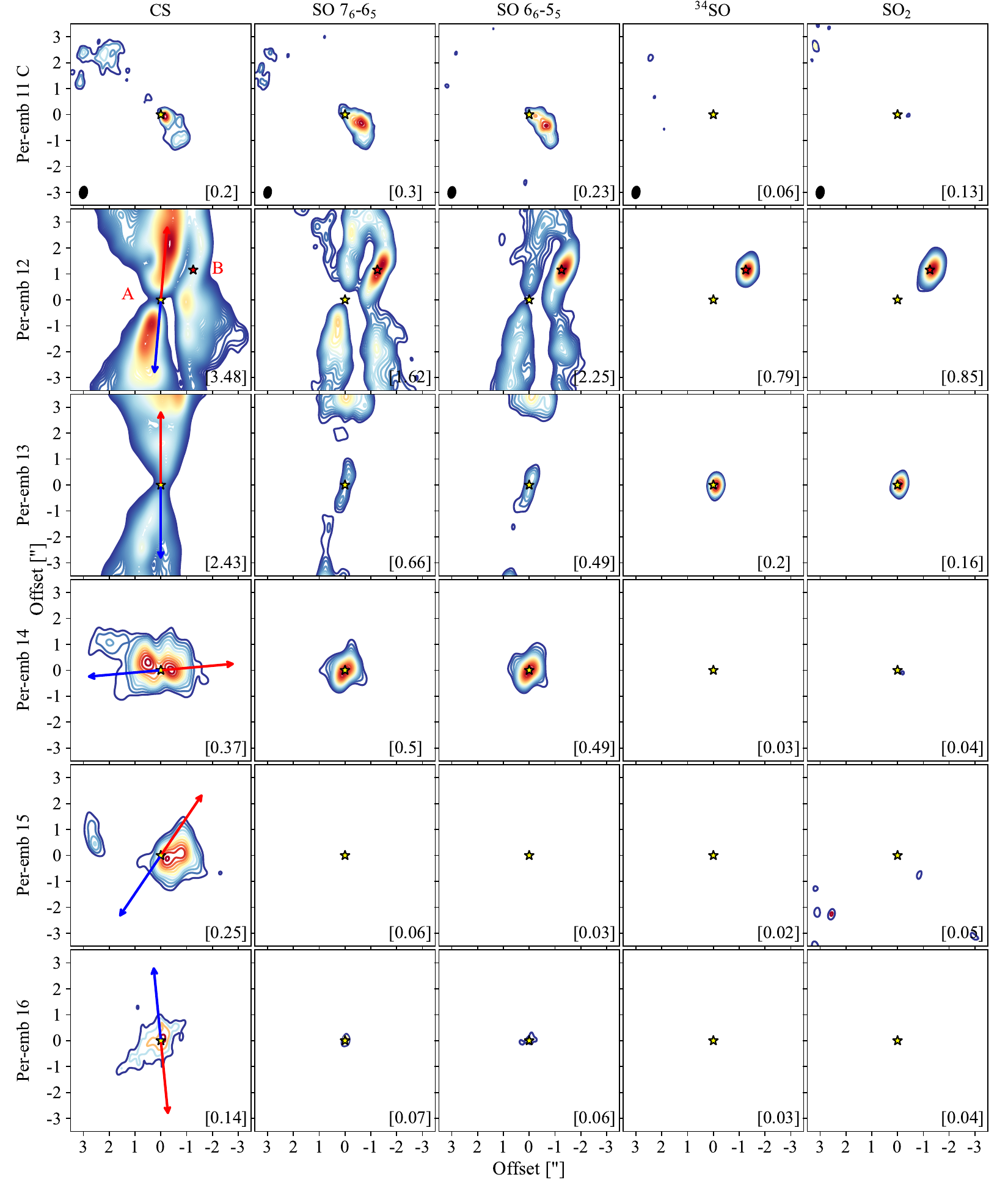}
        \caption[]{\label{fig:Mom0_1}
        (Cont.)
        }
\end{figure*}

\begin{figure*}[h!]
        \ContinuedFloat
        \centering
        \includegraphics[width=.92\textwidth]{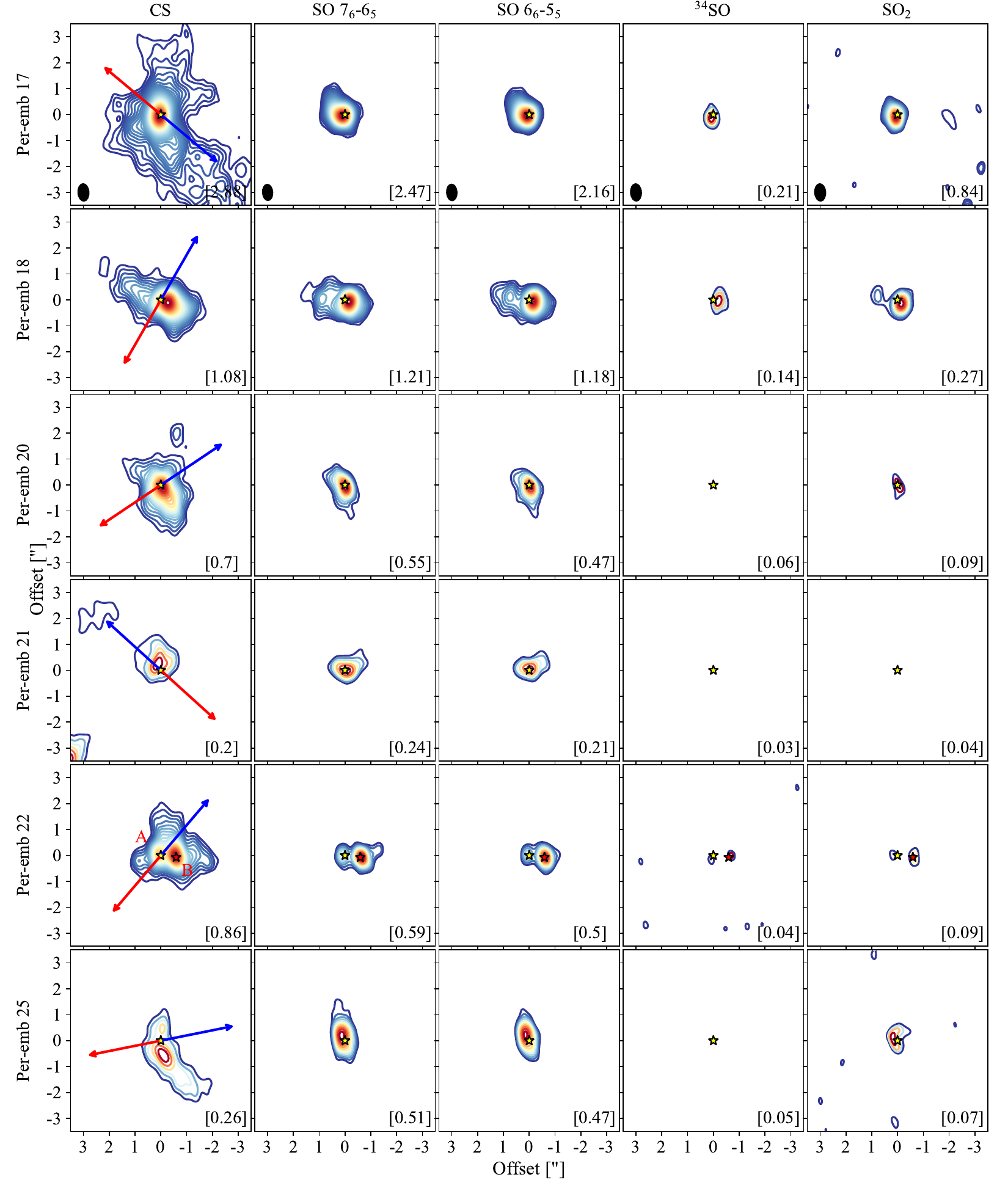}
        \caption[]{\label{fig:Mom0_1}
        (Cont.)
        }
\end{figure*}

\begin{figure*}[h!]
        \ContinuedFloat
        \centering
        \includegraphics[width=.92\textwidth]{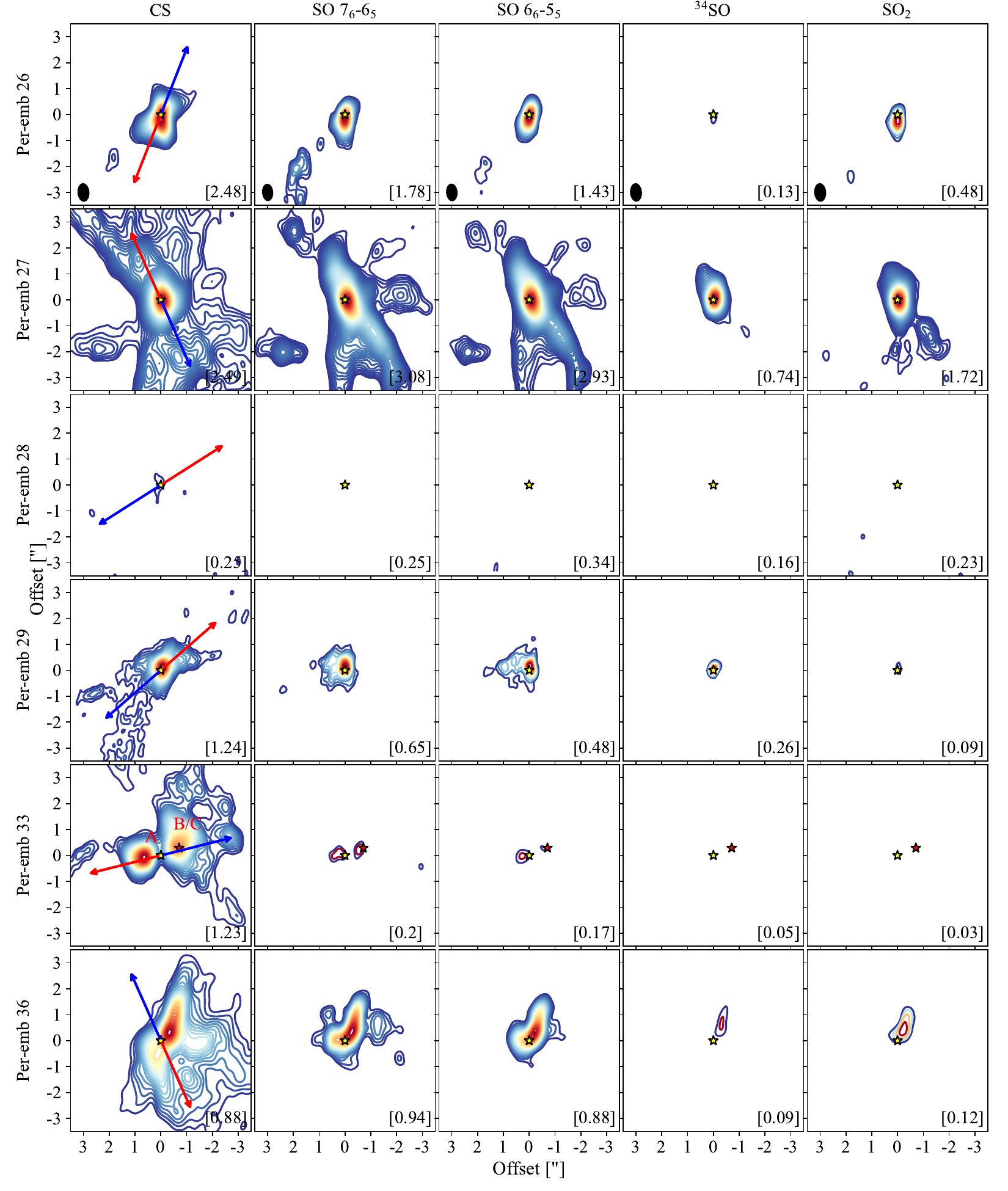}
        \caption[]{\label{fig:Mom0_1}
        (Cont.)
        }
\end{figure*}

\begin{figure*}[h!]
        \ContinuedFloat
        \centering
        \includegraphics[width=.92\textwidth]{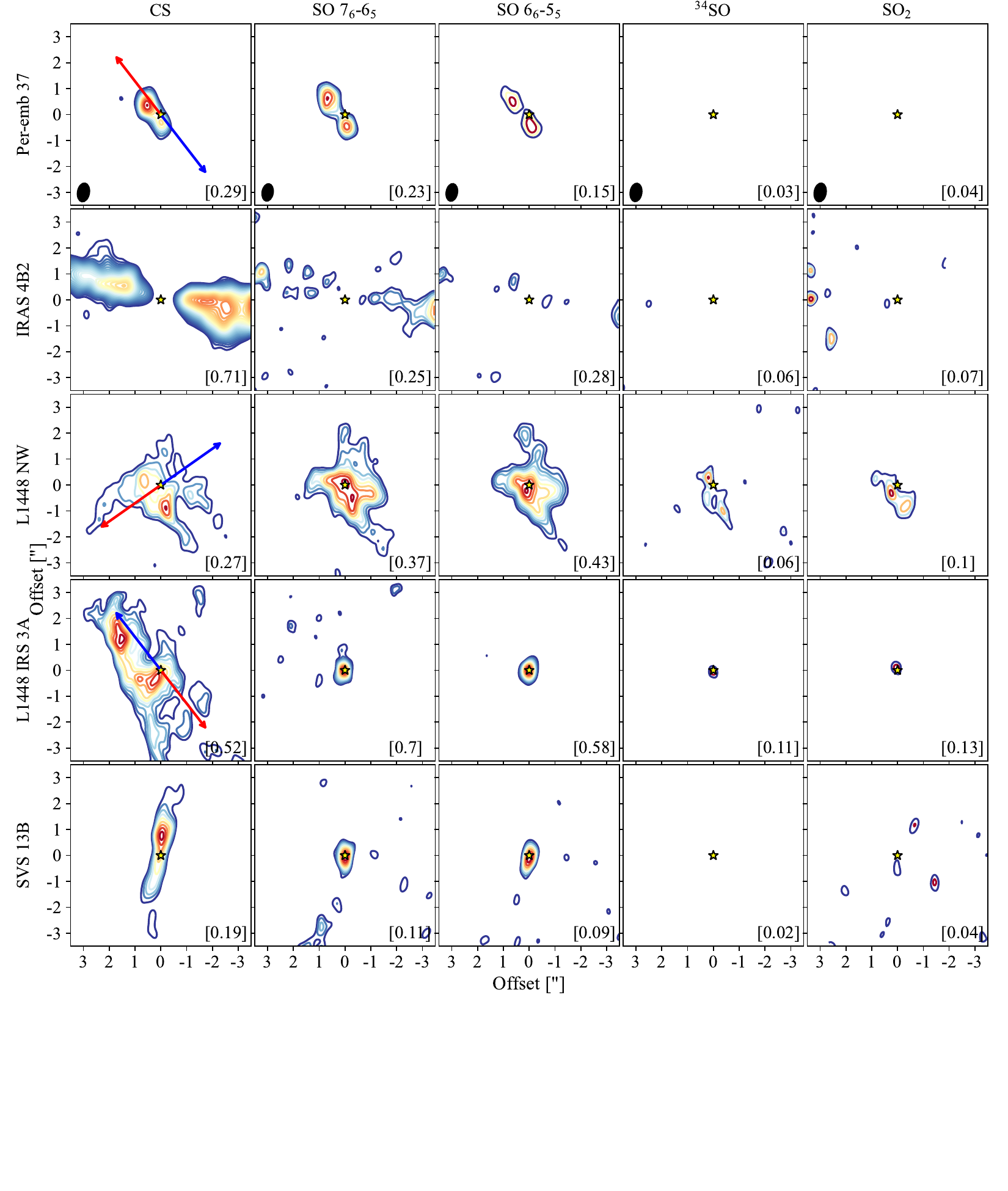}
        \vskip -90pt
        \caption[]{\label{fig:Mom0_1}
        (Cont.)
        }
\end{figure*}

% FIGURE 2
\begin{figure*}[h!]
        \centering
        \includegraphics[width=.92\textwidth]{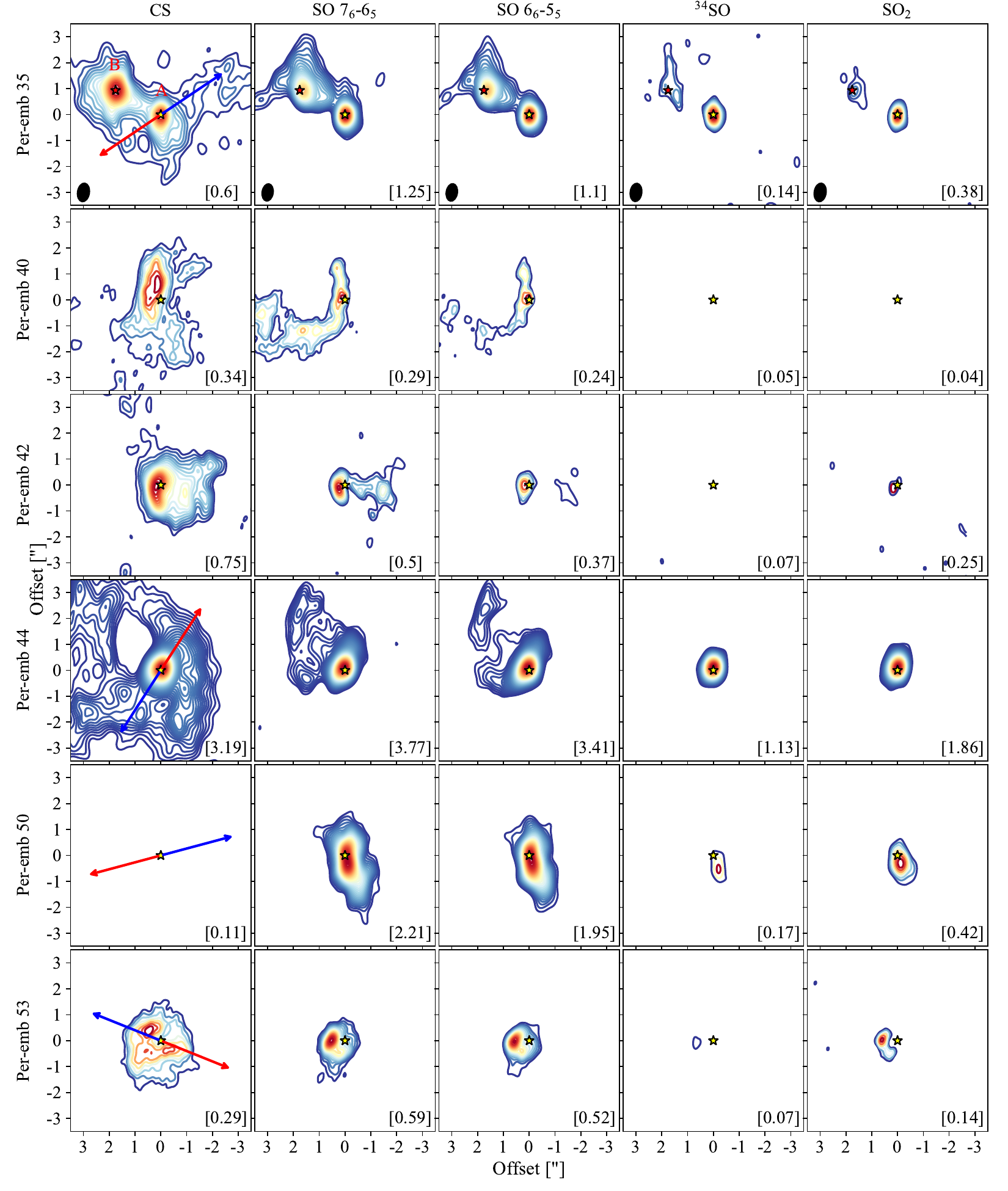}
        \caption[]{\label{fig:Mom0_1_ClassI}
        Class I sources. Moment 0 maps of CS, SO, $^{34}$SO, and SO$_{2}$. The contours start at 3$\sigma$, follow steps of 1$\sigma$, and the emission has been integrated over the velocity ranges listed in Table~\ref{table:integration}. Maximum values (in units of Jy~beam$^{-1}$~km~s$^{-1}$) are indicated in the bottom-right corner of each panel. Red and blue arrows represent the outflow direction and the yellow star indicates the position of the continuum peak. Magenta stars represent binary components and the synthesized beam for each molecular transition is shown in the panels of the first row. Sources with no detections are not shown in this Figure. 
        }
\end{figure*}

\begin{figure*}[h!]
        \ContinuedFloat
        \centering
        \includegraphics[width=.92\textwidth]{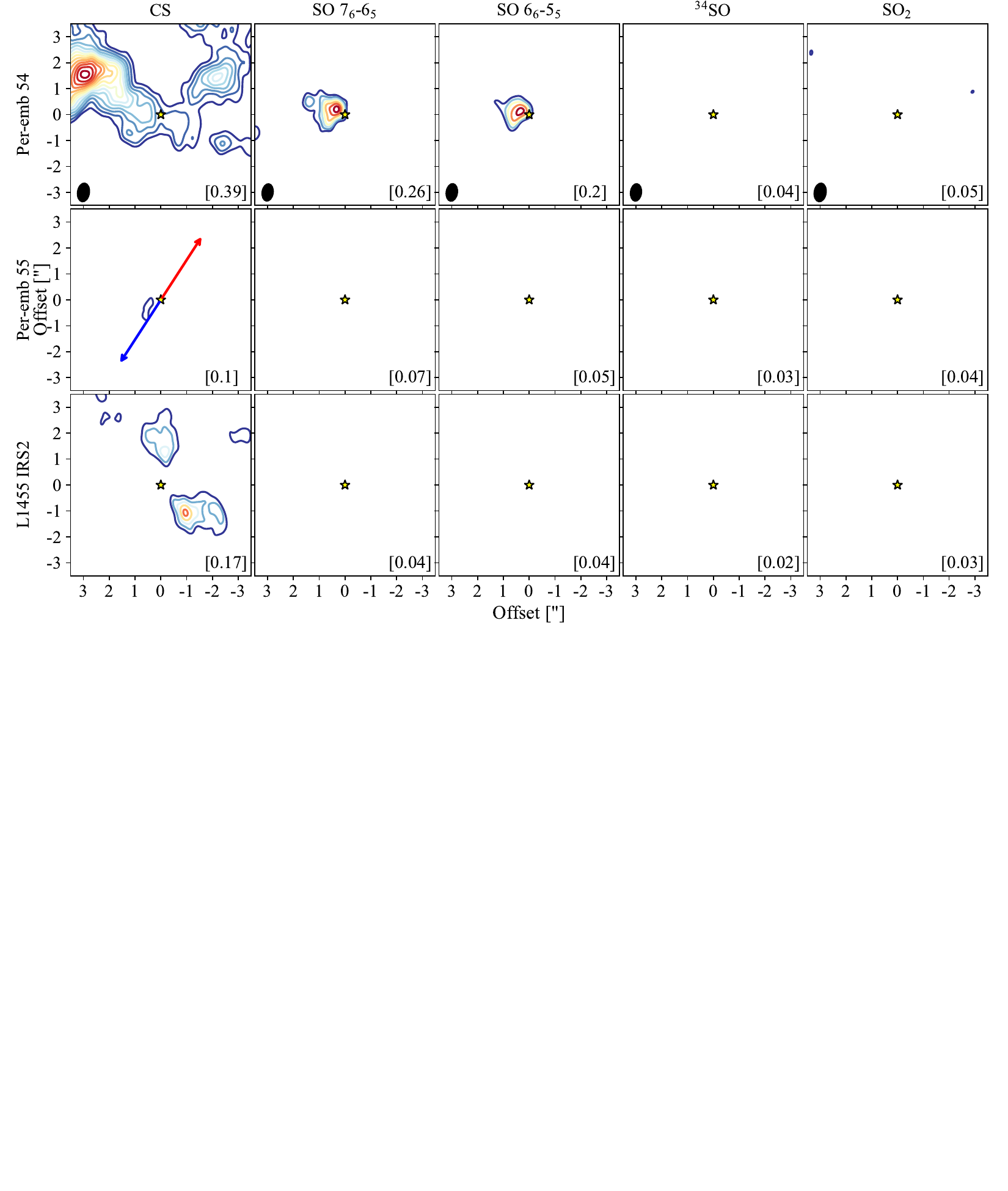}
        \vskip -260pt
        \caption[]{\label{fig:Mom0_1_ClassI}
        (Cont.)
        }
\end{figure*}

\begin{figure*}[h!]
        \centering
        \includegraphics[width=.95\textwidth]{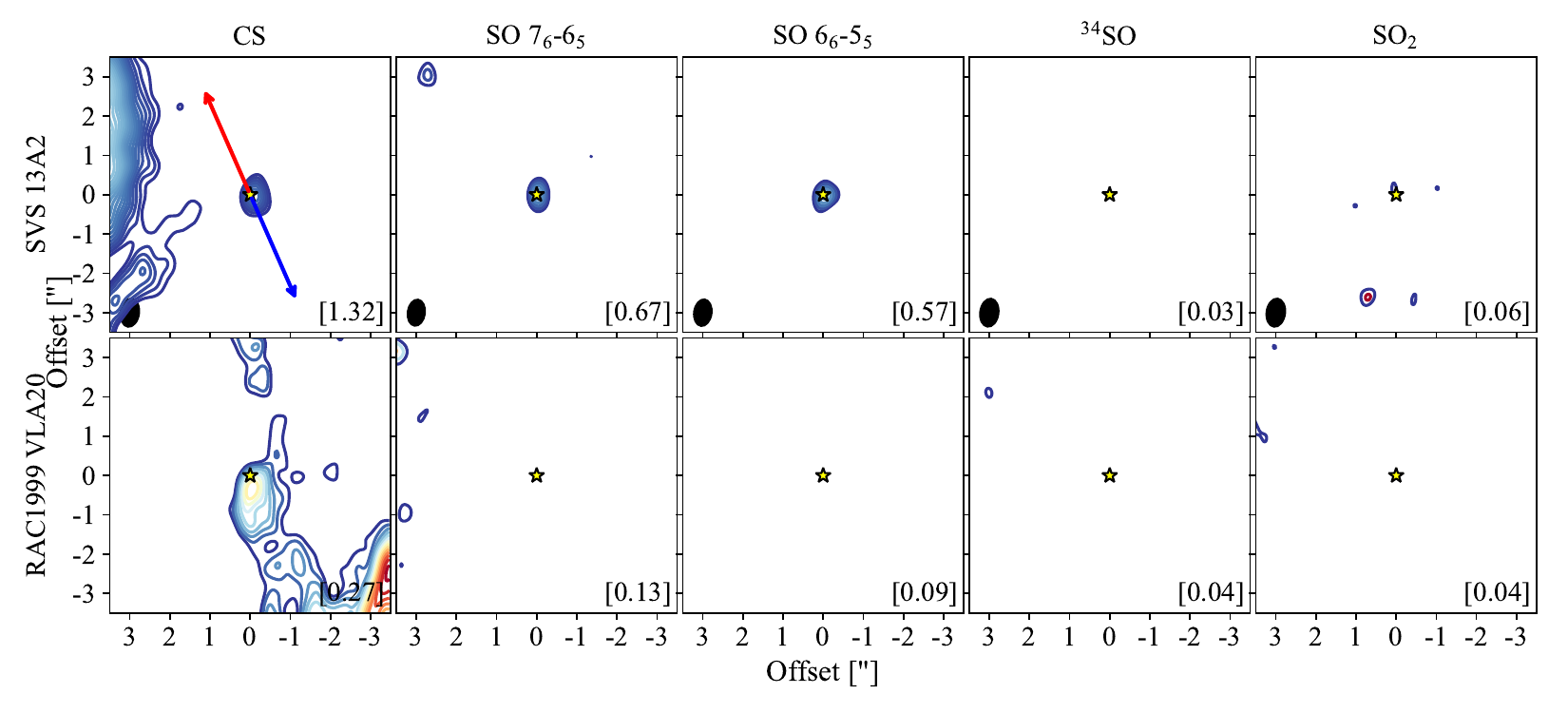}
        \caption[]{\label{fig:Mom0_1_nd}
        Sources with unknown \textit{T$_\mathrm{bol}$}. Moment 0 maps of CS, SO, $^{34}$SO, and SO$_{2}$. The contours start at 3$\sigma$, follow steps of 1$\sigma$, and the emission has been integrated over the velocity ranges listed in Table~\ref{table:integration}. Maximum values (in units of Jy~beam$^{-1}$~km~s$^{-1}$) are indicated in the bottom-right corner of each panel. Red and blue arrows represent the outflow direction and the yellow star indicates the position of the continuum peak. The synthesized beam for each molecular transition is shown in the panels of the first row. Sources with no detections are not shown in this Figure. 
        }
\end{figure*}

From moment 0 maps, the CS line usually shows centrally peaked and extended emission. In some cases, the CS extended emission is clearly tracing the outflow cavities, such as in Per-emb 1, Per-emb 11A, Per-emb 12, Per-emb 13, Per-emb 26, Per-emb 27, and Per-emb 29, while in other cases it seems to be tracing emission perpendicular to the outflow direction, possibly related to a wider outflow cavity wall or an infalling envelope: good examples are Per-emb 8, Per-emb 18, Per-emb 20, Per-emb 25, Per-emb 36, L1448 NW, and L1448 IRS 3A. Only compact emission (comparable to the beam size) is seen around Per-emb 28 and SVS 13 A2, while the emission peaks offset from the continuum peak in the cases of Per-emb 2, Per-emb 12 A and B, Per-emb 13, IRAS 4B2, Per-emb 53, Per-emb 54, Per-emb 55, and L1455 IRS 2. The absence of line emission around the central protostar could be due to optically thick dust, and \cite{DeSimone2020} showed that this is the case for Per-emb 12A (also known as IRAS 4A) where COMs are detected in centimeter wavelengths but not in the millimeter regime. \cite{Yang2021} showed that the other sources with mean continuum brightness temperature (\textit{T$_\mathrm{b,cont}$}) (a proxy of gas column density that correlated with the dust mass of the disk-like structure) higher than 10~K, Per-emb 5, Per-emb 12B, Per-emb 13, Per-emb 11A, Per-emb 33A, and B1-b S, feature detections of several molecular lines, suggesting that the continuum opacity is not significant. For a reference, the \textit{T$_\mathrm{b,cont}$} value of Per-emb 12A is 21.9~K \citep{Yang2021}, and this is the only source in this sample where the dust emission will be considered as optically thick.

Both SO transitions agree with each other and they show a combination of compact and extended emission among the sample. Outflow structures are clearly traced by SO toward Per-emb 1, Per-emb 11A, Per-emb 12, Per-emb 13, Per-emb 26, and Per-emb 27. We note that those sources that have SO tracing the outflow also have CS tracing the outflow -- and they are all Class 0. Additionally, some arc-like structures are seen toward Per-emb 36, Per-emb 40, and Per-emb 44. 

The $^{34}$SO and SO$_{2}$ lines generally show compact emission around the protostar. Only a few exceptions could be seen: the emission extends along the outflow direction toward Per-emb 27, it presents a banana shape perpendicular to the outflow direction in Per-emb 36, it peaks toward the south in Per-emb 50, it peaks toward the east in Per-emb 53, and it is seen perpendicular to the outflow direction in L1448 NW. In addition to the compact emission around the protostar, the SO$_{2}$ transition also shows an east component perpendicular to the outflow direction toward Per-emb 18. It is interesting to note that, when $^{34}$SO and SO$_{2}$ are detected, they trace the same structures. There are a few sources for which SO$_{2}$ is detected but $^{34}$SO is not, such as Per-emb 1, Per-emb 20, Per-emb 22 A and B, Per-emb 25, and Per-emb 42. 

A peculiar case is seen toward Per-emb 50, where SO, $^{34}$SO, and SO$_{2}$ lines are detected but CS is not. One possible reason for the missing CS is a resolving out effect by the interferometer. For two sources, EDJ2009-237 and RAC1999 VLA20, it is unknown whether those are protostars or not because their evolutionary stages have not been determined. No sulfur-bearing species are detected toward EDJ2009-237, but CS emission is seen in RAC1999 VLA20. \cite{Tobin2018} reported that RAC1999 VLA20 might be an extragalactic source, however, given that CS emission is detected toward this sources, we cannot rule out the possibility that RAC1999 VLA20 is a protostar or that the CS emission is associated with nearby protostars (SVS13 sources). Since it is difficult to compare chemical characteristics with other protostars without knowing their physical parameters, we present continuum fluxes, moment 0 maps, and spectra of EDJ2009-237 and RAC1999 VLA20, but they will no longer be discussed in this paper.

In addition to EDJ2009-237, there are other four sources with no detections of sulfur-bearing species: Per-emb 41, Per-emb 60, EDJ2009-172, and EDJ2009-235, all of them Class I sources. Per-emb 41 lies in the border of the field of view that corresponds to B1-b N and B1-b S, and this could prevent weak lines from being detected. Larger scale ($\geq$3$\arcsec$) CS emission is detected toward L1448 IRS 2E and B1-b N, shown in Fig.~\ref{fig:B1bN} in the appendix, and this emission seems to be related to the outflow component. For B1-b S, only CS absorption is seen toward the source position (see Fig.~\ref{fig:spectra}) and given that no emission is seen in the surroundings of the protostar, the absorption feature is likely to be caused by foreground gas and we consider this as a non-detection. 

The detection, or no detection, of CS, SO, $^{34}$SO, and SO$_{2}$ is shown in Table~\ref{table:COM_S} in the Appendix, together with the amount of complex organic molecules (COMs) detected by \cite{Yang2021}. The relationship between detectability of COMs and S-bearing species will be discussed in more detail in Sect. 4.3. In addition, large scale emission of CS is presented in the Appendix C for some sources with complex morphologies: Per-emb 12, Per-emb 13, IRAS 4B2, Per-emb 26, Per-emb 42, Per-emb 27, Per-emb 33, and L1448 IRS 3A (Fig.~\ref{fig:per12_large}).

\subsection{Spectra}

A circular area that corresponds to the major axis of the synthesized beam (see Table~\ref{table:integration} in the Appendix) around the continuum peak position was used to integrate the spectra of the CS, SO 6$_{6}$--5$_{5}$, and SO$_{2}$ lines (Fig.~\ref{fig:spectra}), and SO 6$_{6}$--5$_{5}$, SO 7$_{6}$--6$_{5}$, and $^{34}$SO transitions (Fig.~\ref{fig:spectra_SO}). In most cases, all the sulfur-related species show emission within the same velocity range and the CS line presents absorption features in most of the cases, therefore, integrated CS fluxes calculated in the next Section correspond to lower limits. The spectra of Per-emb 28 and Per-emb 41 look very noisy because they lie close to the edge of the field of view. 

Five of the sources, Per-emb 12B, Per-emb 13, Per-emb 27, Per-emb 29, and Per-emb 44, show an offset emission (about +8~km~s$^{-1}$ from \textit{v$_\mathrm{sys}$}) in the SO 6$_{6}$--5$_{5}$ and SO$_{2}$ spectra (see vertical arrows in Fig.~\ref{fig:spectra}). This offset emission is also seen in $^{34}$SO but not in SO 7$_{6}$--6$_{5}$ (see Fig.~\ref{fig:spectra_SO} in the appendix). As listed in Table~\ref{table:molecules}, both SO transitions have similar \textit{A$_{ij}$} and \textit{E$_\mathrm{up}$} values, therefore, a similar behavior is expected between the two transitions, unless SO 7$_{6}$--6$_{5}$ is much more optically thick than SO 6$_{6}$--5$_{5}$. If this offset component is related with redshifted SO 6$_{6}$--5$_{5}$ and SO$_{2}$ emission, it is intriguing to note that there is no blueshifted counterpart seen around -8~km~s$^{-1}$. Another possibility could be that the offset emission is related with another molecular transition, most likely a COM, given that those five sources are the ones where most of the COMs have been detected in \cite{Yang2021} (see Table~\ref{table:COM_S}). Possible species for this offset emission are methyl cyanate (CH$_{3}$OCN at 244.2471~GHz), formic acid (c--HCOOH at 244.2479~GHz), acetaldehyde (CH$_{3}$CDO at 244.2466~GHz), vinyl cyanide (CH$_{2}$CHC$^{15}$N at 258.2465), or ethylene glycol ((CH$_{2}$OH)$_{2}$ at 258.2506~GHz), which are the strongest lines found close to the offset frequencies (244.2477 and 258.2493~GHz) in SPLATALOGUE\footnote{https://splatalogue.online} and using the molecular data from the Cologne Database of Molecular Spectroscopy \citep[CDMS; ][]{Muller2001, Muller2005, Endres2016}. The origin of this redshifted offset emission is out of the scope of this paper and we focus on the central emission for the rest of this work. 

An interesting spectral shape is seen for Per-emb 50, where the CS line is not detected and the SO spectrum does not show a Gaussian profile, but a broader and increasing spectral shape instead. \cite{Valdivia2022} proposed that a streamer is feeding the envelope-disk system of Per-emb 50 and SO and SO$_{2}$ could be enhanced in shocked regions. The existence of shocks in this source is also supported by \cite{Zhang2023}, who studied the relationship between sulfur-bearing species and dust polarization toward some of the PEACHES sources. Per-emb 50 will be discussed in more detail in Sect. 4.3.

\subsection{Detection rates}

The whole sample consists of 36 Class 0 and 14 Class I sources, along with detection rates for CS, SO, $^{34}$SO, and SO$_{2}$  shown in Fig.~\ref{fig:rate}. A given transition is considered to be detected if its integrated emission is above a 3$\sigma$ value (see Table~\ref{table:fluxes}) or if it shows extended emission in moment 0 maps (see Figs.~\ref{fig:Mom0_1}, \ref{fig:Mom0_1_ClassI}, \ref{fig:Mom0_1_nd} and \ref{fig:B1bN}). Both SO transitions are detected toward the same sources, therefore, we do not distinguish between them when discussing detection rates. As seen in Fig.~\ref{fig:rate}, the CS line is almost ubiquitous toward Class 0 sources and is detected in most of the Class I systems. In this case, the detection rate decreases in around 30$\%$ from Class 0 to Class I sources. SO is detected in most of the Class 0 sources and around half of the Class I ones. The SO detection rate also shows a considerable decrease from 86$\%$ to 57$\%$ from Class 0 to Class I sources, similar to the case of CS. A different behavior is seen for $^{34}$SO and SO$_{2}$, where detection rates are similar for Class 0 and I sources: around 35$\%$ for $^{34}$SO and around 45$\%$ for SO$_{2}$. 

Given that the envelopes of Class 0 sources are more massive than those from Class I systems, the decrease of the CS and SO detections rates from Class 0 to Class I sources seems to be related to a reduction in the envelope mass. On the other hand, the detection rates of $^{34}$SO and SO$_{2}$ do not decrease from Class 0 to Class I sources, suggesting that these $^{34}$SO and SO$_{2}$ lines are not sensitive to the envelope mass, but they seem to be sensitive to \textit{L$_\mathrm{bol}$} (see Sect. 4.1), which can be linked to the mass accretion rate (\textit{$\dot{M}$$_\mathrm{acc}$}). 

The sulfur bearing species, when detected, are tracing multiple components of the system as shown in Fig.~\ref{fig:rate_mol}. The CS transition typically traces extended emission, consistent with the outflow direction and, in a few cases, the emission is seen perpendicular to the outflow direction. In addition, five of the sources (11$\%$) are associated with CS extended structures but no central emission is detected and only one source (B1-b S) shows pure absorption features. Almost half of the SO detections show compact emission (when compared to the beam size) and the rest present extended emission, where two of them (5$\%$) reveal a clear arc-like structure (Per-emb 40 and Per-emb 44). Most of the extended emission detected in SO seems to arise from the outflow component. Compact emission is mainly seen when $^{34}$SO and SO$_{2}$ are detected and, when the emission is extended, most of the sources present a structure perpendicular to the outflow direction. This would be discussed in more detail in Sect. 4.3.

\subsection{Fluxes}

From moment 0 maps, we chose an integrated region that coincides with the beam size (Table~\ref{table:integration}) and employed the casa task \texttt{imfit} to calculate the molecular integrated fluxes listed in Table~\ref{table:fluxes}. Given that the SO 6$_{6}$--5$_{5}$ transition is slightly optically thinner than the SO 7$_{6}$--6$_{5}$ one (see Sect. 4.2.1), integrated fluxes for SO correspond to the 6$_{6}$--5$_{5}$ transition. An upper limit of 3$\sigma$ is set for non detections and lower limits represent optically thick emission.

\subsection{Column densities}

Column densities were calculated for both SO transitions, as well as $^{34}$SO and SO$_{2}$. Given that the CS spectra show absorption features in most of the cases (an indication of optically thick emission and/or self absorption due to infall motion), column densities for CS are lower limits in most of the cases. 

By employing the non-local thermodynamic equilibrium (LTE) radiative transfer code RADEX \citep{vanderTak2007}, we found that the SO 6$_{6}$--5$_{5}$ transition is thermalized for H$_{2}$ number densities (\textit{n$_\mathrm{H}$}) above 10$^{7}$~cm$^{-3}$. Assuming that the SO emission is coming from dense regions (\textit{n$_\mathrm{H}$}~$\geq$~10$^{7}$~cm$^{-3}$), a LTE analysis was employed to calculate the molecular column densities for SO and $^{34}$SO. On the other hand, we found that the SO$_{2}$ 14$_{0,14}$--13$_{1,13}$ transition is thermalized only for very high densities (\textit{n$_\mathrm{H}$}~$\geq$~10$^{9}$~cm$^{-3}$); therefore, the molecular column densities for SO$_{2}$ were estimated from a non-LTE analysis using RADEX. 

\subsubsection{LTE analysis}

Assuming optically thin conditions and that the gas is under LTE, for a given molecular transition, the column density of the upper level (\textit{N$_\mathrm{u}$}) is given by:

% EQUATION 1
\begin{equation} 
    N_\mathrm{u} = \frac{8 \pi k \nu^{2} W}{h c^{3} A_\mathrm{ul}} , \
    \label{eq:Nu}
\end{equation}

\noindent where \textit{k} is the Boltzmann constant, ${\nu}$ is the frequency of the molecular transition, \textit{W} is the integrated line intensity, \textit{h} is the Planck constant, \textit{c} is the speed of light, and \textit{A$_\mathrm{ul}$} is the Einstein coefficient for spontaneous emission. Equation~\ref{eq:Nu} can be rewritten as:

% EQUATION 2
\begin{equation} 
      N_\mathrm{u} = 2375 \times 10{^6} \ \left(\frac{W}{1 \ \mathrm{Jy \ km \ s^{-1}}}\right) \ \left(\frac{1 \ \mathrm{s^{-1}}}{A_\mathrm{ul}}\right) \ \left(\frac{\mathrm{arcsec{^2}}}{\mathrm{\Theta_{area}}} \right) , \
    \label{eq:Nu_bis}
\end{equation}

\noindent where ${\Theta}_\mathrm{area}$ is the area of integration (in arcsec; see Table~\ref{table:integration}) and \textit{N$_\mathrm{u}$} will be obtained in units of cm$^{-2}$. Employing a range of excitation temperatures (\textit{T$_\mathrm{ex}$} from 50 to 150~K), we calculated the total column density of the molecule (\textit{N}) following

% EQUATION 3
\begin{equation} 
    N = \frac{N_\mathrm{u} \ Q(T_\mathrm{ex}) \ \mathrm{exp}(E_\mathrm{up}/T_\mathrm{ex})}{g_\mathrm{u}} , \
    \label{eq:Ntot}
\end{equation}

\noindent where \textit{Q(T$_\mathrm{ex}$)} is the partition function that depends on the excitation temperature, \textit{E$_\mathrm{up}$} is the upper level energy in K, and \textit{g$_\mathrm{u}$} is the level degeneracy. Total column densities for SO and $^{34}$SO are listed in Table~\ref{table:column_density} in the appendix.

% FIGURE 3
\begin{figure*}%[H]
        \centering
        \includegraphics[width=.97\textwidth]{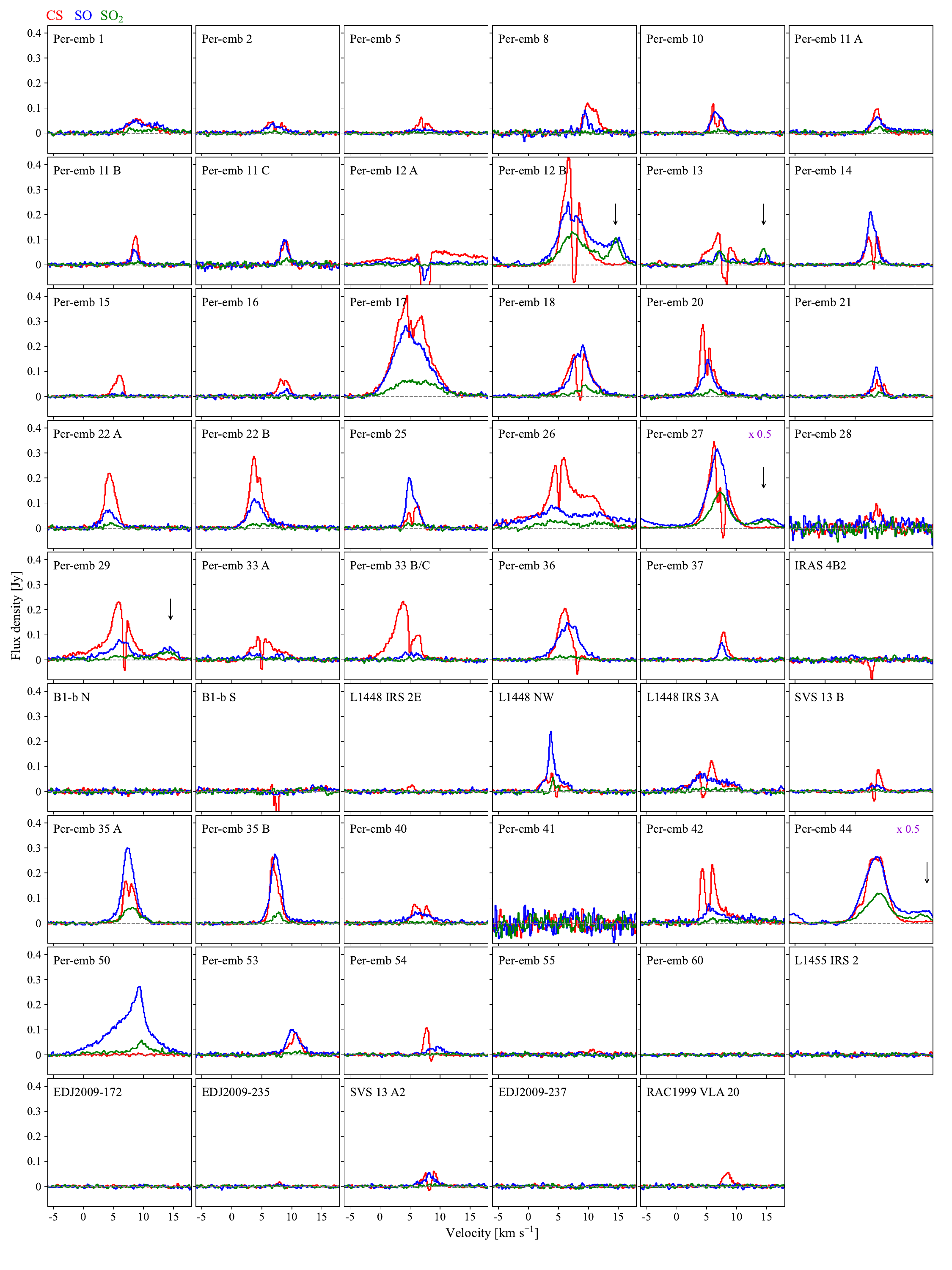}
        \vskip -10pt
        \caption[]{\label{fig:spectra}
        Integrated spectra of CS (red), SO 6$_{6}$--5$_{5}$ (blue), and SO$_{2}$ (green). The integration area for each source is shown in Table~\ref{table:integration}. Black arrows in the panels that correspond to Per-emb 12 B, Per-emb 13, Per-emb 27, Per-emb 29, and Per-emb 44 indicate an offset emission, likely related with a transition from an unidentified COM.
        }
\end{figure*}

% FIGURE 4
\begin{figure*}[h!]
        \centering
        \includegraphics[width=.8\textwidth]{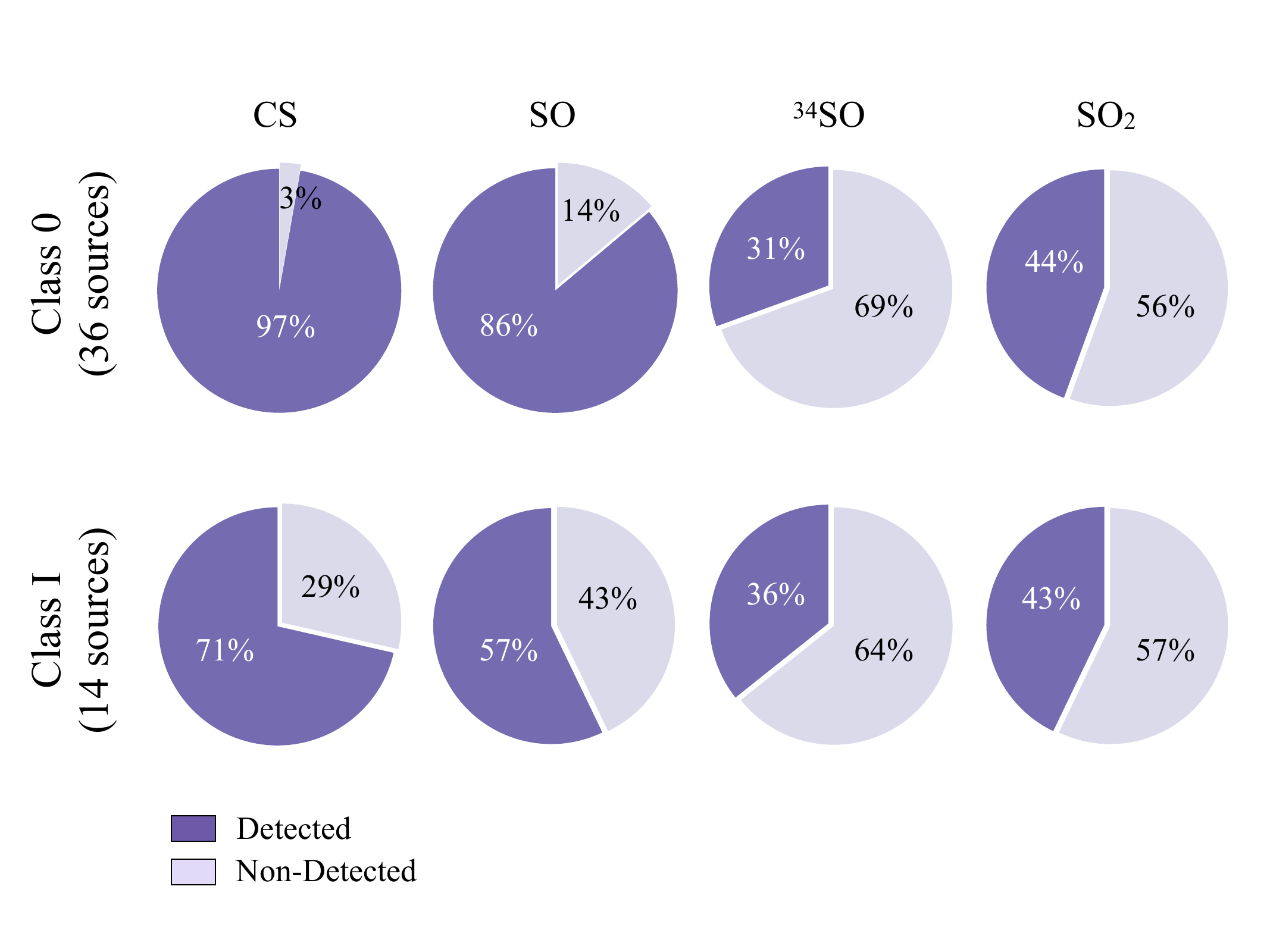}
        \vskip -20pt
        \caption[]{\label{fig:rate}
        Detection rate of CS, SO, $^{34}$SO, and SO$_{2}$ toward Class 0 and Class I sources. Detected refers to emission above 3$\sigma$ that is observed in the vicinity of the source (not necessarily toward the continuum peak).  
        }
\end{figure*}

% FIGURE 5
\begin{figure*}[h!]
        \centering
        \vskip 10pt
        \includegraphics[width=.7\textwidth]{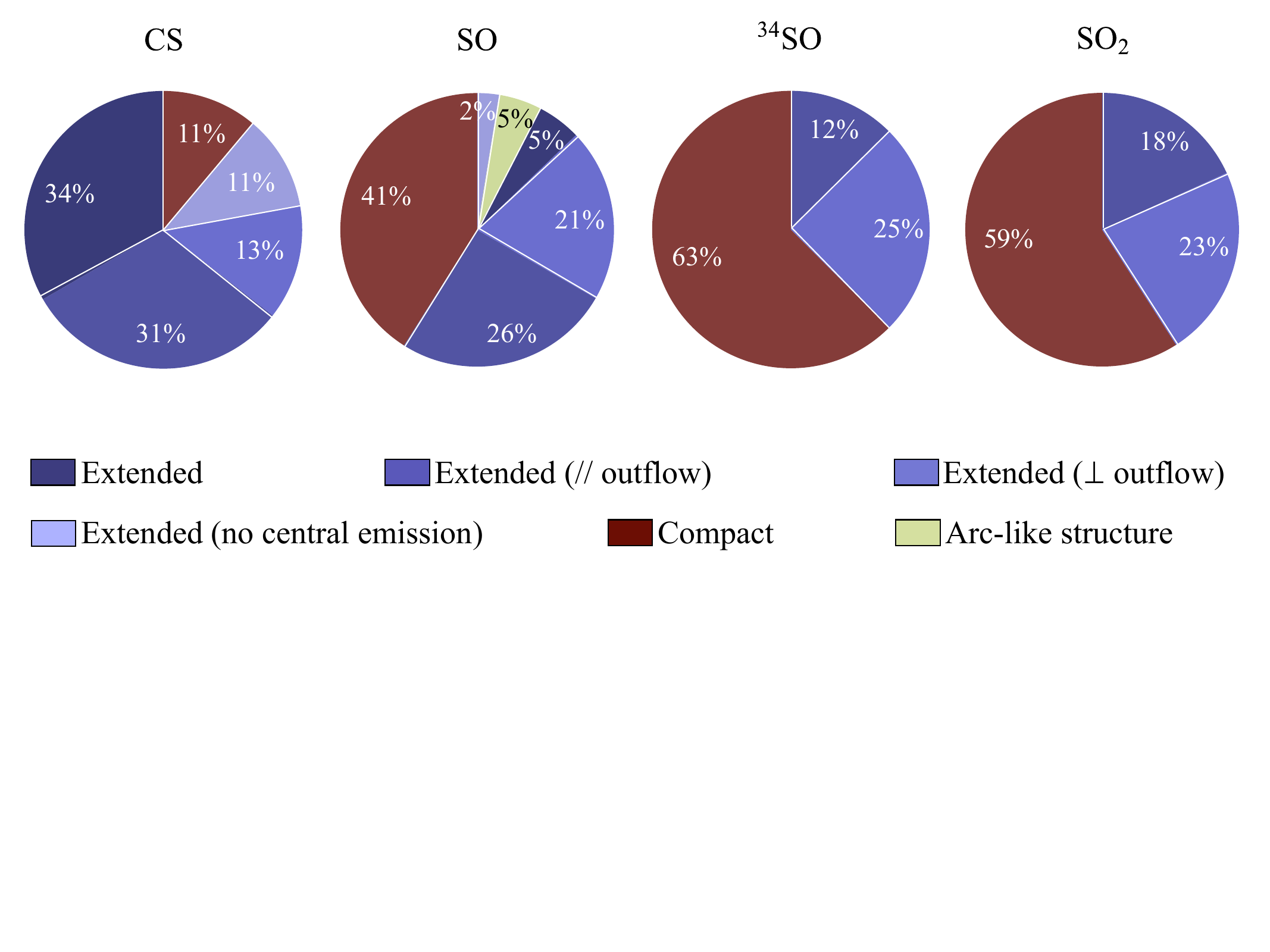}
        \vskip -110pt
        \caption[]{\label{fig:rate_mol}
        Different components where the CS, SO, $^{34}$SO, and SO$_{2}$ transitions are detected. Extended emission is divided into three groups, those sources where the emission is seen perpendicular or parallel to the outflow direction, and those where the structure is unclear (labeled as Extended).
        }
\end{figure*}

\subsubsection{Non-LTE analysis}

Given that the SO$_{2}$ 14$_{0,14}$--13$_{1,13}$ transition is not thermalized for typical inner envelope and disk densities (10$^{7}$ -- 10$^{9}$~cm$^{-3}$), RADEX was employed to estimate SO$_{2}$ molecular column densities (see the example of Per-emb 29 in Fig.~\ref{fig:radex_n}). Table~\ref{table:column_density} lists a range of values for \textit{N$_\mathrm{SO_{2}}$} for \textit{T$_\mathrm{ex}$} between 50 and 150~K and \textit{n$_\mathrm{H}$} between 10$^{7}$ and 10$^{9}$~cm$^{-3}$.  Collisional rates for SO$_{2}$ were taken from the Leiden atomic and molecular database \citep[LAMDA;][]{Balanca2016} and a broadening parameter (\textit{b}) of 2~km~s$^{-1}$ was employed for the calculations. Four sources, Per-emb 12B, Per-emb 17, Per-emb 27, and Per-emb 44, are associated with lower limits of \textit{N$_\mathrm{SO_{2}}$}, given that their molecular column densities can be much higher than 10$^{16}$~cm$^{-2}$ for \textit{n$_\mathrm{H}$}~=~10$^{7}$~cm$^{-3}$ and the SO$_{2}$ emission toward these four sources was found to be optically thick. A more detailed discussion of the optical thickness of the SO$_{2}$ transition is presented in Sect. 4.2.2.

\section{Analysis and discussion}

This section is dedicated to the analysis and discussion of integrated fluxes as a function of \textit{T$_\mathrm{bol}$} and \textit{L$_\mathrm{bol}$}, optically thickness of the lines, molecular column densities, and detectability of COMs and sulfur-bearing molecules.

\subsection{Fluxes as a function of \textit{T$_\mathrm{bol}$} and \textit{L$_\mathrm{bol}$}}

Continuum and molecular line fluxes are compared with the physical properties of the sources, such as \textit{T$_\mathrm{bol}$} and \textit{L$_\mathrm{bol}$}, and are shown in Fig.~\ref{fig:fluxes}. The continuum flux decreases as \textit{T$_\mathrm{bol}$} increases, which is expected, given that the envelope material deceases as the source evolves. A less clear positive trend is seen for the continuum flux as a function of \textit{L$_\mathrm{bol}$}. Multiple sources and unresolved binaries are highlighted in red and blue colors, given that the flux of the components cannot be totally separated in most of the cases. 

Molecular fluxes do not show any trend as a function of \textit{T$_\mathrm{bol}$} (left column of Fig.~\ref{fig:fluxes}); however, a special case must be taken when analyzing CS and SO given that these molecular transitions could be optically thick: CS absorption is seen in the spectra of almost all the sources and (as we explain in Sect. 4.2) when $^{34}$SO is also detected, both SO transitions are optically thick. A slightly positive trend is seen for the SO$_{2}$ flux as a function of \textit{L$_\mathrm{bol}$}, mainly given by the two most luminous sources, Per-emb 27 and Per-emb 44. Without them, the positive trend disappears. Despite the absence of a strong trend, it is clear that $^{34}$SO and SO$_{2}$ are only detected in those sources with \textit{L$_\mathrm{bol}$}~>~1~L$_{\odot}$. This result could be a selection bias, given that 70$\%$ of the sources in this sample are associated with \textit{L$_\mathrm{bol}$}~>~1~L$_{\odot}$, or $^{34}$SO and SO$_{2}$ are hard to detect in sources with low \textit{L$_\mathrm{bol}$}. This is consistent with the results from \cite{Artur2019a}, where they studied a sample of 10 Class 0/I sources and detected SO$_{2}$ emission only in those sources with \textit{L$_\mathrm{bol}$}~$\geq$~1.4~L$_{\odot}$. More sensitive observations of sources with low \textit{L$_\mathrm{bol}$} (<~1~L$_{\odot}$) are needed to test these detection trends.

\subsection{Column densities}

\subsubsection{Optically thick lines}

The CS line shows absorption features in most of the spectra, therefore, the CS integrated fluxes and column densities represent lower limits. For both SO transitions, it is not obvious if they are optically thin or optically thick. Therefore, when the SO and $^{34}$SO lines are detected, the molecular column density of both SO lines are compared to the $^{34}$SO column density (for \textit{T$_\mathrm{ex}$} between 50 and 150~K). The SO/$^{34}$SO ratio should approach a value of 22 \citep[$^{32}$S/$^{34}$S abundance ratio;][]{Wilson1999} if the SO line is optically thin. Figure~\ref{fig:Opt} shows the molecular column density ratios between the two SO transitions and $^{34}$SO, revealing that both SO transitions are considerably optically thick in those sources where $^{34}$SO is also detected. For this reason, future calculations of the SO column density will be performed from the $^{34}$SO integrated flux and later on multiplied by 22. It is important to notice that the $^{34}$SO emission could also be optically thick and, in this case, SO column densities should be considered as lower limits and the SO/$^{34}$SO ratios as upper limits. Unfortunately, we cannot assess the optical thickness of the $^{34}$SO 5$_{6}$--4$_{5}$ transition with the current data set.

% FIGURE 6
\begin{figure*}%[H]
        \centering
        \includegraphics[width=.74\textwidth]{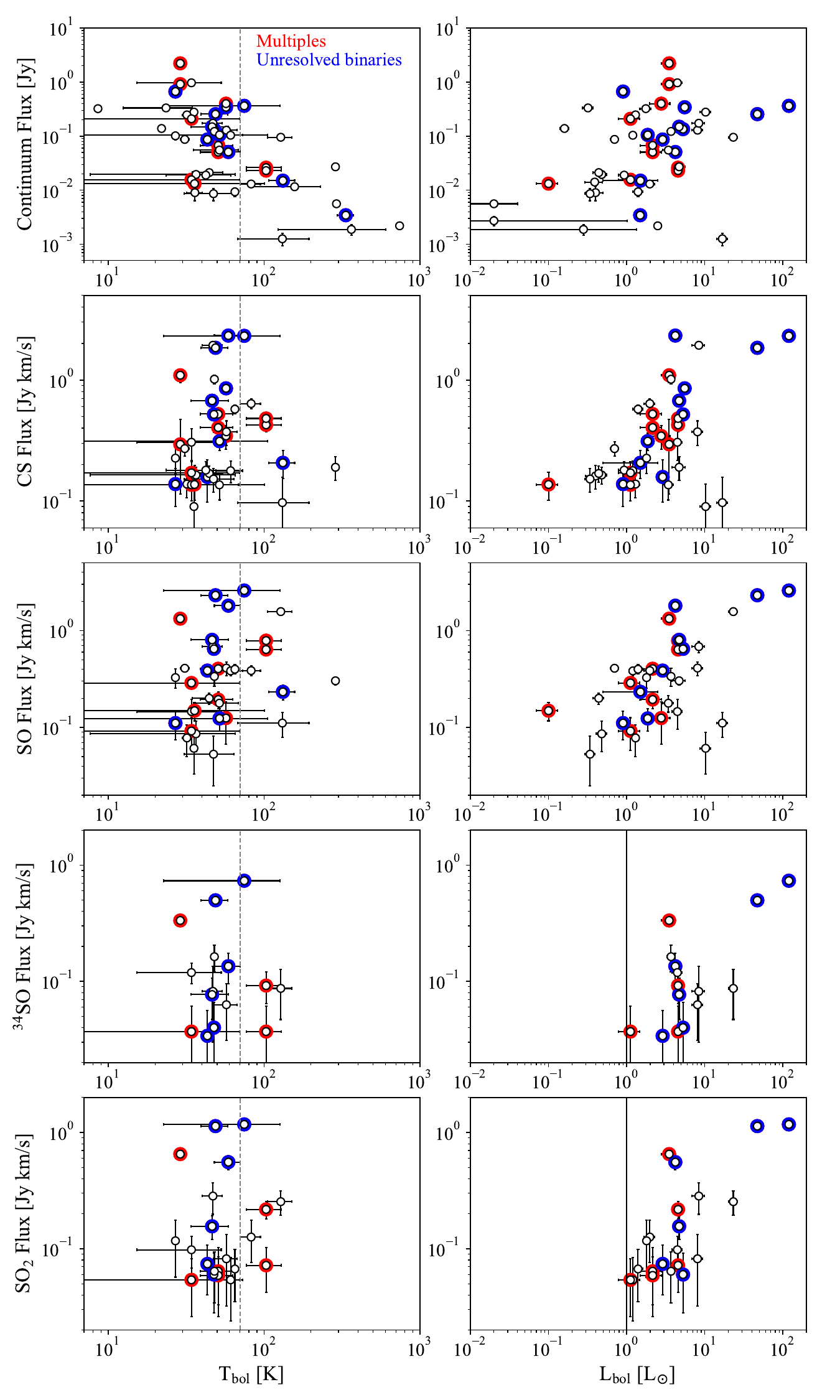}
        \caption[]{\label{fig:fluxes}
        Continuum and molecular fluxes (above 3$\sigma$) as a function of \textit{T$_\mathrm{bol}$} (\textit{left}) and \textit{L$_\mathrm{bol}$} (\textit{right}). The gray dashed line represents \textit{T$_\mathrm{bol}$}~=~70, the separation between Class 0 and Class I sources, while the black solid line in the last two panels of the right column indicates \textit{L$_\mathrm{bol}$}~=~1~L$_{\odot}$. Red circles represent multiple systems (binaries or triples) where the continuum ALMA emission is resolved but they do not have individual \textit{T$_\mathrm{bol}$} and \textit{L$_\mathrm{bol}$} measurements from far IR. Blue circles indicate unresolved binaries in continuum ALMA emission. 
        }
\end{figure*}

Per-emb 13 is the source with the highest detection of COMs \citep{Yang2021}, followed by Per-emb 12 B, Per-emb 27, Per-emb 29, and Per-emb 44. These five sources have at least 14 COMs detected (see Table~\ref{table:COM_S}) and they are associated with the lowest SO/$^{34}$SO ratios (see Fig.~\ref{fig:Opt}), consistent with the most optically thick SO emission. In this sense, the SO/$^{34}$SO ratio might be a good tracer of the inner high-density envelope, and a low value ($\leq$1.5) seems to be related with the detection of multiple COMs.

% FIGURE 7
\begin{figure}[t!]
        \centering
        \includegraphics[width=.49\textwidth]{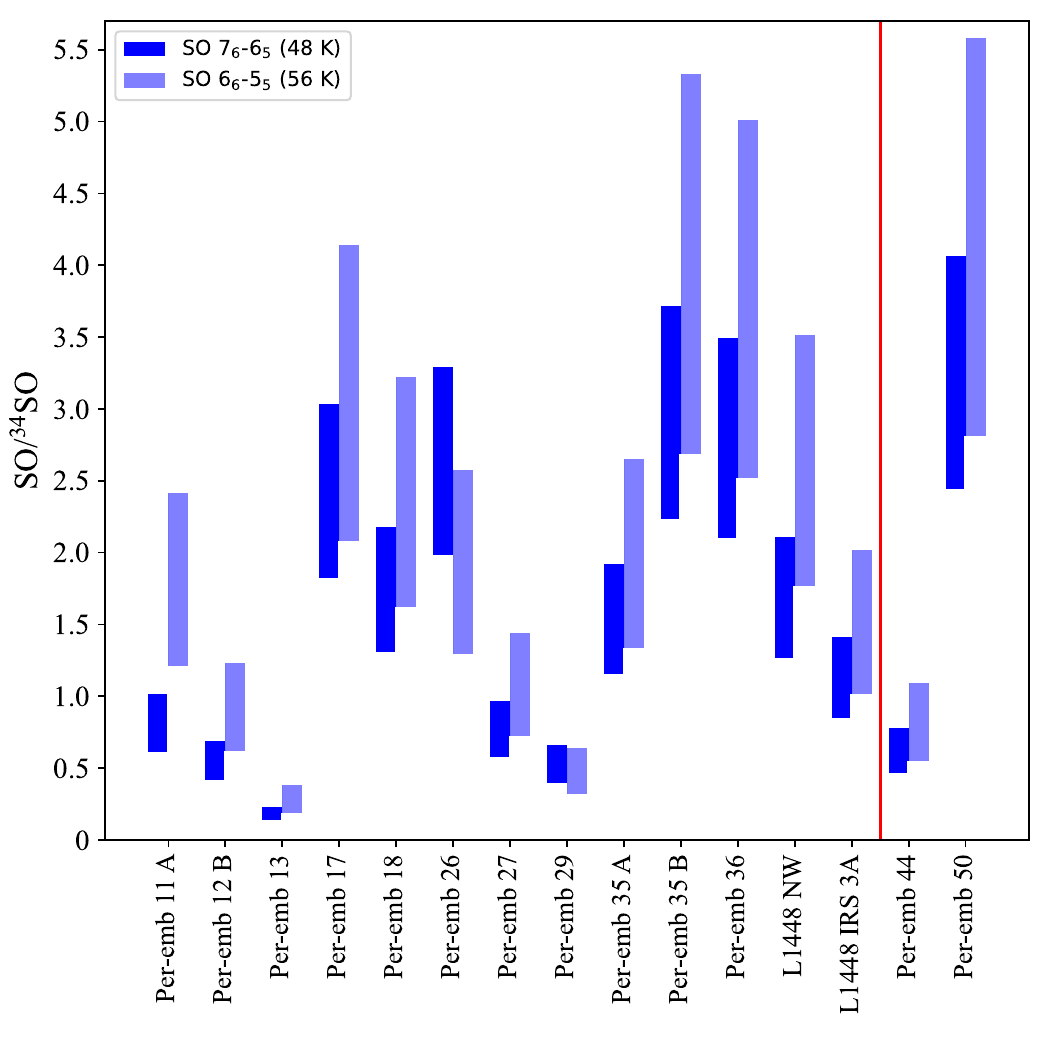}
        \caption[]{\label{fig:Opt}
        Molecular column density ratio of SO over $^{34}$SO, showing that both SO transitions are optically thick. A range of \textit{T$_\mathrm{ex}$}, between 50 and 150~K has been employed for the calculations. Dark blue and light blue bars indicate ratios calculated from SO 7$_{6}$--6$_{5}$ and SO 6$_{6}$--5$_{5}$, respectively. The red vertical line separates Class 0 (left) from Class I sources (right). 
        }
\end{figure}

% FIGURE 8
\begin{figure*}[h!]
        \centering
        \includegraphics[width=.8\textwidth]{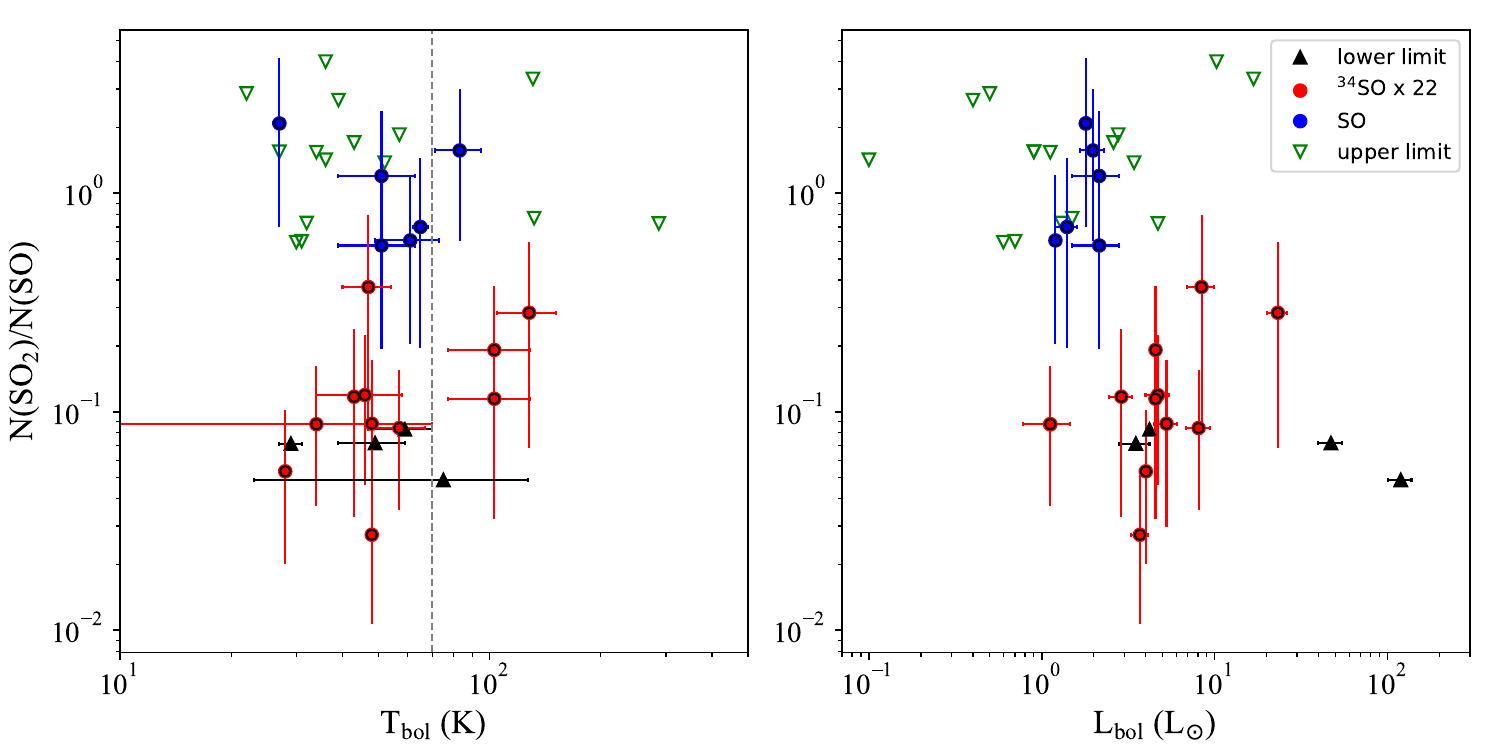}
        \caption[]{\label{fig:SO2_SO}
        Column density ratio between SO$_{2}$ and SO as a function of \textit{T$_\mathrm{bol}$} (\textit{left}) and \textit{L$_\mathrm{bol}$} (\textit{right}). Red dots are values calculated from $^{34}$SO fluxes and later on multiplied by 22 \citep{Wilson1999}, while blue dots indicate those sources where $^{34}$SO is not detected and the ratios are calculated from the SO 6$_{6}$--5$_{5}$ transition. Black triangles represent lower limits (optically thick SO$_{2}$ emission) and green triangles indicate upper limits, given by those sources where SO emission is detected but SO$_{2}$ is not. The gray dashed line in the left panel represents \textit{T$_\mathrm{bol}$}~=~70, the separation between Class 0 and Class I sources. Error bars in the y-axis represent ranges for \textit{T$_\mathrm{ex}$} between 50 and 150~K.
                }
\end{figure*}

% FIGURE 9
\begin{figure*}[h!]
        \centering
        \includegraphics[width=.85\textwidth]{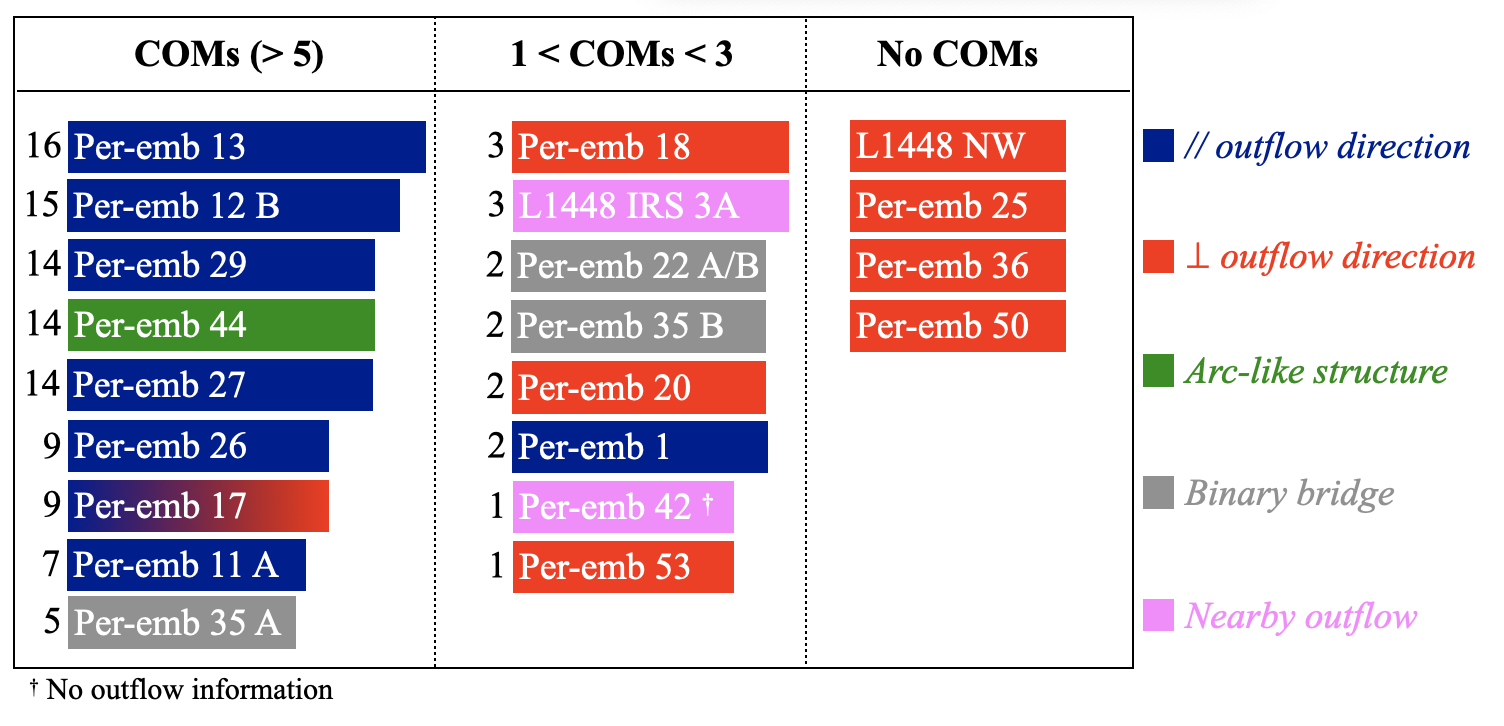}
        \caption[]{\label{fig:COMs}
        S-bearing species vs. COMs detection. Sources with detection of SO$_{2}$ are clustered by the amount of detected COMs in \cite{Yang2021}. The numbers on the left of the source names indicate the amount of different COMs detected. The different colors indicate where the emission of CS or SO is seen. Those sources where the CS or SO emission is parallel to the outflow direction are usually rich in COMs, while sources where the CS or SO emission is perpendicular to the outflow direction do not show detection of COMs.
                }
\end{figure*}

% FIGURE 10
\begin{figure}[h!]
        \centering
        \includegraphics[width=.45\textwidth]{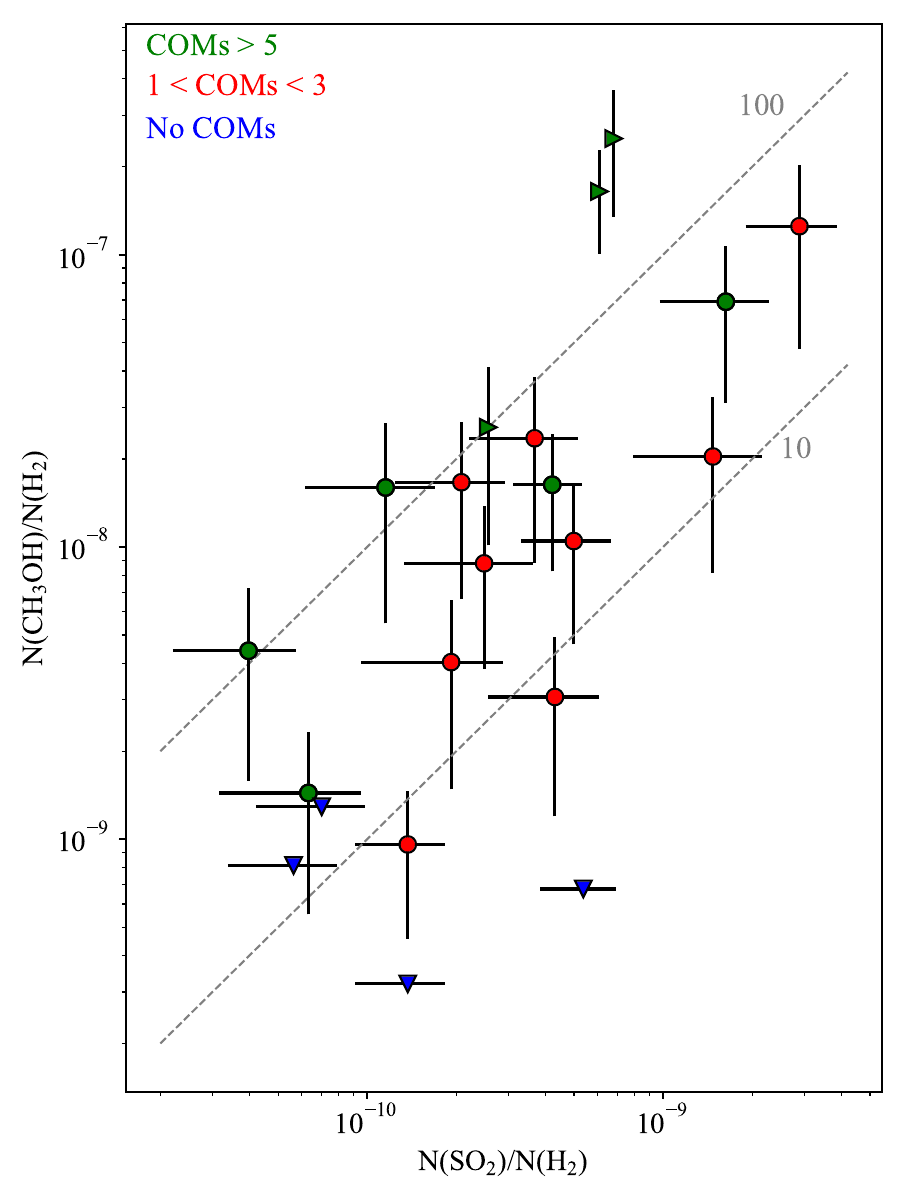}
        \caption[]{\label{fig:SO2_CH3OH}
        CH$_{3}$OH abundance as a function of the SO$_{2}$ abundance for the sources in Fig.~\ref{fig:COMs}. The different colors represent those sources where more than five different COMs were detected (green), between one and three different COMs were detected (red), and no COMs (blue) were detected in \cite{Yang2021}. Gray dashed lines show a 10:1 and 100:1 linear correlation.
        }
\end{figure}

% FIGURE 11
\begin{figure*}[h!]
        \centering
        \includegraphics[width=.95\textwidth]{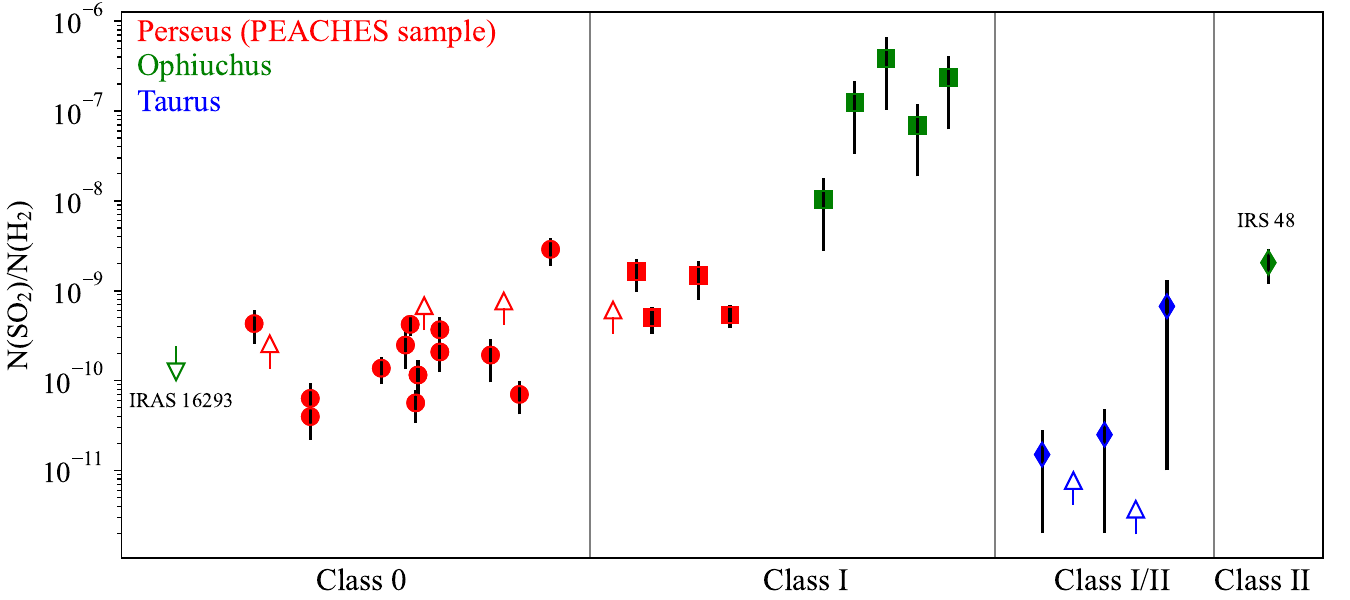}
        \caption[]{\label{fig:Ntot}
        Column density ratio between SO$_{2}$ and H$_{2}$ for sources in different evolutionary stages. Red, green, and blue symbols represent sources from Perseus, Ophiuchus, and Taurus, respectively. Dots indicate Class 0 sources, squares represent Class I sources, and diamonds show Class I/II and Class II sources. Empty triangles indicate upper or lower limits. Values for IRAS 16295 were taken from \cite{Drozdovskaya2018} and \cite{Jorgensen2016}. Perseus sources are from this work, Ophiuchus Class I sources correspond to IRS 67, GSS30-IRS1, IRS 44, IRS 43, and Elias 29, from left to right, studied by \cite{Artur2019a}, and Taurus sources are taken from \cite{Garufi2022} and represent DG Tau (outer disk), HL Tau (inner disk), HL Tau (outer disk), IRAS 04302, and T Tau S, from left to right. The upper limit for IRAS 16293 is given by the optically thick continuum emission and lower limits for Perseus and Taurus sources are given by optically thick SO$_{2}$ emission. The PEACHES sample and Class I Ophiuchus sources are ordered by \textit{T$_\mathrm{bol}$}.
        }
\end{figure*}

\subsubsection{Optical thickness of the SO$_{2}$ transition}

Given that the PEACHES survey did not target any $^{34}$SO$_{2}$ transition, the same analysis as the one carried out for SO cannot be replicated for SO$_{2}$. The SO$_{2}$ emission could, therefore, be optically thick. To assess the optical depth of the SO$_{2}$ 14$_{0,14}$--13$_{1,13}$ transition, we employed RADEX \citep{vanderTak2007} and generated models for H$_{2}$ number densities (\textit{n$_\mathrm{H}$}) between 10$^{3}$ and 10$^{9}$~cm$^{-3}$ and three different excitation temperatures: 50, 100, and 150~K. A broadening parameter (\textit{b}) of 2~km~s$^{-1}$ was employed for the calculations. RADEX results are presented in Fig.~\ref{fig:radex} in the appendix and  compared with calculated SO$_{2}$ column densities (see Sect. 3.5.2 and Table~\ref{table:fluxes}). For a value of \textit{N$_\mathrm{SO_{2}}$} below 10$^{15}$~cm$^{-2}$, the SO$_{2}$ emission could be interpreted as optically thin emission (assuming that the emission fills the beam uniformly) and most of the sources lie within this range. It should be noticed that the SO$_{2}$ emission might originate in a smaller region compared to the beam size and, in this case, the emission could be optically thick. Higher angular resolution observations would be needed to assess the optical depth of the SO$_{2}$ emission at smaller scales. On the other hand, there are four sources, Per-emb 12A, Per-emb 17, Per-emb 27, and Per-emb 44, associated with lower limits of \textit{N$_\mathrm{SO_{2}}$} ($\geq$~2~$\times$~10$^{15}$~cm$^{-2}$); therefore, they have high probabilities for the SO$_{2}$ emission to be optically thick (see Fig.~\ref{fig:radex}).

\subsubsection{\textit{N$_\mathrm{SO_{2}}$}/\textit{N$_\mathrm{SO}$}}

Assuming that the SO$_{2}$ transition is optically thin in most of the sources, the column density ratio between SO$_{2}$ and SO is shown in Fig.~\ref{fig:SO2_SO} as a function of \textit{T$_\mathrm{bol}$} and \textit{L$_\mathrm{bol}$}. No clear trend is seen for \textit{N$_\mathrm{SO_{2}}$}/\textit{N$_\mathrm{SO}$} as a function of \textit{T$_\mathrm{bol}$}. On the other hand, \textit{N$_\mathrm{SO_{2}}$}/\textit{N$_\mathrm{SO}$} shows a tentative increasing trend for \textit{L$_\mathrm{bol}$} between 1 and 10~L$_{\odot}$ (red dots). Those sources where SO emission is detected but $^{34}$SO emission is not, are shown with blue dots and represent upper limits, given that the SO emission could be optically thick. For \textit{L$_\mathrm{bol}$} higher than 10~L$_{\odot}$, a tentative decreasing trend is seen for the \textit{N$_\mathrm{SO_{2}}$}/\textit{N$_\mathrm{SO}$} ratio, however, caution must be taken given that the SO$_{2}$ emission is optically thick toward the most luminous sources. As seen in Fig.~\ref{fig:fluxes}, both SO and $^{34}$SO, as well as SO$_{2}$ fluxes increase as a function of \textit{L$_\mathrm{bol}$}; however, the UV and X-ray radiation field of the protostar could efficiently photo-dissociate SO$_{2}$ to form SO and decreases the SO$_{2}$/SO ratio for higher luminosities \citep{Booth2021, vanGelder2021}. Future observations of SO, $^{34}$SO, SO$_{2}$, and $^{34}$SO$_{2}$ toward other sources, within a broad range of bolometric luminosities, are needed to assess the \textit{N$_\mathrm{SO_{2}}$}/\textit{N$_\mathrm{SO}$} ratio for \textit{L$_\mathrm{bol}$} higher than 10~L$_{\odot}$ and confirm (or otherwise) whether SO$_{2}$ is being efficiently destroyed in those luminous sources.

\subsubsection{\textit{N$_\mathrm{H_{2}}$}}
To calculate molecular abundances, the column density of molecular hydrogen (\textit{N$_\mathrm{H_{2}}$}) is estimated from the continuum peak flux and following Equation (A.27) from \cite{Kauffmann2008}, as follows:

% EQUATION 4
\begin{equation} 
\begin{split}
    N_\mathrm{H_{2}} = 2.02 \times 10^{20} \mathrm{cm^{-2}} \left( \mathrm{e}^{1.439 \left( \lambda /\mathrm{mm} \right)^{-1} \left( T_\mathrm{dust} /10 \ \mathrm{K} \right)^{-1}} - 1 \right) \left( \frac{\lambda}{\mathrm{mm}} \right)^{3} \\
    \cdot \left( \frac{\kappa_{\nu}}{0.01 \ \mathrm{cm}^{2} \mathrm{g}^{-1}} \right)^{-1} \left( \frac{F_{\nu}^{\mathrm{beam}}}{\mathrm{mJy} \ \mathrm{beam}^{-1}} \right) \left( \frac{\theta_{\mathrm{HPBW}}}{10 \ \mathrm{arcsec}} \right)^{-2} , \
    \label{eq:NH2}
\end{split}
\end{equation}

\noindent where $\lambda$ is the wavelength, \textit{T$_\mathrm{dust}$} is the dust temperature, $\kappa_{\nu}$ is the dust opacity, \textit{F$_{\nu}^{\mathrm{beam}}$} is the observed flux per beam, and $\theta_\mathrm{HPBW}$ corresponds to the beam size. For $\lambda$~=~1.2~mm, we employed $\kappa_{\nu}$~=~0.01056~cm$^{2}$~g$^{-1}$ \citep[for a gas-to-dust radio of 100; ][]{Ossenkopf1994}, and a dust temperature of 30~K \citep[found to be appropriate for Class 0 sources; ][]{Dunham2014b}. Values for \textit{F$_{\nu}^{\mathrm{beam}}$} and $\theta_\mathrm{HPBW}$ are listed in Tables~\ref{table:observations} and \ref{table:integration}, respectively. Column densities of H$_{2}$ are listed in Table~\ref{table:column_density} and they will be used in the following sections to infer the abundances of SO$_{2}$ and CH$_{3}$OH.

\subsection{S-bearing species versus COMs detection}

In the study by \cite{Yang2021}, the authors present the detection (or non-detection) of COMs toward the PEACHES sample. From a total of 16 COMs, the most frequently detected ones were CH$_{3}$OH and CH$_{3}$CN. The rest are NH$_{2}$CHO, C$_{2}$H$_{5}$CN, CH$_{2}$DCN, HCOCH$_{2}$OH, CH$_{3}$CHO, C$_{2}$H$_{5}$OH, CH$_{3}$COCH$_{3}$, CH$_{3}$OCH$_{3}$, CH$_{3}$OCHO \textit{v}~=~1, CH$_{3}$OCHO, \textit{t}-HCOOH, $^{13}$CH$_{3}$OH, CH$_{3}$$^{18}$OH, and CH$_{2}$DOH. We note that none of these COMs contain sulfur.

The molecule SO$_{2}$ has been associated with warm chemistry, given that its desorption temperature is above 60~K \citep{vanGelder2021}, and it is efficiently formed in the gas phase for temperatures between 100 and 200~K \citep{Charnley1997}. Therefore, some correlation between the detection of COMs and SO$_{2}$ would be expected. To test this correlation, we separated those sources for which SO$_{2}$ emission is detected into three groups, following the detection rate of COMs from \citep{Yang2021}: (\textit{i}) sources very rich in COMs (at least five different COMs); (\textit{ii}) sources with only a few COMs detected, between one and three; and (\textit{iii}) sources with no COMs detected. These three groups are presented in Fig.~\ref{fig:COMs}, where the origin of the emission from the sulfur-bearing species is also indicated. 

Most of the sources that show CS and SO emission that is consistent with the outflow component are rich in COMs, with the exception of Per-emb 44, which shows an arc-like structure, Per-emb 35 A, where the emission seems to connect the binary system, and Per-emb 17, whose CS emission is seen in both directions: parallel and perpendicular to the outflow direction. The only source rich in COMs, where SO$_{2}$ (in addition to SO and $^{34}$SO) has not been detected, is B1-b S. Nine different COMs were detected toward this source \citep{Yang2021} and only CS absorption is seen in this dataset. Nevertheless, the non detection of $^{34}$SO and SO$_{2}$ is consistent with the low \textit{L$_\mathrm{bol}$} value of B1-b S (0.32~L$_{\odot}$). The detection of COMs in the gas-phase seems to be related with the presence of collimated outflows, where a high column density of warm gas is present. This is consistent with the results from \cite{Yang2021} and \cite{vanGelder2020}, where they found that the detectability of COMs depends neither on the continuum brightness temperature, nor the bolometric luminosity and the bolometric temperature. The detection of SO$_{2}$ around the source position, together with CS and SO emission tracing high collimated outflows, could be an important indicator of the presence of multiple COMs in the gas-phase. 

When only a few COMs have been detected (mainly CH$_{3}$OH and CH$_{3}$CN), the emission of S-bearing species is seen toward multiple components. The emission could be perpendicular to the outflow direction, tracing the material between resolved binaries, consistent with the outflow component and, for two sources L1448 IRS 3A and Per-emb 42, it is not clear if the emission is related with those sources or it is being affected by nearby outflows (see Fig.~\ref{fig:per12_large} in the Appendix). 

For those sources with SO$_{2}$ detection and CS and SO emission perpendicular to the outflow direction (L1448 NW, Per-emb 25, Per-emb 36, and Per-emb 50), no COMs have been detected (see Fig.~\ref{fig:COMs}). These sources might be associated with wider outflow cavity walls and, consequently, with a lower column density of warm gas, or with infalling material from the envelope to the disk-protostar system. It is important to notice that, at disk scales, the outflow cavity walls could show a flattened structure in the disk direction, given that these regions very close to the protostar belong to the base of the outflow structure. The two scenarios, infalling material or wider cavity walls, require a more detailed dynamic study and it is beyond the scope of this paper. The SO$_{2}$ emission toward these sources with no detection of COMS could be originating in regions that are exposed to UV radiation from the protostar, in disk winds \citep{Tabone2017}, shocked regions in the inner regions of the outflow cavity walls \citep{Podio2015}, or accretion shocks at the disk-envelope interface \citep{Artur2022, Zhang2023}. Observations with higher angular resolution and a more detailed study of these four scenarios are needed to assess the origin of the SO$_{2}$ emission.

Per-emb 50, one of the sources where COMs have not been detected, has been associated with an infalling streamer by \cite{Valdivia2022}. In addition, \cite{Zhang2023} proposed that the $^{34}$SO and SO$_{2}$ emission toward Per-emb 50 is related with shocks at the centrifugal barrier ($\sim$0$\farcs$4), which is located beyond the radius of the disk ($\sim$0$\farcs$1) derived from dust continuum maps \citep{Agurto2019}. Apart from Per-emb 50, there are another three sources (Per-emb 2, Per-emb 27, and Per-emb 44) that have been associated with infalling streamers. \cite{Pineda2020} proposed that large-scale accretion flows funnel material down to disk scales in Per-emb 2 by observing molecular transitions from HC$_{3}$N, CCS, and $^{13}$CS. The binary system Per-emb 27 has been associated with accretion streamers caused by the interaction of the binaries by \cite{Jorgensen2022}, and \cite{Hsieh2023} proposed that the system Per-emb 44 might be fed by an infalling streamer from envelope scales ($\sim$700 au). Toward Per-emb 44, the arc-like structure that we observe in SO is consistent with the streamer suggested by \cite{Hsieh2023}. 

From this data set, the sources that show CS and SO emission perpendicular to the outflow direction (Per-emb 18, Per-emb 20, Per-emb 25, Per-emb 36, Per-emb 53, and L1448 NW) seem to be the perfect candidates for studying their large-scale structure in more detail and to search for shocks or disk winds at disk scales.

Figure~\ref{fig:SO2_CH3OH} shows the CH$_{3}$OH abundance as a function of the SO$_{2}$ abundance, for those sources where SO$_{2}$ has been detected. The CH$_{3}$OH column densities were taken from \cite{Yang2021}, and blue triangles show the upper limits, given that CH$_{3}$OH is not detected toward these sources. As seen in Fig.~\ref{fig:SO2_CH3OH}, the abundance of CH$_{3}$OH is not correlated with the detectability of multiple COMs: both groups (detected COMs between one and three and more than five COMs detected) present CH$_{3}$OH abundances between 10$^{-9}$ and 10$^{-7}$. Nevertheless, a positive trend is seen between the abundances of SO$_{2}$ and CH$_{3}$OH, and CH$_{3}$OH is between 10 and 100 times more abundant than SO$_{2}$. This positive trend may suggest that both species, SO$_{2}$ and CH$_{3}$OH, are coming from warm gas and, for those sources where COMs are not detected, the SO$_{2}$ emission might originate from a different physical process, such as shocked regions or disk winds. Values for SO$_{2}$ and CH$_{3}$OH abundances are listed in Table~\ref{table:column_density}.

\subsubsection{SO$_{2}$ detection in other sources}

SO$_{2}$ emission has been detected in a handful of other sources, including the Class 0 sources IRAS 16293-2422 \citep{Jorgensen2016, Drozdovskaya2018}, a sample of five Class I sources in the Ophiuchus star forming region \citep{Artur2019a}, in different components of four Class I/II sources in Taurus \citep{Garufi2022}, and toward the Class II disk IRS 48 \citep{Booth2021}. The ratio between \textit{N$_\mathrm{SO_{2}}$} and \textit{N$_\mathrm{H_{2}}$} for these sources and for the PEACHES sample is shown in Fig.~\ref{fig:Ntot}, highlighting the different star-forming regions and evolutionary classes. For the PEACHES sample, an excitation temperature (\textit{T$_\mathrm{ex}$}) between 50 and 150~K has been employed for the \textit{N$_\mathrm{SO_{2}}$} calculations and the region of integration corresponds to the beam size ($\sim$0$\farcs$5 or 150~au). Assumptions and calculations made with values from the literature are presented in the Appendix F and Table~\ref{table:SO2_H2_lit}.

The SO$_{2}$ abundance of the PEACHES sample seems to increase from Class 0 to Class I sources, probably due to a decrease in the continuum flux (and therefore \textit{N$_\mathrm{H_{2}}$}) as the system evolves. Most of the Ophiuchus Class I sources show that the SO$_{2}$ abundance is between one and two orders of magnitude higher than the PEACHES Class 0/I sample. One of the Ophiuchus Class I sources, IRS 44, has been associated with the presence of accretion shocks, from the analysis of multiple SO$_{2}$ and $^{34}$SO$_{2}$ transitions \citep{Artur2022}. IRS 48, also located in Ophiuchus, shows higher values than less evolved Class I/II sources in Taurus. IRS 48 is the only Class II disk where prominent SO$_{2}$ has been detected so far and \cite{Booth2021} proposed that the SO$_{2}$ emission is originating from the sublimation of ices at the dust cavity wall, observed as an asymmetric dust trap \citep{vanderMarel2013}. Therefore, the high SO$_{2}$ column densities toward the Ophiuchus Class I sources and toward IRS 48 seem to be related to the sublimation of SO$_{2}$ from the dust grains, either through accretion shocks or the exposure to the UV radiation from the protostar. Intermediate Class I/II sources in Taurus show a large spread in SO$_{2}$ abundances, mainly given by two lower limits and T Tau S, a source with bright SO$_{2}$ emission \citep[$\sim$60~mJy~beam$^{-1}$~km~s$^{-1}$; ][]{Garufi2022} and weak continuum emission \citep[$\sim$10~mJy~beam$^{-1}$; ][]{Garufi2021}. Finally, the hot-corino source IRAS 16293 presents an upper limit in the SO$_{2}$ abundance, given by the optically thick continuum, and the fluxes were taken in an offset position \citep{Jorgensen2016, Drozdovskaya2018}.

A possibility for the observed SO$_{2}$ abundances in Fig.~\ref{fig:Ntot} is that the Ophiuchus star-forming region could be enriched in sulfur-bearing species, in comparison with Perseus (the PEACHES sample) and Taurus (Class I/II sources). In this scenario, the sulfur enrichment will be present in the solid phase and, when some physical mechanism heats the dust grains, S-bearing species get desorbed from the dust to the gas-phase. The heating mechanism (outflows, accretion shocks, or exposure to the protostellar UV radiation) is essential, given that there are other Class I sources in Ophiuchus (associated with low \textit{L$_\mathrm{bol}$}) where SO$_{2}$ is not detected \citep{Artur2019a}. This interpretation is consistent with resent results from \cite{Fuente2023}, where they studied the sulfur elemental abundance in Taurus, Perseus, and Orion A, finding that Taurus and Perseus are more depleted in sulfur than Orion A. They suggest that dust grains are negatively charged in regions where S$^{+}$ is abundant toward Taurus and Perseus, promoting the adsorption of S$^{+}$ to the grain surface. With this adsorption of ions, \cite{Laas2019} and \cite{Shingledecker2020} predicted that organo-sulfur compounds and S$_{8}$ could be an important sulfur reservoir in the solid phase. 

The main difference between Taurus/Perseus and Orion/Ophiuchus is the UV intensity, \textit{G$_\mathrm{0}$}.  This value is $\sim$30 toward Perseus and $\leq$100 in Taurus, while it could reach values from 100 to 1000 in Orion A and Ophiuchus \citep{Xia2022}. Given that Ophiuchus is more similar to Orion A in terms of UV intensity, it could be possible that Ophiuchus is also undepleted in sulfur, supporting its high SO$_{2}$ column density observed in Fig.~\ref{fig:Ntot}. Future observations of SO$_{2}$ transitions in a representative number of sources located in different star-forming regions would clarify if the Ophiuchus molecular cloud is, indeed, particularly enriched in sulfur-bearing species. In addition, solid-phase observations could demonstrate if the sulfur reservoir varies between star-forming regions that are exposed to different UV radiation fields.

The linear trend between the SO$_{2}$ and CH$_{3}$OH abundances shown in Fig.~\ref{fig:SO2_CH3OH} suggests that the SO$_{2}$ emission toward most of the Class 0 sources in Fig.~\ref{fig:Ntot} arises from the high column density of warm material. In the Ophiuchus Class I sources, CH$_{3}$OH emission has not been detected \citep{Artur2019a}, and SO$_{2}$ emission toward these sources and more evolved Class I/II might originate in accretion shocks. Finally, the more evolved Class II disk IRS 48 presents SO$_{2}$ emission that originates from a dust trap exposed to the UV radiation of the protostar. Therefore, the SO$_{2}$ emission detected in different evolutionary stages seems to arise from different physical mechanisms.

\section{Summary}

This work presents high-resolution ($\sim$0$\farcs$6, 180~au) ALMA observations of a sample of 50 Class 0/I sources located in the Perseus star-forming region. A detailed study of the sulfur-bearing species (CS, SO, $^{34}$SO, and SO$_{2}$) has been carried out and the main results are summarized below:

\begin{itemize}

\item Moment 0 maps show that the CS emission traces extended structures and follows the outflow direction in most of the cases. The SO emission is also seen toward extended structures, but an important percentage of sources ($\sim$40$\%$) show compact emission, while the majority of the $^{34}$SO and SO$_{2}$ detections are compact.  

\item A comparison of the detection rates between Class 0 and Class I sources indicates that the CS and SO lines are sensitive to the evolutionary state of the source and, therefore, to the amount of mass present in the envelope. On the other hand, the $^{34}$SO and SO$_{2}$ transitions seem to be more sensitive to the bolometric luminosity, which can be linked to the mass accretion rate.

\item The $^{34}$SO and SO$_{2}$ transitions are only detected in those sources with bolometric luminosities above 1~L$_{\odot}$, which could be a real finding or a selection bias, given that 70$\%$ of the sources have  \textit{L$_\mathrm{bol}$}~$\geq$~1~L$_{\odot}$.

\item When $^{34}$SO is detected, the ratio SO/$^{34}$SO is lower than the canonical value of 22 in all the cases, suggesting that the SO emission is optically thick toward these sources. The lowest values ($\leq$~1.5) are associated with those sources rich in COMs, where at least 14 different COMs were detected. Therefore, the SO/$^{34}$SO ratio seems to be a good tracer of the inner high-density envelope and it could be used in the future to infer the presence of multiple COMs. 

\item The column density ratio between SO$_{2}$ and SO shows a positive trend for sources with \textit{L$_\mathrm{bol}$}~$\leq$~10~L$_{\odot}$ and a tentative decreasing trend for sources with \textit{L$_\mathrm{bol}$}~$\geq$~10~L$_{\odot}$, possibly related to the photo-dissociation of SO$_{2}$ and enhancement of SO in those very luminous sources.

\item The comparison between the morphology of S-bearing species and the detectability of COMs suggests that the detection of COMs is linked to the presence of high collimated outflows. The detection of SO$_{2}$ around the source position, together with CS and SO emission tracing high collimated outflows, could be an important indicator of the presence of multiple COMs in the gas-phase. 

\item When the emission of CS and SO is seen perpendicular to the outflow direction, no COMs (or only a few of them) are detected. In these cases, the CS and SO emission could be tracing wider outflow cavity walls or infalling material, and higher angular resolution observations are needed to asses the origin of the SO$_{2}$ emission: UV irradiated regions, disk winds, accretion shocks at the disk-envelope interface, or shocked regions in the outflow cavity walls. 

\item Per-emb 18, Per-emb 20, Per-emb 25, Per-emb 36, Per-emb 53, and L1448 NW seem to be ideal candidates to investigate in more detail their large scale structures and to search for the origin of the SO$_{2}$ emission close to the protostar. 

\item In those sources where SO$_{2}$ and CH$_{3}$OH are detected, a positive trend is seen between the CH$_{3}$OH and the SO$_{2}$ abundance, suggesting that both species are tracing warm material in those sources. The CH$_{3}$OH abundance is between 10 and 100 times higher than the SO$_{2}$ abundance and sources with multiple COMs detection do not show any trend with the CH$_{3}$OH abundance.

\item When SO$_{2}$ is detected, the SO$_{2}$ abundances toward the sources in Perseus seem to be comparable to that in Taurus sources and lower than Ophiuchus sources. The depletion of sulfur seems to depend on the external UV radiation field of the molecular cloud and Perseus would be more depleted in sulfur than Ophiuchus. 

\item The SO$_{2}$ emission toward young Class 0 sources seem to be associated with a high column density of warm material, while accretion shocks might desorb SO$_{2}$ molecules to the gas-phase in Class I and Class I/II. Finally, the exposure to the protostellar UV radiation is the preferred mechanism to explain the SO$_{2}$ in more evolved Class II disks. 

\end{itemize}
 
This paper shows the power of a large survey with multiple targeted molecular transition, making it possible to compare the emission of sulfur-bearing species between them and with the detection of COMs. Future observations of sulfur-bearing species toward Class 0/I sources in other star-forming regions are essential to confirm the observed values and trends in detection rates, SO$_{2}$ column densities and abundances, ratios between SO$_{2}$ and SO, and the possible connection with the detection of COMs.

\begin{acknowledgements}

We thank the anonymous referee for a number of good suggestions that helped us to improve this work. This paper makes use of the following ALMA data: ADS/JAO.ALMA$\#$2016.0.00391.S. ALMA is a partnership of ESO (representing its member states), NSF (USA) and NINS (Japan), together with NRC (Canada), MOST and ASIAA (Taiwan), and KASI (Republic of Korea), in cooperation with the Republic of Chile. The Joint ALMA Observatory is operated by ESO, AUI/NRAO and NAOJ. The National Radio Astronomy Observatory is a facility of the National Science Foundation operated under cooperative agreement by Associated Universities, Inc. E.A.dlV. acknowledges financial support provided by FONDECYT grant 3200797. V.V.G. gratefully acknowledges support from FONDECYT Regular 1221352, ANID BASAL projects ACE210002 and FB210003, and ANID, -- Millennium Science Initiative Program -- NCN19\_171. Y.-L.Y. acknowledges support from a Grant-in-Aid from the Ministry of Education, Culture, Sports, Science, and Technology of Japan (22K20389). Y.-L.Y. and N.S. acknowledge support from a Grant-in-Aid from the Ministry of Education, Culture, Sports, Science, and Technology of Japan (20H05845, 20H05844), and a pioneering project in RIKEN (Evolution of Matter in the Universe).

\end{acknowledgements}

\bibliographystyle{aa} 
\bibliography{References}

\begin{thebibliography}{67}
\expandafter\ifx\csname natexlab\endcsname\relax\def\natexlab#1{#1}\fi

\bibitem[{{Agurto-Gangas} {et~al.}(2019){Agurto-Gangas}, {Pineda},
  {Sz{\H{u}}cs}, {Testi}, {Tazzari}, {Miotello}, {Caselli}, {Dunham},
  {Stephens}, \& {Bourke}}]{Agurto2019}
{Agurto-Gangas}, C., {Pineda}, J.~E., {Sz{\H{u}}cs}, L., {et~al.} 2019, \aap,
  623, A147

\bibitem[{{Artur de la Villarmois} {et~al.}(2022){Artur de la Villarmois},
  {Guzm{\'a}n}, {J{\o}rgensen}, {Kristensen}, {Bergin}, {Harsono}, {Sakai},
  {van Dishoeck}, \& {Yamamoto}}]{Artur2022}
{Artur de la Villarmois}, E., {Guzm{\'a}n}, V.~V., {J{\o}rgensen}, J.~K.,
  {et~al.} 2022, \aap, 667, A20

\bibitem[{{Artur de la Villarmois} {et~al.}(2019){Artur de la Villarmois},
  {J{\o}rgensen}, {Kristensen}, {Bergin}, {Harsono}, {Sakai}, {van Dishoeck},
  \& {Yamamoto}}]{Artur2019a}
{Artur de la Villarmois}, E., {J{\o}rgensen}, J.~K., {Kristensen}, L.~E.,
  {et~al.} 2019, \aap, 626, A71

\bibitem[{{Balan{\c c}a} {et~al.}(2016){Balan{\c c}a}, {Spielfiedel}, \&
  {Feautrier}}]{Balanca2016}
{Balan{\c c}a}, C., {Spielfiedel}, A., \& {Feautrier}, N. 2016, \mnras, 460,
  3766

\bibitem[{{Boogert} {et~al.}(1997){Boogert}, {Schutte}, {Helmich}, {Tielens},
  \& {Wooden}}]{Boogert1997}
{Boogert}, A.~C.~A., {Schutte}, W.~A., {Helmich}, F.~P., {Tielens},
  A.~G.~G.~M., \& {Wooden}, D.~H. 1997, \aap, 317, 929

\bibitem[{{Booth} {et~al.}(2023){Booth}, {Ilee}, {Walsh}, {Kama}, {Keyte}, {van
  Dishoeck}, \& {Nomura}}]{Booth2023}
{Booth}, A.~S., {Ilee}, J.~D., {Walsh}, C., {et~al.} 2023, \aap, 669, A53

\bibitem[{{Booth} {et~al.}(2021){Booth}, {van der Marel}, {Leemker}, {van
  Dishoeck}, \& {Ohashi}}]{Booth2021}
{Booth}, A.~S., {van der Marel}, N., {Leemker}, M., {van Dishoeck}, E.~F., \&
  {Ohashi}, S. 2021, \aap, 651, L6

\bibitem[{{Calmonte} {et~al.}(2016){Calmonte}, {Altwegg}, {Balsiger},
  {Berthelier}, {Bieler}, {Cessateur}, {Dhooghe}, {van Dishoeck}, {Fiethe},
  {Fuselier}, {Gasc}, {Gombosi}, {H{\"a}ssig}, {Le Roy}, {Rubin}, {S{\'e}mon},
  {Tzou}, \& {Wampfler}}]{Calmonte2016}
{Calmonte}, U., {Altwegg}, K., {Balsiger}, H., {et~al.} 2016, \mnras, 462, S253

\bibitem[{{Cazaux} {et~al.}(2022){Cazaux}, {Carrascosa}, {Mu{\~n}oz Caro},
  {Caselli}, {Fuente}, {Navarro-Almaida}, \&
  {Rivi{\'e}re-Marichalar}}]{Cazaux2022}
{Cazaux}, S., {Carrascosa}, H., {Mu{\~n}oz Caro}, G.~M., {et~al.} 2022, \aap,
  657, A100

\bibitem[{{Charnley}(1997)}]{Charnley1997}
{Charnley}, S.~B. 1997, \apj, 481, 396

\bibitem[{{Cridland} {et~al.}(2022){Cridland}, {Rosotti}, {Tabone},
  {Tychoniec}, {McClure}, {Nazari}, \& {van Dishoeck}}]{Cridland2022}
{Cridland}, A.~J., {Rosotti}, G.~P., {Tabone}, B., {et~al.} 2022, \aap, 662,
  A90

\bibitem[{{De Simone} {et~al.}(2020){De Simone}, {Ceccarelli}, {Codella},
  {Svoboda}, {Chandler}, {Bouvier}, {Yamamoto}, {Sakai}, {Caselli}, {Favre},
  {Loinard}, {Lefloch}, {Liu}, {L{\'o}pez-Sepulcre}, {Pineda}, {Taquet}, \&
  {Testi}}]{DeSimone2020}
{De Simone}, M., {Ceccarelli}, C., {Codella}, C., {et~al.} 2020, \apjl, 896, L3

\bibitem[{{Drozdovskaya} {et~al.}(2018){Drozdovskaya}, {van Dishoeck},
  {J{\o}rgensen}, {Calmonte}, {van der Wiel}, {Coutens}, {Calcutt},
  {M{\"u}ller}, {Bjerkeli}, {Persson}, {Wampfler}, \&
  {Altwegg}}]{Drozdovskaya2018}
{Drozdovskaya}, M.~N., {van Dishoeck}, E.~F., {J{\o}rgensen}, J.~K., {et~al.}
  2018, \mnras, 476, 4949

\bibitem[{{Dunham} {et~al.}(2014){Dunham}, {Vorobyov}, \& {Arce}}]{Dunham2014b}
{Dunham}, M.~M., {Vorobyov}, E.~I., \& {Arce}, H.~G. 2014, \mnras, 444, 887

\bibitem[{{Endres} {et~al.}(2016){Endres}, {Schlemmer}, {Schilke}, {Stutzki},
  \& {M{\"u}ller}}]{Endres2016}
{Endres}, C.~P., {Schlemmer}, S., {Schilke}, P., {Stutzki}, J., \&
  {M{\"u}ller}, H. S.~P. 2016, Journal of Molecular Spectroscopy, 327, 95

\bibitem[{{Fuente} {et~al.}(2023){Fuente}, {Rivi{\`e}re-Marichalar},
  {Beitia-Antero}, {Caselli}, {Wakelam}, {Esplugues}, {Rodr{\'\i}guez-Baras},
  {Navarro-Almaida}, {Gerin}, {Kramer}, {Bachiller}, {Goicoechea},
  {Jim{\'e}nez-Serra}, {Loison}, {Ivlev}, {Mart{\'\i}n-Dom{\'e}nech},
  {Spezzano}, {Roncero}, {Mu{\~n}oz-Caro}, {Cazaux}, \&
  {Marcelino}}]{Fuente2023}
{Fuente}, A., {Rivi{\`e}re-Marichalar}, P., {Beitia-Antero}, L., {et~al.} 2023,
  \aap, 670, A114

\bibitem[{{Garufi} {et~al.}(2021){Garufi}, {Podio}, {Codella}, {Fedele},
  {Bianchi}, {Favre}, {Bacciotti}, {Ceccarelli}, {Mercimek}, {Rygl}, {Teague},
  \& {Testi}}]{Garufi2021}
{Garufi}, A., {Podio}, L., {Codella}, C., {et~al.} 2021, \aap, 645, A145

\bibitem[{{Garufi} {et~al.}(2022){Garufi}, {Podio}, {Codella}, {Segura-Cox},
  {Vander Donckt}, {Mercimek}, {Bacciotti}, {Fedele}, {Kasper}, {Pineda},
  {Humphreys}, \& {Testi}}]{Garufi2022}
{Garufi}, A., {Podio}, L., {Codella}, C., {et~al.} 2022, \aap, 658, A104

\bibitem[{{Geballe} {et~al.}(1985){Geballe}, {Baas}, {Greenberg}, \&
  {Schutte}}]{Geballe1985}
{Geballe}, T.~R., {Baas}, F., {Greenberg}, J.~M., \& {Schutte}, W. 1985, \aap,
  146, L6

\bibitem[{{Harsono} {et~al.}(2014){Harsono}, {J{\o}rgensen}, {van Dishoeck},
  {Hogerheijde}, {Bruderer}, {Persson}, \& {Mottram}}]{Harsono2014}
{Harsono}, D., {J{\o}rgensen}, J.~K., {van Dishoeck}, E.~F., {et~al.} 2014,
  \aap, 562, A77

\bibitem[{{Hsieh} {et~al.}(2023){Hsieh}, {Segura-Cox}, {Pineda}, {Caselli},
  {Bouscasse}, {Neri}, {Lopez-Sepulcre}, {Valdivia-Mena}, {Maureira},
  {Henning}, {Smirnov-Pinchukov}, {Semenov}, {M{\"o}ller}, {Cunningham},
  {Fuente}, {Marino}, {Dutrey}, {Tafalla}, {Chapillon}, {Ceccarelli}, \&
  {Zhao}}]{Hsieh2023}
{Hsieh}, T.~H., {Segura-Cox}, D.~M., {Pineda}, J.~E., {et~al.} 2023, \aap, 669,
  A137

\bibitem[{{Huang} {et~al.}(2023){Huang}, {Bergin}, {Bae}, {Benisty}, \&
  {Andrews}}]{Huang2023}
{Huang}, J., {Bergin}, E.~A., {Bae}, J., {Benisty}, M., \& {Andrews}, S.~M.
  2023, \apj, 943, 107

\bibitem[{{Jim{\'e}nez-Escobar} \& {Mu{\~n}oz Caro}(2011)}]{Jimenez2011}
{Jim{\'e}nez-Escobar}, A. \& {Mu{\~n}oz Caro}, G.~M. 2011, \aap, 536, A91

\bibitem[{{J{\o}rgensen} {et~al.}(2022){J{\o}rgensen}, {Kuruwita}, {Harsono},
  {Haugb{\o}lle}, {Kristensen}, \& {Bergin}}]{Jorgensen2022}
{J{\o}rgensen}, J.~K., {Kuruwita}, R.~L., {Harsono}, D., {et~al.} 2022, \nat,
  606, 272

\bibitem[{{J{\o}rgensen} {et~al.}(2016){J{\o}rgensen}, {van der Wiel},
  {Coutens}, {Lykke}, {M{\"u}ller}, {van Dishoeck}, {Calcutt}, {Bjerkeli},
  {Bourke}, {Drozdovskaya}, {Favre}, {Fayolle}, {Garrod}, {Jacobsen},
  {{\"O}berg}, {Persson}, \& {Wampfler}}]{Jorgensen2016}
{J{\o}rgensen}, J.~K., {van der Wiel}, M.~H.~D., {Coutens}, A., {et~al.} 2016,
  \aap, 595, A117

\bibitem[{{Kama} {et~al.}(2019){Kama}, {Shorttle}, {Jermyn}, {Folsom},
  {Furuya}, {Bergin}, {Walsh}, \& {Keller}}]{Kama2019}
{Kama}, M., {Shorttle}, O., {Jermyn}, A.~S., {et~al.} 2019, \apj, 885, 114

\bibitem[{{Kauffmann} {et~al.}(2008){Kauffmann}, {Bertoldi}, {Bourke}, {Evans},
  \& {Lee}}]{Kauffmann2008}
{Kauffmann}, J., {Bertoldi}, F., {Bourke}, T.~L., {Evans}, II, N.~J., \& {Lee},
  C.~W. 2008, \aap, 487, 993

\bibitem[{{Laas} \& {Caselli}(2019)}]{Laas2019}
{Laas}, J.~C. \& {Caselli}, P. 2019, \aap, 624, A108

\bibitem[{{Le Gal} {et~al.}(2019){Le Gal}, {{\"O}berg}, {Loomis}, {Pegues}, \&
  {Bergner}}]{LeGal2019}
{Le Gal}, R., {{\"O}berg}, K.~I., {Loomis}, R.~A., {Pegues}, J., \& {Bergner},
  J.~B. 2019, \apj, 876, 72

\bibitem[{{McClure} {et~al.}(2023){McClure}, {Rocha}, {Pontoppidan}, {Crouzet},
  {Chu}, {Dartois}, {Lamberts}, {Noble}, {Pendleton}, {Perotti}, {Qasim},
  {Rachid}, {Smith}, {Sun}, {Beck}, {Boogert}, {Brown}, {Caselli}, {Charnley},
  {Cuppen}, {Dickinson}, {Drozdovskaya}, {Egami}, {Erkal}, {Fraser}, {Garrod},
  {Harsono}, {Ioppolo}, {Jim{\'e}nez-Serra}, {Jin}, {J{\o}rgensen},
  {Kristensen}, {Lis}, {McCoustra}, {McGuire}, {Melnick}, {{\~A}-berg},
  {Palumbo}, {Shimonishi}, {Sturm}, {van Dishoeck}, \&
  {Linnartz}}]{McClure2023}
{McClure}, M.~K., {Rocha}, W.~R.~M., {Pontoppidan}, K.~M., {et~al.} 2023,
  Nature Astronomy, 7, 431

\bibitem[{{M{\"u}ller} {et~al.}(2005){M{\"u}ller}, {Schl{\"o}der}, {Stutzki},
  \& {Winnewisser}}]{Muller2005}
{M{\"u}ller}, H. S.~P., {Schl{\"o}der}, F., {Stutzki}, J., \& {Winnewisser}, G.
  2005, Journal of Molecular Structure, 742, 215

\bibitem[{{M{\"u}ller} {et~al.}(2001){M{\"u}ller}, {Thorwirth}, {Roth}, \&
  {Winnewisser}}]{Muller2001}
{M{\"u}ller}, H.~S.~P., {Thorwirth}, S., {Roth}, D.~A., \& {Winnewisser}, G.
  2001, \aap, 370, L49

\bibitem[{{Murillo} {et~al.}(2016){Murillo}, {van Dishoeck}, {Tobin}, \&
  {Fedele}}]{Murillo2016}
{Murillo}, N.~M., {van Dishoeck}, E.~F., {Tobin}, J.~J., \& {Fedele}, D. 2016,
  \aap, 592, A56

\bibitem[{{Ossenkopf} \& {Henning}(1994)}]{Ossenkopf1994}
{Ossenkopf}, V. \& {Henning}, T. 1994, \aap, 291, 943

\bibitem[{{Oya} {et~al.}(2019){Oya}, {L{\'o}pez-Sepulcre}, {Sakai}, {Watanabe},
  {Higuchi}, {Hirota}, {Aikawa}, {Sakai}, {Ceccarelli}, {Lefloch}, {Caux},
  {Vastel}, {Kahane}, \& {Yamamoto}}]{Oya2019}
{Oya}, Y., {L{\'o}pez-Sepulcre}, A., {Sakai}, N., {et~al.} 2019, \apj, 881, 112

\bibitem[{{Palumbo} {et~al.}(1995){Palumbo}, {Tielens}, \&
  {Tokunaga}}]{Palumbo1995}
{Palumbo}, M.~E., {Tielens}, A.~G.~G.~M., \& {Tokunaga}, A.~T. 1995, \apj, 449,
  674

\bibitem[{{Phuong} {et~al.}(2018){Phuong}, {Chapillon}, {Majumdar}, {Dutrey},
  {Guilloteau}, {Pi{\'e}tu}, {Wakelam}, {Diep}, {Tang}, {Beck}, \&
  {Bary}}]{Phuong2018}
{Phuong}, N.~T., {Chapillon}, E., {Majumdar}, L., {et~al.} 2018, \aap, 616, L5

\bibitem[{{Phuong} {et~al.}(2021){Phuong}, {Dutrey}, {Chapillon}, {Guilloteau},
  {Bary}, {Beck}, {Coutens}, {Denis-Alpizar}, {Di Folco}, {Diep}, {Majumdar},
  {Melisse}, {Lee}, {Pietu}, {Stoecklin}, \& {Tang}}]{Phuong2021}
{Phuong}, N.~T., {Dutrey}, A., {Chapillon}, E., {et~al.} 2021, \aap, 653, L5

\bibitem[{{Pineda} {et~al.}(2020){Pineda}, {Segura-Cox}, {Caselli},
  {Cunningham}, {Zhao}, {Schmiedeke}, {Maureira}, \& {Neri}}]{Pineda2020}
{Pineda}, J.~E., {Segura-Cox}, D., {Caselli}, P., {et~al.} 2020, Nature
  Astronomy, 4, 1158

\bibitem[{{Podio} {et~al.}(2015){Podio}, {Codella}, {Gueth}, {Cabrit},
  {Bachiller}, {Gusdorf}, {Lee}, {Lefloch}, {Leurini}, {Nisini}, \&
  {Tafalla}}]{Podio2015}
{Podio}, L., {Codella}, C., {Gueth}, F., {et~al.} 2015, \aap, 581, A85

\bibitem[{{Pudritz} {et~al.}(2007){Pudritz}, {Ouyed}, {Fendt}, \&
  {Brandenburg}}]{Pudritz2007}
{Pudritz}, R.~E., {Ouyed}, R., {Fendt}, C., \& {Brandenburg}, A. 2007,
  Protostars and Planets V, 277

\bibitem[{{Reipurth} \& {Bally}(2001)}]{Reipurth2001}
{Reipurth}, B. \& {Bally}, J. 2001, ARAA, 39, 403

\bibitem[{{Rivi{\`e}re-Marichalar} {et~al.}(2022){Rivi{\`e}re-Marichalar},
  {Fuente}, {Esplugues}, {Wakelam}, {le Gal}, {Baruteau}, {Ribas},
  {Mac{\'\i}as}, {Neri}, \& {Navarro-Almaida}}]{Riviere2022}
{Rivi{\`e}re-Marichalar}, P., {Fuente}, A., {Esplugues}, G., {et~al.} 2022,
  \aap, 665, A61

\bibitem[{{Rivi{\`e}re-Marichalar} {et~al.}(2019){Rivi{\`e}re-Marichalar},
  {Fuente}, {Goicoechea}, {Pety}, {Le Gal}, {Gratier}, {Guzm{\'a}n}, {Roueff},
  {Loison}, {Wakelam}, \& {Gerin}}]{Riviere2019}
{Rivi{\`e}re-Marichalar}, P., {Fuente}, A., {Goicoechea}, J.~R., {et~al.} 2019,
  \aap, 628, A16

\bibitem[{{Rivi{\`e}re-Marichalar} {et~al.}(2021){Rivi{\`e}re-Marichalar},
  {Fuente}, {Le Gal}, {Arabhavi}, {Cazaux}, {Navarro-Almaida}, {Ribas},
  {Mendigut{\'\i}a}, {Barrado}, \& {Montesinos}}]{Riviere2021}
{Rivi{\`e}re-Marichalar}, P., {Fuente}, A., {Le Gal}, R., {et~al.} 2021, \aap,
  652, A46

\bibitem[{{Rivi{\`e}re-Marichalar} {et~al.}(2020){Rivi{\`e}re-Marichalar},
  {Fuente}, {Le Gal}, {Baruteau}, {Neri}, {Navarro-Almaida},
  {Trevi{\~n}o-Morales}, {Mac{\'\i}as}, {Bachiller}, \& {Osorio}}]{Riviere2020}
{Rivi{\`e}re-Marichalar}, P., {Fuente}, A., {Le Gal}, R., {et~al.} 2020, \aap,
  642, A32

\bibitem[{{Ruffle} {et~al.}(1999){Ruffle}, {Hartquist}, {Caselli}, \&
  {Williams}}]{Ruffle1999}
{Ruffle}, D.~P., {Hartquist}, T.~W., {Caselli}, P., \& {Williams}, D.~A. 1999,
  \mnras, 306, 691

\bibitem[{{Sakai} {et~al.}(2014){Sakai}, {Sakai}, {Hirota}, {Watanabe},
  {Ceccarelli}, {Kahane}, {Bottinelli}, {Caux}, {Demyk}, {Vastel}, {Coutens},
  {Taquet}, {Ohashi}, {Takakuwa}, {Yen}, {Aikawa}, \& {Yamamoto}}]{Sakai2014}
{Sakai}, N., {Sakai}, T., {Hirota}, T., {et~al.} 2014, \nat, 507, 78

\bibitem[{{Shingledecker} {et~al.}(2020){Shingledecker}, {Lamberts}, {Laas},
  {Vasyunin}, {Herbst}, {K{\"a}stner}, \& {Caselli}}]{Shingledecker2020}
{Shingledecker}, C.~N., {Lamberts}, T., {Laas}, J.~C., {et~al.} 2020, \apj,
  888, 52

\bibitem[{{Tabone} {et~al.}(2017){Tabone}, {Cabrit}, {Bianchi}, {Ferreira},
  {Pineau des For{\^e}ts}, {Codella}, {Gusdorf}, {Gueth}, {Podio}, \&
  {Chapillon}}]{Tabone2017}
{Tabone}, B., {Cabrit}, S., {Bianchi}, E., {et~al.} 2017, \aap, 607, L6

\bibitem[{{Tobin} {et~al.}(2016){Tobin}, {Looney}, {Li}, {Chandler}, {Dunham},
  {Segura-Cox}, {Sadavoy}, {Melis}, {Harris}, {Kratter}, \&
  {Perez}}]{Tobin2016a}
{Tobin}, J.~J., {Looney}, L.~W., {Li}, Z.-Y., {et~al.} 2016, \apj, 818, 73

\bibitem[{{Tobin} {et~al.}(2018){Tobin}, {Looney}, {Li}, {Sadavoy}, {Dunham},
  {Segura-Cox}, {Kratter}, {Chandler}, {Melis}, {Harris}, \&
  {Perez}}]{Tobin2018}
{Tobin}, J.~J., {Looney}, L.~W., {Li}, Z.-Y., {et~al.} 2018, \apj, 867, 43

\bibitem[{{Tychoniec} {et~al.}(2020){Tychoniec}, {Manara}, {Rosotti}, {van
  Dishoeck}, {Cridland}, {Hsieh}, {Murillo}, {Segura-Cox}, {van Terwisga}, \&
  {Tobin}}]{Tychoniec2020}
{Tychoniec}, {\L}., {Manara}, C.~F., {Rosotti}, G.~P., {et~al.} 2020, \aap,
  640, A19

\bibitem[{{Tychoniec} {et~al.}(2021){Tychoniec}, {van Dishoeck}, {van't Hoff},
  {van Gelder}, {Tabone}, {Chen}, {Harsono}, {Hull}, {Hogerheijde}, {Murillo},
  \& {Tobin}}]{Tychoniec2021}
{Tychoniec}, {\L}., {van Dishoeck}, E.~F., {van't Hoff}, M. L.~R., {et~al.}
  2021, \aap, 655, A65

\bibitem[{{Valdivia-Mena} {et~al.}(2022){Valdivia-Mena}, {Pineda},
  {Segura-Cox}, {Caselli}, {Neri}, {L{\'o}pez-Sepulcre}, {Cunningham},
  {Bouscasse}, {Semenov}, {Henning}, {Pi{\'e}tu}, {Chapillon}, {Dutrey},
  {Fuente}, {Guilloteau}, {Hsieh}, {Jim{\'e}nez-Serra}, {Marino}, {Maureira},
  {Smirnov-Pinchukov}, {Tafalla}, \& {Zhao}}]{Valdivia2022}
{Valdivia-Mena}, M.~T., {Pineda}, J.~E., {Segura-Cox}, D.~M., {et~al.} 2022,
  \aap, 667, A12

\bibitem[{{van der Marel} {et~al.}(2021){van der Marel}, {Bosman}, {Krijt},
  {Mulders}, \& {Bergner}}]{vanderMarel2021}
{van der Marel}, N., {Bosman}, A.~D., {Krijt}, S., {Mulders}, G.~D., \&
  {Bergner}, J.~B. 2021, \aap, 653, L9

\bibitem[{{van der Marel} {et~al.}(2013){van der Marel}, {Kristensen},
  {Visser}, {Mottram}, {Y{\i}ld{\i}z}, \& {van Dishoeck}}]{vanderMarel2013}
{van der Marel}, N., {Kristensen}, L.~E., {Visser}, R., {et~al.} 2013, \aap,
  556, A76

\bibitem[{{van der Tak} {et~al.}(2007){van der Tak}, {Black}, {Sch{\"o}ier},
  {Jansen}, \& {van Dishoeck}}]{vanderTak2007}
{van der Tak}, F.~F.~S., {Black}, J.~H., {Sch{\"o}ier}, F.~L., {Jansen}, D.~J.,
  \& {van Dishoeck}, E.~F. 2007, \aap, 468, 627

\bibitem[{{van Gelder} {et~al.}(2020){van Gelder}, {Tabone}, {Tychoniec}, {van
  Dishoeck}, {Beuther}, {Boogert}, {Caratti o Garatti}, {Klaassen}, {Linnartz},
  {M{\"u}ller}, \& {Taquet}}]{vanGelder2020}
{van Gelder}, M.~L., {Tabone}, B., {Tychoniec}, {\L}., {et~al.} 2020, \aap,
  639, A87

\bibitem[{{van Gelder} {et~al.}(2021){van Gelder}, {Tabone}, {van Dishoeck}, \&
  {Godard}}]{vanGelder2021}
{van Gelder}, M.~L., {Tabone}, B., {van Dishoeck}, E.~F., \& {Godard}, B. 2021,
  \aap, 653, A159

\bibitem[{{Wilson}(1999)}]{Wilson1999}
{Wilson}, T.~L. 1999, Reports on Progress in Physics, 62, 143

\bibitem[{{Xia} {et~al.}(2022){Xia}, {Tang}, {Zhi}, {Jiao}, {Xie}, {Fuller},
  {Goldsmith}, \& {Li}}]{Xia2022}
{Xia}, J., {Tang}, N., {Zhi}, Q., {et~al.} 2022, Research in Astronomy and
  Astrophysics, 22, 085017

\bibitem[{{Yang} {et~al.}(2022){Yang}, {Green}, {Pontoppidan}, {Bergner},
  {Cleeves}, {Evans}, {Garrod}, {Jin}, {Kim}, {Kim}, {Lee}, {Sakai},
  {Shingledecker}, {Shope}, {Tobin}, \& {van Dishoeck}}]{Yang2022}
{Yang}, Y.-L., {Green}, J.~D., {Pontoppidan}, K.~M., {et~al.} 2022, \apjl, 941,
  L13

\bibitem[{{Yang} {et~al.}(2021){Yang}, {Sakai}, {Zhang}, {Murillo}, {Zhang},
  {Higuchi}, {Zeng}, {L{\'o}pez-Sepulcre}, {Yamamoto}, {Lefloch}, {Bouvier},
  {Ceccarelli}, {Hirota}, {Imai}, {Oya}, {Sakai}, \& {Watanabe}}]{Yang2021}
{Yang}, Y.-L., {Sakai}, N., {Zhang}, Y., {et~al.} 2021, \apj, 910, 20

\bibitem[{{Zhang} {et~al.}(2018){Zhang}, {Higuchi}, {Sakai}, {Oya},
  {L{\'o}pez-Sepulcre}, {Imai}, {Sakai}, {Watanabe}, {Ceccarelli}, {Lefloch},
  \& {Yamamoto}}]{Zhang2018}
{Zhang}, Y., {Higuchi}, A.~E., {Sakai}, N., {et~al.} 2018, \apj, 864, 76

\bibitem[{{Zhang} {et~al.}(2023){Zhang}, {Yang}, {Zhang}, {Cox}, {Zeng},
  {Murillo}, {Ohashi}, \& {Sakai}}]{Zhang2023}
{Zhang}, Z.~E., {Yang}, Y.-l., {Zhang}, Y., {et~al.} 2023, \apj, 946, 113

\bibitem[{{Zucker} {et~al.}(2020){Zucker}, {Speagle}, {Schlafly}, {Green},
  {Finkbeiner}, {Goodman}, \& {Alves}}]{Zucker2020}
{Zucker}, C., {Speagle}, J.~S., {Schlafly}, E.~F., {et~al.} 2020, \aap, 633,
  A51

\end{thebibliography}

\begin{appendix}

\section{Coordinates and continuum fluxes}

Table~\ref{table:observations} lists the PEACHES sample, divided into Class 0 sources, Class I sources, and those sources where \textit{T$_\mathrm{bol}$} is unknown. A 2D Gaussian fit was employed to the continuum emission to estimate the coordinates and continuum fluxes. 

% TABLE 2
\begin{table*}[t!]
        \caption{Properties of the observed sources: results from a 2D Gaussian fit, bolometric temperatures, and bolometric luminosities.}
        \label{table:observations}
        \centering
        \begin{tabular}{l l l l c c c}
                \hline\hline
                Source                          	& Common names          	& R. A. (J2000)   	& Dec (J2000)   	& \textit{F$_\mathrm{1.2~mm}$}	& \textit{T$_\mathrm{bol}$} $^{(a)}$	& \textit{L$_\mathrm{bol}$} $^{(a)}$	\\
                                                        	&                                 	& (hh:mm:ss)           	& (dd:mm:ss)        	& (mJy~beam$^{-1}$)                     	& (K)                                             	& (L$_{\odot}$)                         		\\
                \hline
                \multicolumn{7}{c}{Class 0 (\textit{T$_\mathrm{bol}$}~$\leq$~70~K)}                                                                                                                                                                                                                     	\\              
                \hline
                Per-emb 1                         	& HH 211 MMS               	& 03:43:56.81       	& +32:00:50.2        	& 42.9~$\pm$~2.7            			& 27~$\pm$~1                              	& 1.80~$\pm$~0.10                       	\\
                Per-emb 2                       	& IRAS 03292+3039      	& 03:32:17.92      	& +30:49:47.8        	& 66.3~$\pm$~2.8                       	& 27~$\pm$~1                              	& 0.90~$\pm$~0.07                       	\\
                Per-emb 5                         	& IRAS 03282+3035         	& 03:31:20.94     	& +30:45:30.2        	& 138.7~$\pm$~1.7            		& 32~$\pm$~2                              	& 1.30~$\pm$~0.10                       	\\
                Per-emb 8                      	&                                       	& 03:44:43.98         	& +32:01:35.2        	& 79.9~$\pm$~1.7             		& 43~$\pm$~6                              	& 2.60~$\pm$~0.50                       	\\
                Per-emb 10                      	& B1-d                          	& 03:33:16.43     	& +31:06:52.0        	& 18.5~$\pm$~0.2                     	& 30~$\pm$~2                              	& 0.60~$\pm$~0.05                       	\\
                Per-emb 11 A $^{(b)}$	& IC 348 MMS                 	& 03:43:57.07         	& +32:03:04.8        	& 97.3~$\pm$~2.7                      	& 34~$\pm$~36                     		& 1.12~$\pm$~0.34                       	\\
                Per-emb 11 B $^{(b)}$  	& IC 348 MMS                 	& 03:43:56.88         	& +32:03:03.0        	& 4.2~$\pm$~0.4                          	& 34~$\pm$~36                    		& 1.12~$\pm$~0.34                       	\\
                Per-emb 11 C                    	& IC 348 MMS                 	& 03:43:57.70      	& +32:03:09.9        	& 4.0~$\pm$~0.4                         	& 36~$\pm$~65                     		& 0.10~$\pm$~0.03                       	\\
                Per-emb 12 A $^{(b)}$   	& NGC 1333 IRAS4 A2   	& 03:29:10.54      	& +31:13:30.9         	& 411~$\pm$~39 $^{(e)}$                 	& 29~$\pm$~2                            	& 3.50~$\pm$~0.70                 		\\
                Per-emb 12 B $^{(b)}$   	& NGC 1333 IRAS4 A1   	& 03:29:10.44        	& +31:13:32.1         	& 209~$\pm$~17                          	& 29~$\pm$~2                            	& 3.50~$\pm$~0.70                 		\\
                Per-emb 13                      	& NGC 1333 IRAS4 B1   	& 03:29:12.02      	& +31:13:08.0         	& 303.8~$\pm$~8.8                 		& 28~$\pm$~1                              	& 4.00~$\pm$~0.30                       	\\
                Per-emb 14                      	& NGC 1333 IRAS4 C      	& 03:29:13.55       	& +31:13:58.1         	& 66.4~$\pm$~0.7                         	& 31~$\pm$~2                              	& 0.70~$\pm$~0.08                       	\\
                Per-emb 15                      	&                                       	& 03:29:04.06        	& +31:14:46.2         	& 3.2~$\pm$~0.3                       	& 36~$\pm$~4                              	& 0.40~$\pm$~0.10                       	\\
                Per-emb 16                      	&                                       	& 03:43:50.98        	& +32:03:24.1         	& 3.3~$\pm$~0.3                          	& 39~$\pm$~2                              	& 0.40~$\pm$~0.04                       	\\
                Per-emb 17                      	& L1455 IRS1                  	& 03:27:39.11      	& +30:13:03.0         	& 32.9~$\pm$~0.9                     	& 59~$\pm$~11                     		& 4.20~$\pm$~0.10                       	\\
                Per-emb 18                      	& NGC 1333 IRAS7 SM1 	& 03:29:11.27		& +31:18:31.1         	& 70.2~$\pm$~1.1                   		& 46~$\pm$~12                     		& 4.70~$\pm$~0.72                       	\\
                Per-emb 20                      	& L1455 IRS4                   	& 03:27:43.28       	& +30:12:28.9         	& 2.6~$\pm$~0.2                         	& 65~$\pm$~3                              	& 1.40~$\pm$~0.20                       	\\
                Per-emb 21                      	& NGC 1333 IRAS7 SM2	& 03:29:10.67     	& +31:18:20.2         	& 44.2~$\pm$~1.1                     	& 52~$\pm$~17                    		& 3.42~$\pm$~0.53                       	\\
                Per-emb 22 A $^{(b)}$   	& L1448 IRS2              	& 03:25:22.41        	& +30:45:13.3         	& 30.3~$\pm$~1.3                      	& 51~$\pm$~12                     		& 2.15~$\pm$~0.66                       	\\
                Per-emb 22 B $^{(b)}$   	& L1448 IRS2              	& 03:25:22.37		& +30:45:13.2         	& 17.1~$\pm$~0.1                    		& 51~$\pm$~12                     		& 2.15~$\pm$~0.66                       	\\
                Per-emb 25                      	& IRAS 03235+3004       	& 03:26:37.51         	& +30:15:27.8         	& 89.9~$\pm$~0.2                     	& 61~$\pm$~12                     		& 1.20~$\pm$~0.02                       	\\
                Per-emb 26                      	& L1448-mm                  	& 03:25:38.88        	& +30:44:05.3         	& 137.5~$\pm$~2.7                   	& 47~$\pm$~7                              	& 8.40~$\pm$~1.50                       	\\
                Per-emb 27                      	& NGC 1333 IRAS 2A      	& 03:28:55.57        	& +31:14:37.0         	& 118.1~$\pm$~8.2                          	& 49~$\pm$~10                     		& 47.06~$\pm$~7.21                      	\\
                Per-emb 28                      	&                                      	& 03:43:51.01      	& +32:03:08.0         	& 13.8~$\pm$~0.4                       	& 45~$\pm$~2                              	& 0.70~$\pm$~0.08                       	\\
                Per-emb 29                      	& B1-c                          	& 03:33:17.88   	& +31:09:31.7         	& 73.4~$\pm$~3.6                     	& 48~$\pm$~1                              	& 3.70~$\pm$~0.40                       	\\
                Per-emb 33 A $^{(b)}$   	& L1448 IRS 3B          	& 03:25:36.38           	& +30:45:14.8        	& 156.0~$\pm$~8.8                        	& 57~$\pm$~3                              	& 2.77~$\pm$~0.80                       	\\
                Per-emb 33 B/C $^{(b)}$ 	& L1448 IRS 3B          	& 03:25:36.32           	& +30:45:15.0         	& 84.3~$\pm$~1.2                  		& 57~$\pm$~3                              	& 5.53~$\pm$~0.80                       	\\
                Per-emb 36                      	& NGC 1333 IRAS 2B     	& 03:28:57.38       	& +31:14:15.8          	& 118.9~$\pm$~0.6                    	& 48~$\pm$~1                              	& 5.27~$\pm$~0.81                       	\\
                Per-emb 37                      	&                                      	& 03:29:18.97   	& +31:23:14.3         	& 11.6~$\pm$~0.6                         	& 22~$\pm$~1                              	& 0.50~$\pm$~0.10                       	\\
                IRAS 4B2                   	& NGC 1333 IRAS4 B2   	& 03:29:12.84           	& +31:13:06.9          	& 146.2~$\pm$~3.9                        	& 9~$\pm$~1                             	& 1.76~$\pm$~0.28                 		\\
                B1-b N                          	&                                     	& 03:33:21.21        	& +31:07:43.6         	& 69.3~$\pm$~1.7                        	& 22~$\pm$~1                              	& 0.16~$\pm$~0.02                       	\\
                B1-b S                          	&                                     	& 03:33:21.36        	& +31:07:26.3         	& 134.1~$\pm$~1.4                     	& 24~$\pm$~11                    		& 0.32~$\pm$~0.05                       	\\
                L1448 IRS 2E $^{(c)}$   	&                                     	& 03:25:25.66     	& +30:44:56.7          	& --                                                    	& 15                                              	& 0.05~$\pm$~0.05                       	\\
                L1448 NW                        	& L1448 IRS 3C          	& 03:25:35.67        	& +30:45:34.2         	& 55.2~$\pm$~1.8                     	& 43~$\pm$~1                              	& 2.89~$\pm$~0.44                       	\\
                L1448 IRS 3A                    	&                                     	& 03:25:36.50         	& +30:45:21.9         	& 52.5~$\pm$~2.5                      	& 57~$\pm$~10                     		& 8.08~$\pm$~1.27                       	\\
                SVS 13 B                        	&                                     	& 03:29:03.08         	& +31:15:51.7         	& 135.5~$\pm$~5.2                         	& 36~$\pm$~1                              	& 10.26~$\pm$~1.57                      	\\
                \hline
                \multicolumn{7}{c}{Class I (\textit{T$_\mathrm{bol}$}~$\geq$~70~K)}                                                                                                                                                                                                     			\\              
                \hline
                Per-emb 35 A $^{(b)}$   	& NGC 1333 IRAS1      	& 03:28:37.10     	& +31:13:30.8         	& 20.6~$\pm$~0.6                       	& 103~$\pm$~26                    		& 4.55~$\pm$~0.30                      	\\
                Per-emb 35 B $^{(b)}$   	& NGC 1333 IRAS1 		& 03:28:37.22  		& +31:13:31.7         	& 16.1~$\pm$~0.7                   		& 103~$\pm$~26                    		& 4.55~$\pm$~0.30                       	\\
                Per-emb 40                      	& B1-a                          	& 03:33:16.67      	& +31:07:54.9         	& 13.7~$\pm$~0.2                      	& 132~$\pm$~25                    		& 1.50~$\pm$~1.00                       	\\
                Per-emb 41 $^{(d)}$        	& B1-b W                      	& 03:33:20.34      	& +31:07:21.3         	& 11.4~$\pm$~0.5                        	& 157~$\pm$~72                    		& 0.17~$\pm$~0.36                       	\\
                Per-emb 42                      	& L1448C-S                 	& 03:25:39.14         	& +30:43:57.9         	& 11.1~$\pm$~0.3                        	& 83~$\pm$~12                     		& 1.98~$\pm$~0.31                       	\\
                Per-emb 44                      	& NGC 1333 SVS 13A     	& 03:29:03.76       	& +31:16:03.7         	& 132.0~$\pm$~7.4                    	& 75~$\pm$~52                     		& 119.28~$\pm$~18.31            		\\
                Per-emb 50                      	& IRAS 03260+3111 A     	& 03:29:07.77         	& +31:21:57.1         	& 87.4~$\pm$~0.3                       	& 128~$\pm$~23                    		& 23.20~$\pm$~3.00                      	\\
                Per-emb 53                      	& B5 IRS1                    	& 03:47:41.59         	& +32:51:43.6         	& 14.8~$\pm$~0.2                      	& 287~$\pm$~8                     		& 4.70~$\pm$~0.90                       	\\
                Per-emb 54                      	& NGC 1333 IRAS6    	& 03:29:01.55         	& +31:20:20.5         	& 1.5~$\pm$~0.1                           	& 131~$\pm$~63                    		& 16.80~$\pm$~2.60                      	\\
                Per-emb 55                      	& IRAS 03415+3152       	& 03:44:43.30         	& +32:01:31.2         	& 3.4~$\pm$~0.1                          	& 334~$\pm$~340                   		& 1.49~$\pm$~0.25                       	\\
                Per-emb 60                      	&                                     	& 03:29:20.05         	& +31:24:07.4         	& 1.6~$\pm$~0.1                             	& 363~$\pm$~240                   		& 0.28~$\pm$~1.05                       	\\
                L1455 IRS 2                     	&                                     	& 03:27:47.69         	& +30:12:04.4         	& 2.1~$\pm$~0.1                         	& 740                                     		& 2.50~$\pm$~0.10                       	\\
                EDJ2009-172                   	&                                     	& 03:28:56.65         	& +31:18:35.4         	& 13.6~$\pm$~0.1                         	& 1100                                    		& 0.39~$\pm$~0.10                       	\\
                EDJ2009-235                  	&                                     	& 03:29:18.26          	& +31:23:19.7         	& 5.5~$\pm$~0.1                         	& 291~$\pm$~14                    		& 0.02~$\pm$~0.02                       	\\
                \hline
                \multicolumn{7}{c}{Unknown \textit{T$_\mathrm{bol}$}}                                                                                                                                                                                                                   				\\              
                \hline
                SVS 13 A2                       	& VLA3                          	& 03:29:03.39       	& +31:16:01.6         	& 13.9~$\pm$~0.8                        	& -                                               	& -                                             		\\
                EDJ2009-237                 	&                                   	& 03:29:18.74      	& +31:23:25.2         	& 2.5~$\pm$~0.2                          	& -                                               	& 0.02                                  		\\
                RAC1999 VLA20           	&                            		& 03:29:04.25         	& +31:16:09.1          	& 5.8~$\pm$~0.2                        	& -                                               	& -                                             		\\
                \hline\hline
        \end{tabular}
        \tablefoot{$^{(a)}$From \cite{Tobin2016a} and \cite{Murillo2016}. $^{(b)}$\textit{L$_\mathrm{bol}$} values divided by the amount of components (unresolved binaries and triple systems). $^{(c)}$No continuum detection. $^{(d)}$Edge of the field of view. $^{(e)}$Optically thick continuum emission.}
\end{table*}

\section{Integrated values, detection of S-bearing species, fluxes, and column densities}

Table~\ref{table:integration} lists the systemic velocity (\textit{v$_\mathrm{sys}$}), estimated from the CS spectra (see Fig.~\ref{fig:spectra}), the outflow position angle (PA; measured from north to east), and the integration areas and velocity ranges used to create moment 0 maps and to extract molecular fluxes, which are listed in Table~\ref{table:fluxes}. Later on, molecular fluxes were employed to calculate molecular column densities using \textit{T$_\mathrm{ex}$} between 50 and 150~K, and are presented in Table~\ref{table:column_density}. We note that the SO column density for those sources that have also detection of $^{34}$SO need to be considered as lower limits, given that they are optically thick (see Sect. 4.2.1).

Table~\ref{table:COM_S} lists the number of COMs detected from \cite{Yang2021} and the detection (or not) of sulfur-bearing species. Detection could refer to emission toward the source position (above 3$\sigma$) or to extended emission that seems to be related with the source.

% TABLE 3
\begin{table*}[t!]
        \caption{Multiplicity of the sources, outflow direction (for the blueshifted lobe), and integration area and ranges used for the calculation of fluxes.}
        \label{table:integration}
        \centering
        \begin{tabular}{l c c c c c}
                \hline\hline
                Source      			& Multiplicity $^{(a)}$           	& \textit{v$_\mathrm{sys}$} $^{(b)}$	& Outflow PA $^{(c)}$	& Int. area $^{(d)}$                        & Int. range 		\\
                                        		&                                               	& (km~s$^{-1}$)                           	& ($\degr$)                     	& ($\arcsec$)                             & (km~s$^{-1}$)      	\\
                \hline
                \multicolumn{6}{c}{Class 0 (\textit{T$_\mathrm{bol}$}~$\leq$~70~K)}                                                                                                                                                     		\\              
                \hline
                Per-emb 1        		&                                               	& 9.5                                          		& 120                           	& 0.49                                    & [3, 18]               		\\
                Per-emb 2             	& Unresolved binary          	& 7.7                                            	& 128                           	& 0.45                                    & [2, 12]               		\\
                Per-emb 5         		&                                               	& 7.0                                       		& 125                           	& 0.45                                    & [3, 10]               		\\
                Per-emb 8           	& Unresolved binary             	& 11.0                                            	& 314                           	& 0.50                                    & [6, 14]               		\\
                Per-emb 10              	&                                               	& 7.0                                  		& 230                           	& 0.46                                    & [4, 10]               		\\
                Per-emb 11 A            	&                                               	& 9.0                                        		& 340                           	& 0.50                                    & [5, 12]               		\\
                Per-emb 11 B          	&                                               	& 9.0                                               	& 165                           	& 0.50                                    & [5, 11]               		\\
                Per-emb 11 C          	&                                               	& 9.0                                              	& -                                     	& 0.50                                    & [6, 11]               		\\
                Per-emb 12              	& Binary                                  	& 7.5                                        		& 175                           	& 0.66                                    & [1, 13]               		\\
                Per-emb 13              	&                                               	& 8.0                                                   & 180                          	& 0.66                                    & [2, 12]               		\\
                Per-emb 14              	&                                               	& 8.0                                               	& 95                                	& 0.66                                    & [3, 11]               		\\
                Per-emb 15              	&                                               	& 5.5                                               	& 145                           	& 0.66                                    & [2, 7]                		\\
                Per-emb 16              	&                                               	& 9.0                                             	& 5                                 	& 0.50                                    & [5, 11]               		\\
                Per-emb 17              	& Unresolved binary               	& 6.0                                                	& 240                           	& 0.64                                    & [-4, 15]              		\\
                Per-emb 18              	& Unresolved binary             	& 8.5                                                	& 150                           	& 0.66                                    & [0, 15]               		\\
                Per-emb 20              	&                                               	& 5.0                                                	& 305                           	& 0.64                                    & [0, 10]               		\\
                Per-emb 21              	&                                               	& 9.0                                              	& 48                                 	& 0.66                                    & [5, 12]               		\\
                Per-emb 22              	& Binary                                  	& 4.5                                                 	& 306                           	& 0.64                                    & [0, 10]               		\\
                Per-emb 25              	&                                               	& 5.5                                               	& 280                           	& 0.64                                    & [2, 9]                		\\
                Per-emb 26              	&                                               	& 5.5                                              	& 340                           	& 0.64                                    & [-8, 20]              		\\
                Per-emb 27              	& Unresolved binary               	& 7.5                                           	& 204, 285                      	& 0.66                                    & [1, 12]               		\\
                Per-emb 28              	&                                               	& 9.0                                             	& 120                           	& 0.50                                    & [5, 11]               		\\
                Per-emb 29              	&                                               	& 6.5                                                	& 130                           	& 0.46                                    & [-5, 11]              		\\
                Per-emb 33 A            	& Triple (with Per-emb 33 B/C)	& 5.0                                             	& 284                           	& 0.64                                    & [0, 12]               		\\
                Per-emb 33 B/C		& Unresolved binary               	& 5.0                                             	& 284                           	& 0.64                                    & [-5, 8]               		\\
                Per-emb 36              	& Unresolved binary               	& 8.0                                           	& 24                                 	& 0.66                                    & [0, 12]               		\\
                Per-emb 37              	&                                               	& 8.0                                               	& 220                           	& 0.66                                    & [4, 11]               		\\
                IRAS 4B2           	&                                               	& 7.5                                          		& -                                   	& 0.66                                    & [5, 10]               		\\
                B1-b N                  	&                                               	& ?                                                	& -                                   	& 0.46                                    & [6, 10]               		\\
                B1-b S                  	&                                               	& 7.5                                             	& 120                           	& 0.46                                    & [4, 9]                		\\
                L1448 IRS 2E      	&                                               	& 5.0                                             	& 310                           	& 0.64                                    & [3, 7]                		\\
                L1448 NW                	& Unresolved binary                 	& 4.5                                               	& 305                           	& 0.64                                    & [0, 8]                		\\
                L1448 IRS 3A     	&                                               	& 4.5                                           	& 38                                 	& 0.64                                    & [-2, 12]              		\\
                SVS 13 B          		&                                               	& 8.0                                           	& 160                           	& 0.66                                    & [6, 11]               		\\
                \hline
                \multicolumn{6}{c}{Class I (\textit{T$_\mathrm{bol}$}~$\geq$~70~K)}                                                                                                                                                     		\\              
                \hline
                Per-emb 35              	& Binary                                    	& 7.5                                            	& 292                           	& 0.66                                    & [1, 12]               		\\
                Per-emb 40              	& Unresolved binary               	& 7.0                                            	& -                               	& 0.46                                    & [2, 12]               		\\
                Per-emb 41              	&                                               	& ?                                             		& 210                           	& 0.46                                    & [4.5, 8.5]    		\\
                Per-emb 42              	&                                               	& 5.5                                              	& -                                 	& 0.64                                    & [2, 13]               		\\
                Per-emb 44              	& Unresolved binary               	& 9.0                                            	& 147                           	& 0.66                                    & [1, 14]               		\\
                Per-emb 50              	&                                               	& 9.0 $^{(e)}$                                    	& 285                           	& 0.66                                    & [-6, 19]              		\\
                Per-emb 53              	&                                               	& 11.0                                           	& 70                          		& 0.51                                    & [5, 14]               		\\
                Per-emb 54              	&                                               	& 8.5                                      		& -                            		& 0.66                                    & [5, 13]               		\\
                Per-emb 55              	& Unresolved binary              	& 11.0                                            	& -                               	& 0.50                                    & [5, 13]               		\\
                Per-emb 60              	&                                               	& ?                                                 	& -                             		& 0.66                                    & [5, 9]                		\\
                L1455 IRS 2             	&                                               	& ?                                                 	& -                             		& 0.64                                    & [3, 7]                		\\
                EDJ2009-172         	&                                               	& ?                                                   	& -                             		& 0.66                                    & [5.5, 9.5]    		\\
                EDJ2009-235      	&                                               	& 8.0                                       		& -                                 	& 0.66                                    & [5, 9]                		\\
                \hline
                \multicolumn{6}{c}{Unknown \textit{T$_\mathrm{bol}$}}                                                                                                                                                                   			\\              
                \hline
                SVS 13 A2               	&                                               	& 8.0                                           	& 205                           	& 0.66                                    & [5, 11]               		\\
                EDJ2009-237       	&                                               	& ?                                            		& -                               	& 0.66                                    & [5, 9]               		\\
                \hline\hline
        \end{tabular}
        \tablefoot{$^{(a)}$Unresolved binaries in these data, with individual coordinates, can be found in Tobin et al. (2018). $^{(b)}$Estimated from CS spectra. $^{(c)}$Values taken from Tobin et al. (2018) and estimated from Stephens et al. (2019). $^{(d)}$Diameter of a circular region, which coincides with the angular resolution of the observations \citep[see ][]{Yang2021}. $^{(e)}$Estimated from SO emission.}
\end{table*}

% TABLE 5
\begin{table*}[t!]
        \caption{Integrated fluxes over the region and velocity ranges from Table~\ref{table:integration} for CS, SO, $^{34}$SO, and SO$_{2}$.}
        \label{table:fluxes}
        \centering
        \begin{tabular}{l l l l l}
                \hline\hline
                Source            		& CS                             	& SO $^{a}$       		& $^{34}$SO         		& SO$_{2}$            		\\
                                          		& (Jy~km~s$^{-1}$)       	& (Jy~km~s$^{-1}$)      	& (Jy~km~s$^{-1}$)          	& (Jy~km~s$^{-1}$)    	\\
                \hline
                \multicolumn{5}{c}{Class 0 (\textit{T$_\mathrm{bol}$}~$\leq$~70~K)}                                                                                 		\\              
                \hline
                Per-emb 1            	& $\geq$~0.23             	& 0.33~$\pm$~0.04         	& <~0.05                        	& 0.12~$\pm$~0.03    	\\
                Per-emb 2        		& $\geq$~0.14                  	& 0.11~$\pm$~0.02         	& <~0.04                       	& <~0.05                        	\\
                Per-emb 5     		& $\geq$~0.14                 	& 0.08~$\pm$~0.01         	& <~0.02                       	& <~0.02                     	\\
                Per-emb 8              	& $\geq$~0.31                	& 0.12~$\pm$~0.02         	& <~0.04                       	& <~0.05                    	\\
                Per-emb 10              	& $\geq$~0.17                	& 0.20~$\pm$~0.01         	& <~0.03                         	& <~0.04                       	\\
                Per-emb 11 A            	& $\geq$~0.17                	& $\geq$~0.29 $^{c}$      	& 0.04~$\pm$~0.01          	& 0.05~$\pm$~0.02        	\\
                Per-emb 11 B            	& $\geq$~0.14                	& 0.09~$\pm$~0.02         	& <~0.03                         	& <~0.04                        	\\
                Per-emb 11 C          	& $\geq$~0.14                	& 0.15~$\pm$~0.02         	& <~0.03                     	& <~0.05                      	\\
                Per-emb 12 A $^{b}$	& $\geq$~0.29                	& 0.15~$\pm$~0.04         	& <~0.08                      	& <~0.06                   		\\
                Per-emb 12 B           	& $\geq$~1.09                	& $\geq$~1.33 $^{c}$      	& 0.33~$\pm$~0.02          	& 0.65~$\pm$~0.03        	\\
                Per-emb 13              	& $\geq$~0.31                	& $\geq$~0.15 $^{c}$      	& 0.13~$\pm$~0.01       	& 0.10~$\pm$~0.02          	\\
                Per-emb 14              	& $\geq$~0.27                	& 0.41~$\pm$~0.02         	& <~0.03                    	& <~0.04                   		\\
                Per-emb 15              	& $\geq$~0.16                	& <~0.04                          	& <~0.04            		& <~0.05                          	\\
                Per-emb 16              	& $\geq$~0.15               	& 0.05~$\pm$~0.01         	& <~0.03                    	& <~0.04                    	\\
                Per-emb 17              	& $\geq$~2.33                	& $\geq$~1.80 $^{c}$      	& 0.14~$\pm$~0.02        	& 0.56~$\pm$~0.04     	\\
                Per-emb 18              	& $\geq$~0.67                	& $\geq$~0.80 $^{c}$      	& 0.08~$\pm$~0.02         	& 0.16~$\pm$~0.02         	\\
                Per-emb 20              	& $\geq$~0.57                	& 0.40~$\pm$~0.02         	& <~0.07                         	& 0.07~$\pm$~0.02          	\\
                Per-emb 21              	& $\geq$~0.14                	& 0.18~$\pm$~0.02         	& <~0.03                        	& <~0.03                           	\\
                Per-emb 22 A         	& $\geq$~0.52                	& 0.19~$\pm$~0.02         	& <~0.04                       	& 0.06~$\pm$~0.01          	\\
                Per-emb 22 B         	& $\geq$~0.40                	& 0.40~$\pm$~0.02         	& <~0.04                        	& 0.06~$\pm$~0.02          	\\
                Per-emb 25              	& $\geq$~0.18                	& 0.38~$\pm$~0.02         	& <~0.04                       	& 0.05~$\pm$~0.02          	\\
                Per-emb 26              	& $\geq$~1.94                	& $\geq$~0.68 $^{c}$      	& 0.08~$\pm$~0.03     	& 0.28~$\pm$~0.04          	\\
                Per-emb 27              	& $\geq$~1.84                	& $\geq$~2.31 $^{c}$      	& 0.50~$\pm$~0.01          	& 1.14~$\pm$~0.03          	\\
                Per-emb 28              	& $\geq$~0.18                	&  <~0.04                        	& <~0.03                    	& <~0.04                          	\\
                Per-emb 29              	& $\geq$~1.02                	& $\geq$~0.34 $^{c}$      	& 0.16~$\pm$~0.02        	& 0.06~$\pm$~0.02       	\\
                Per-emb 33 A         	& $\geq$~0.35                	& 0.13~$\pm$~0.03         	& <~0.05                         	& <~0.03                          	\\
                Per-emb 33 B/C  	& $\geq$~0.85                	& <~0.09                  		& <~0.05                          	& <~0.03                       	\\
                Per-emb 36              	& $\geq$~0.52                	& $\geq$~0.65 $^{c}$      	& 0.04~$\pm$~0.01      	& 0.06~$\pm$~0.02    	\\
                Per-emb 37              	& $\geq$~0.16                	& 0.09~$\pm$~0.02         	& <~0.03                     	& <~0.04                    	\\
                IRAS 4B2             	& -0.05~$\pm$~0.03        	& <~0.07                          	& <~0.03                       	& <~0.04                          	\\
                B1-b N                  	& <~0.06   $^{d}$         	& <~0.03                          	& <~0.02                      	& <~0.03                          	\\
                B1-b S                  	& -0.06~$\pm$~0.02       	& <~0.03                          	& <~0.03               		& <~0.03                          	\\
                L1448 IRS 2E    	& <~0.05 $^{d}$         	& <~0.17                         	& <~0.04                       	& <~0.03                          	\\
                L1448 NW                	& $\geq$~0.16                	& $\geq$~0.39 $^{c}$      	& 0.03~$\pm$~0.01       	& 0.07~$\pm$~0.02    	\\
                L1448 IRS 3A        	& $\geq$~0.37                	& $\geq$~0.41 $^{c}$      	& 0.06~$\pm$~0.02        	& 0.08~$\pm$~0.03       	\\
                SVS 13 B                 	& $\geq$~0.09                	& 0.06~$\pm$~0.01        	& <~0.03                     	& <~0.04                          	\\
                \hline
                \multicolumn{5}{c}{Class I (\textit{T$_\mathrm{bol}$}~$\geq$~70~K)}                                                                                		\\              
                \hline          
                Per-emb 35 A            & $\geq$~0.43             	& $\geq$~0.79 $^{c}$     	& 0.09~$\pm$~0.01      	& 0.22~$\pm$~0.02     	\\
                Per-emb 35 B            & $\geq$~0.48          		& $\geq$~0.64 $^{c}$    	& 0.04~$\pm$~0.01       	& 0.07~$\pm$~0.02  		\\
                Per-emb 40              & $\geq$~0.21                   	& 0.23~$\pm$~0.02       	& <~0.04                          	& <~0.05                 		\\
                Per-emb 41              & <~0.05                  		& <~0.05                         	& <~0.05                     	& <~0.05                          	\\
                Per-emb 42              & $\geq$~0.64                   	& 0.39~$\pm$~0.02         	& <~0.05                       	& 0.13~$\pm$~0.03        	\\
                Per-emb 44              & $\geq$~2.32                   	& $\geq$~2.59 $^{c}$      	& 0.73~$\pm$~0.02       	& 1.17~$\pm$~0.02      	\\
                Per-emb 50              & <~0.08                 		& $\geq$~1.57 $^{c}$      	& 0.09~$\pm$~0.02        	& 0.25~$\pm$~0.03        	\\
                Per-emb 53              & $\geq$~0.19                   	& 0.30~$\pm$~0.02         	& <~0.04 $^{d}$         	& <~0.05 $^{d}$        	\\
                Per-emb 54              & $\geq$~0.10                   	& 0.11~$\pm$~0.02        	& <~0.06                        	& <~0.05                         	 \\
                Per-emb 55              & $\geq$~0.05                   	& <~0.05                          	& <~0.04                      	& <~0.05                          	\\
                Per-emb 60              & <~0.03                     	& <~0.03                          	& <~0.04                         	& <~0.03                          	\\
                L1455 IRS 2             & <~0.05 $^{d}$         		& <~0.03                         	& <~0.04                  		& <~0.03                          	\\
                EDJ2009-172             & <~0.03                        	& <~0.03                          	& <~0.04                         	& <~0.08                          	\\
                EDJ2009-235             & <~0.03                        	& <~0.03                          	& <~0.03                        	& <~0.03                          	\\
                \hline
                \multicolumn{5}{c}{Unknown \textit{T$_\mathrm{bol}$}}                                                                                                          		\\              
                \hline
                SVS 13 A2               & $\geq$~0.11                   	& 0.12~$\pm$~0.02     	& <~0.03                           	& <~0.05                          	\\
                EDJ2009-237             & <~0.03                  		& <~0.03                          	& <~0.03                         	& <~0.04                          	\\
                \hline\hline
        \end{tabular}
        \tablefoot{$^{(a)}$SO 6$_{6}$--5$_{5}$. $^{(b)}$Optically thick continuum emission. $^{(c)}$Optically thick line emission. $^{(d)}$Only extended emission. Non detections present upper limits of 3$\sigma$. }
\end{table*}

% TABLE 6
\begin{table*}[t!]
        \caption{Molecular column densities for SO, $^{34}$SO, SO$_{2}$, and H$_{2}$, together with SO$_{2}$ and CH$_{3}$OH abundances}
        \label{table:column_density}
        \centering
        \begin{tabular}{l c c c | r r r r}
                \hline\hline
                Source               	&  \textit{N}(SO) $^{a,b}$ 	&  \textit{N}($^{34}$SO) $^{b}$    		& \textit{N}(SO$_{2}$) $^{c}$      	&       & \textit{N}(H$_{2}$) $^{d}$  		& [SO$_{2}$/H$_{2}$] 	& [CH$_{3}$OH/H$_{2}$]	\\
                                                	& ($\times$10$^{14}$~cm$^{-2}$) & ($\times$10$^{14}$~cm$^{-2}$)	& ($\times$10$^{14}$~cm$^{-2}$) 	&       & ($\times$10$^{24}$~cm$^{-2}$)	& ($\times$10$^{-10}$) 	& ($\times$10$^{-9}$)      	\\
                \hline
                \multicolumn{7}{c}{Class 0 (\textit{T$_\mathrm{bol}$}~$\leq$~70~K)}                                                                                                     																		\\              
                \hline
                Per-emb 1              	& [4.83 -- 7.16]                	& <~1.5                                                   	& [5 -- 20]             				&       & 2.9     0~$\pm$~0.18            		& 4.3~$\pm$~1.7 		& 3.1~$\pm$~1.9         	\\
                Per-emb 2               	& [1.94 -- 2.88]                	& <~1.4                                                   	& <~4.9         					&       & 5.32~$\pm$~0.22         			& <~0.9                 		& 0.6~$\pm$~0.3         	\\
                Per-emb 5     		& [1.37 -- 2.02]                	& <~0.7                                                   	& <~2.0        					&       & 11.13~$\pm$~0.14                		& <~0.2                 		& 0.9~$\pm$~0.5         	\\
                Per-emb 8             	& [1.76 -- 2.61]                	& <~1.1                                                   	& <~4.0         					&       & 5.19~$\pm$~0.11         			& <~0.8                 		& <~0.7                 		\\
                Per-emb 10              	& [3.35 -- 4.97]                	& <~1.0                                                   	& <~3.8         					&       & 1.42~$\pm$~0.02         			& <~2.7                 		& 1.9~$\pm$~1.1         	\\
                Per-emb 11 A        	& $\geq$~4.09           	& [1.69 -- 2.46]                                          	& [2 -- 6]              				&       & 6.32~$\pm$~0.18         			& 0.6~$\pm$~0.3 		& 1.4~$\pm$~0.9         	\\
                Per-emb 11 B       	& [1.30 -- 1.93]                	& <~0.8                                                   	& <~3.2         					&       & 0.27~$\pm$~0.02         			& <~11.9                        	& <~9.5                 		\\
                Per-emb 11 C        	& [2.11 -- 3.13]                	& <~0.8                                                   	& <~4.0         					&       & 0.26~$\pm$~0.02         			& <~15.4                        	& 24.6~$\pm$~15.5         	\\
                Per-emb 12 A $^{e}$	& [1.20 -- 1.80]                	& <~1.3                                                   	& <~2.8         					&       & 15.33~$\pm$~1.45                	& <~0.2                 		& 0.5~$\pm$~0.3         	\\
                Per-emb 12 B       	& $\geq$~10.79  		& [8.77 -- 12.74]                                       	& $\geq$~20       				&       & 7.79~$\pm$~0.63               		& >~2.6                 		& 25.7~$\pm$~15.5         	\\
                Per-emb 13           	& $\geq$~1.19           	& [3.13 -- 4.54]                                        	& [2 -- 7]              				&       & 11.33~$\pm$~0.33                		& 0.4~$\pm$~0.2 		& 4.4~$\pm$~2.8         	\\
                Per-emb 14              	& [3.32 -- 4.92]                	& <~0.5                                                   	& <~1.8         					&       & 2.47~$\pm$~0.03         			& <~0.7                 		& <~1.3                 		\\
                Per-emb 15              	& <~0.5                 		& <~0.6                                                 	& <~2.3           					&       & 0.12~$\pm$~0.01               		& <~19.2                        	& <~23.3                  		\\
                Per-emb 16              	& [0.75 -- 1.11]                	& <~0.8                                                   	& <~3.2         					&       & 0.22~$\pm$~0.02         			& <~14.5                        	& <~9.3                 		\\
                Per-emb 17              	& $\geq$~15.61  		& [3.77 -- 5.48]                                           	& $\geq$~10       				&       & 1.30~$\pm$~0.04               		& >~7.7                 		& --                              	\\
                Per-emb 18              	& $\geq$~6.52           	& [2.02 -- 2.94]                                           	& [3 -- 10]             				&       & 2.62~$\pm$~0.04         			& 2.5~$\pm$~1.1 		& 8.8~$\pm$~5.0 		\\
                Per-emb 20              	& [3.45 -- 5.12]                	& <~1.2                                                   	& [1 -- 5]              				&       & 0.10~$\pm$~0.01         			& 28.9~$\pm$~9.9        	& 125.1~$\pm$~77.6      	\\
                Per-emb 21              	& [1.45 -- 2.15]                	& <~0.5                                                   	& <~1.4         					&       & 1.65~$\pm$~0.04         			& <~0.8                 		& 10.3~$\pm$~6.1        	\\
                Per-emb 22 A      	& [1.68 -- 2.49]                	& <~0.7                                                   	& [1 -- 4]             				&       & 1.20~$\pm$~0.05         			& 2.1~$\pm$~0.8         	& 16.7~$\pm$~10.0       	\\
                Per-emb 22 B      	& [3.50 -- 5.18]                	& <~0.7                                                   	& [1 -- 4]              				&       & 0.68~$\pm$~0.05         			& 3.7~$\pm$~1.5 		& 23.5~$\pm$~14.7       	\\
                Per-emb 25              	& [3.32 -- 4.91]                	& <~0.7                                                   	& [1 -- 4]              				&       & 3.56~$\pm$~0.09         			& 0.7~$\pm$~0.3 		& <~1.3                 		\\
                Per-emb 26              	& $\geq$~5.90           	& [2.29 -- 3.33]                                        	& [6 -- 40]             				&       & 5.45~$\pm$~0.11         			& 4.2~$\pm$~1.1 		& 16.3~$\pm$~8.1        	\\
                Per-emb 27              	& $\geq$~18.81  		& [13.08 -- 19.00]                         		& $\geq$~30       				&       & 4.40~$\pm$~0.31               		& >~6.8                 		& 249.8~$\pm$~114.9	\\
                Per-emb 28              	& <~0.8                 		& <~0.8                                                 	& <~3.2           					&       & 0.89~$\pm$~0.03               		& <~3.6                 		& <~19.0                  		\\
                Per-emb 29              	& $\geq$~5.63           	& [8.82 -- 12.80]                                  	& [3 -- 10]             				&       & 5.63~$\pm$~0.28         			& 1.2~$\pm$~0.5 		& 16.0~$\pm$~10.5       	\\
                Per-emb 33 A       	& [1.08 -- 1.60]                	& <~0.9                                                   	& <~1.5         					&       & 6.19~$\pm$~0.35         			& <~0.2                 		& 0.9~$\pm$~0.5 		\\
                Per-emb 33 B/C  	& <~1.2                 		& <~0.9                                                 	& <~1.5           					&       & 3.34~$\pm$~0.05               		& <~0.4                 		& <~1.2                   		\\
                Per-emb 36            	& $\geq$~5.27           	& [1.05 -- 1.53]                             		& [1 -- 4]              				&       & 4.43~$\pm$~0.02         			& 0.6~$\pm$~0.2 		& <~0.8                 		\\
                Per-emb 37           	& [0.70 -- 1.04]                	& <~0.5                                                   	& <~1.8         					&       & 0.43~$\pm$~0.02         			&  <~4.2                        	& <~3.0                 		\\
                IRAS 4B2        		& <~0.8                 		& <~0.5                                                   	& <~1.8         					&       & 5.45~$\pm$~0.15         			& <~0.3                 		& <~0.06                        	\\
                B1-b N                  	& <~0.7                 		& <~0.7                                                 	& <~2.8          					&       & 5.32~$\pm$~0.13               		& <~0.5                 		& <~0.7                   		\\
                B1-b S                  	& <~0.7                 		& <~1.0                                                 	& <~2.8           					&       & 10.29~$\pm$~0.11              		& <~0.3                 		& 1.5~$\pm$~0.9   		\\
                L1448 IRS 2E    	& <~2.2                 		& <~0.7                                                 	& <~1.5           					&       & --                                    		& --                            	& --                              	\\
                L1448 NW                	& $\geq$~3.34           	& [0.95 -- 1.38]                                		& [1 -- 5]              				&       & 2.19~$\pm$~0.07         			& 1.4~$\pm$~0.5 		& 1.0~$\pm$~0.5 		\\
                L1448 IRS 3A      	& $\geq$~3.55           	& [1.76 -- 2.56]                                 		& [2 -- 6]              				&       & 2.08~$\pm$~0.10         			& 1.9~$\pm$~1.0 		& 4.0~$\pm$~2.6 		\\
                SVS 13 B              	& [0.50 -- 0.74]                	& <~0.5                                                   	& <~1.8         					&       & 5.05~$\pm$~0.19         			& <~0.4                 		& <~0.5                 		\\
                \hline
                \multicolumn{7}{c}{Class I (\textit{T$_\mathrm{bol}$}~$\geq$~70~K)}                                                                                                                                                                     										\\              
                \hline
                Per-emb 35 A        	& $\geq$~6.41           	& [2.42 -- 3.51]                                		& [5 -- 20]             				&       & 0.77~$\pm$~0.02         			& 16.3~$\pm$~6.5        	& 69.0~$\pm$~37.8       	\\
                Per-emb 35 B        	& $\geq$~5.19           	& [0.97 -- 1.41]                                    	& [1 -- 5]              				&       & 0.60~$\pm$~0.03         			& 5.0~$\pm$~1.7 		& 10.5~$\pm$~5.8        	\\
                Per-emb 40       		& [3.90 -- 5.79]                	& <~1.3                                                   	& <~4.7         					&       & 1.05~$\pm$~0.01         			& <~4.5                 		& <~4.6                 		\\
                Per-emb 41              	& <~1.2                 		& <~1.7                                                 	& <~4.7           					&       & 0.88~$\pm$~0.04               		& <~5.3                 		& --                              	\\
                Per-emb 42              	& [3.34 -- 4.95]                	& <~0.9                                                   	& [3 -- 10]              				&       & 0.44~$\pm$~0.01               		& 14.7~$\pm$~6.8        	& 20.4~$\pm$~12.2 		\\
                Per-emb 44              	& $\geq$~21.05 		& [19.28 -- 28.00]                                      	& $\geq$~30       				&       & 4.92~$\pm$~0.28               		& >~6.1                 		& 164.6~$\pm$~63.7       	\\
                Per-emb 50              	& $\geq$~12.76  		& [2.29 -- 3.32]                                  		& [5 -- 30]               				&       & 3.26~$\pm$~0.01               		& 5.4~$\pm$~1.5 		& <~0.7                   		\\
                Per-emb 53              	& [4.13 -- 6.12]                	& <~1.1                                                   	& <~3.8         					&       & 0.92~$\pm$~0.01         			& <~4.1                 		& 5.5~$\pm$~3.5 		\\
                Per-emb 54              	& [0.90 -- 1.34]                	& <~1.0                                                   	& <~2.3         					&       & 0.05~$\pm$~0.01         			& <~46.0                        	& <~41.5                        	\\
                Per-emb 55              	& <~1.0                 		& <~1.1                                                 	& <~4.0           					&       & 0.22~$\pm$~0.06               		& <~18.2                        	& <~17.6                  		\\
                Per-emb 60              	& <~0.4                 		& <~0.6                                                 	& <~1.4           					&       & 0.06~$\pm$~0.01               		& <~23.3                        	& --                              	\\
                L1455 IRS 2             	& <~0.4                 		& <~0.7                                                 	& <~1.5           					&       & 0.08~$\pm$~0.01               		& <~18.8                        	& <~54.0                  		\\
                EDJ2009-172         	& <~0.4                 		& <~0.6                                                 	& <~3.7           					&       & 0.51~$\pm$~0.01               		& <~7.3                		& --                              	\\
                EDJ2009-235         	& <~0.4                		& <~0.5                                                 	& <~1.4           					&       & 0.21~$\pm$~0.01               		& <~6.7                 		& <~19.4                  		\\
                \hline
                \multicolumn{7}{c}{Unknown \textit{T$_\mathrm{bol}$}}                                                                                                                                                                                   											\\              
                \hline
                SVS 13 A2          	& [0.98 -- 1.45]                	& <~0.5                                                   	& <~2.3         					&       & 0.52~$\pm$~0.03         			& <~4.4                 		& 13.4~$\pm$~8.9        	\\
                EDJ2009-237        	& <~0.4                 		& <~0.5                                                	& <~1.8           					&       & 0.09~$\pm$~0.01              		& <~20.0                        	& --                              	\\
                \hline\hline
        \end{tabular}
        \tablefoot{$^{(a)}$SO 6$_{6}$--5$_{5}$. $^{(b)}$Calculated from a LTE analysis. $^{(c)}$Estimated from RADEX (non-LTE) for \textit{n$_\mathrm{H}$} between 10$^{7}$ and 10$^{9}$~cm$^{-3}$. $^{(d)}$Assuming \textit{T$_\mathrm{dust}$}~=~30~K. $^{(e)}$Optically thick continuum emission.}
\end{table*}

% TABLE 4
\begin{table}[h!]
        \caption{Number of COMs detected, from \cite{Yang2021}, and detection of sulfur-bearing species toward each source.}
        \label{table:COM_S}
        \centering
        \begin{tabular}{l c c c c c c}
                \hline\hline
                Source 			& COMs	& CS          		& SO            		& $^{34}$SO  		& SO$_{2}$ 		& Class	\\
                \hline
                Per-emb 13           	& 16      	& $\checkmark$  	& $\checkmark$    	& $\checkmark$  	& $\checkmark$  	& 0    	\\
                Per-emb 12 B       	& 15    	& $\checkmark$  	& $\checkmark$    	& $\checkmark$  	& $\checkmark$  	& 0   		\\
                Per-emb 29              	& 14   	& $\checkmark$  	& $\checkmark$    	& $\checkmark$  	& $\checkmark$  	& 0   		\\
                Per-emb 44              	& 14     	& $\checkmark$  	& $\checkmark$    	& $\checkmark$  	& $\checkmark$  	& I   		\\
                Per-emb 27              	& 14       	& $\checkmark$  	& $\checkmark$    	& $\checkmark$  	& $\checkmark$  	& 0   		\\
                B1-b S                  	& 9        	& --                         	& --                      	& --                        	& --                       	& 0    	\\
                Per-emb 26              	& 9          	& $\checkmark$  	& $\checkmark$    	& $\checkmark$  	& $\checkmark$  	& 0     	\\
                Per-emb 17              	& 9         	& $\checkmark$  	& $\checkmark$    	& $\checkmark$  	& $\checkmark$  	& 0     	\\
                Per-emb 11 A         	& 7          	& $\checkmark$  	& $\checkmark$    	& $\checkmark$  	& $\checkmark$  	& 0     	\\
                Per-emb 35 A         	& 5           	& $\checkmark$  	& $\checkmark$    	& $\checkmark$  	& $\checkmark$  	& I     	\\
                Per-emb 33 A         	& 3     	& $\checkmark$  	& $\checkmark$    	& --                       	& --                         	& 0       	\\
                Per-emb 18              	& 3          	& $\checkmark$  	& $\checkmark$    	& $\checkmark$  	& $\checkmark$  	& 0     	\\
                L1448 IRS 3A        	& 3       	& $\checkmark$  	& $\checkmark$    	& $\checkmark$  	& $\checkmark$  	& 0     	\\
                SVS 13B                 	& 3       	& $\checkmark$  	& $\checkmark$    	& --                     	& --                       	& 0       	\\
                Per-emb 22 B         	& 2           	& $\checkmark$  	& $\checkmark$    	& --                       	& $\checkmark$  	& 0     	\\
                Per-emb 22 A         	& 2           	& $\checkmark$  	& $\checkmark$    	& --                      	& $\checkmark$  	& 0     	\\
                Per-emb 20              	& 2          	& $\checkmark$  	& $\checkmark$    	& --                       	& $\checkmark$  	& 0     	\\
                Per-emb 35 B         	& 2  		& $\checkmark$  	& $\checkmark$    	& $\checkmark$  	& $\checkmark$  	& I     	\\
                Per-emb 12 A         	& 2      	& $\checkmark$  	& $\checkmark$    	& --                    	& --                      	& 0       	\\
                Per-emb 5              	& 2    	& $\checkmark$    	& --                       	& --                  		& --                     	& 0     	\\
                Per-emb 1                	& 2           	& $\checkmark$    	& $\checkmark$	& --                 		& $\checkmark$  	& 0       	\\
                Per-emb 42              	& 1   		& $\checkmark$  	& $\checkmark$    	& --                 		& $\checkmark$  	& I     	\\
                Per-emb 21              	& 1           	& $\checkmark$  	& $\checkmark$    	& --              		& --                  		& 0       	\\
                Per-emb 2              	& 1          	& $\checkmark$    	& $\checkmark$  	& --                     	& --                   		& 0       	\\
                Per-emb 10              	& 1         	& $\checkmark$  	& $\checkmark$    	& --                  		& --                		& 0       	\\
                Per-emb 11 C         	& 1        	& $\checkmark$  	& $\checkmark$    	& --                    	& --                  		& 0       	\\
                Per-emb 53              	& 1         	& $\checkmark$  	& $\checkmark$    	& $\checkmark$  	& $\checkmark$  	& I     	\\
                SVS 13 A2 $^{a}$   	& 1        	& $\checkmark$  	& $\checkmark$    	& --            		& --                  		& 0/I?    	\\                      
                L1448 NW                	& 0           	& $\checkmark$  	& $\checkmark$    	& $\checkmark$  	& $\checkmark$  	& 0     	\\      
                Per-emb 33 B/C  	& 0           	& $\checkmark$  	& $\checkmark$    	& --                  		& --                    	& 0       	\\
                Per-emb 25              	& 0          	& $\checkmark$  	& $\checkmark$    	& --                       	& $\checkmark$  	& 0     	\\
                L1455 IRS2              	& 0        	& $\checkmark$  	& --                          	& --                      	& --               		& I       	\\
                Per-emb 36              	& 0         	& $\checkmark$  	& $\checkmark$    	& $\checkmark$  	& $\checkmark$  	& 0     	\\
                Per-emb 54              	& 0         	& $\checkmark$  	& $\checkmark$    	& --              		& --                    	& I       	\\
                Per-emb 15              	& 0         	& $\checkmark$  	& --                  		& --              		& --               		& 0       	\\
                Per-emb 50              	& 0           	& --                        	& $\checkmark$    	& $\checkmark$  	& $\checkmark$  	& I     	\\
                IRAS 4B2               	& 0     	& $\checkmark$    	& $\checkmark$  	& --           			& --                		& 0       	\\
                Per-emb 14              	& 0          	& $\checkmark$  	& $\checkmark$    	& --             		& --                     	& 0       	\\
                Per-emb 37              	& 0        	& $\checkmark$  	& $\checkmark$    	& --                      	& --                 		& 0       	\\
                Per-emb 40              	& 0        	& $\checkmark$  	& $\checkmark$    	& --                 		& --                   		& I       	\\
                B1-b N                  	& 0         	& $\checkmark$  	& --                         	& --              		& --                  		& 0       	\\
                Per-emb 16              	& 0         	& $\checkmark$  	& --                        	& --                 		& --                     	& 0       	\\
                Per-emb 28              	& 0        	& $\checkmark$  	& --                     	& --                  		& --             		& 0       	\\
                Per-emb 11 B        	& 0         	& $\checkmark$  	& $\checkmark$    	& --             		& --                     	& 0       	\\
                Per-emb 55              	& 0         	& $\checkmark$  	& --                        	& --               		& --            		& I       	\\
                Per-emb 8             	& 0       	& $\checkmark$    	& $\checkmark$  	& --             		& --        			& 0       	\\      
                Per-emb 60              	& 0           	& --                        	& --                         	& --                		& --                   		& I       	\\      
                EDJ2009-172         	& 0         	& --                      	& --                  		& --                     	& --                   		& I       	\\
                EDJ2009-235         	& 0          	& --                		& --                   		& --               		& --                       	& I       	\\
                Per-emb 41 $^{b}$  	& ?          	& ?                  		& ?                         	& ?                 		& ?                		& I       	\\
                \hline\hline
        \end{tabular}
        \tablefoot{$^{(a)}$Source with unknown \textit{T$_\mathrm{bol}$} but, given the clustered nature, we could speculate that it is a protostar. $^{(b)}$Source at the edge of the field of view.
        }
\end{table}

\section{Large-scale CS emission}

\begin{figure*}%[H]
        \centering
        \includegraphics[width=.49\textwidth]{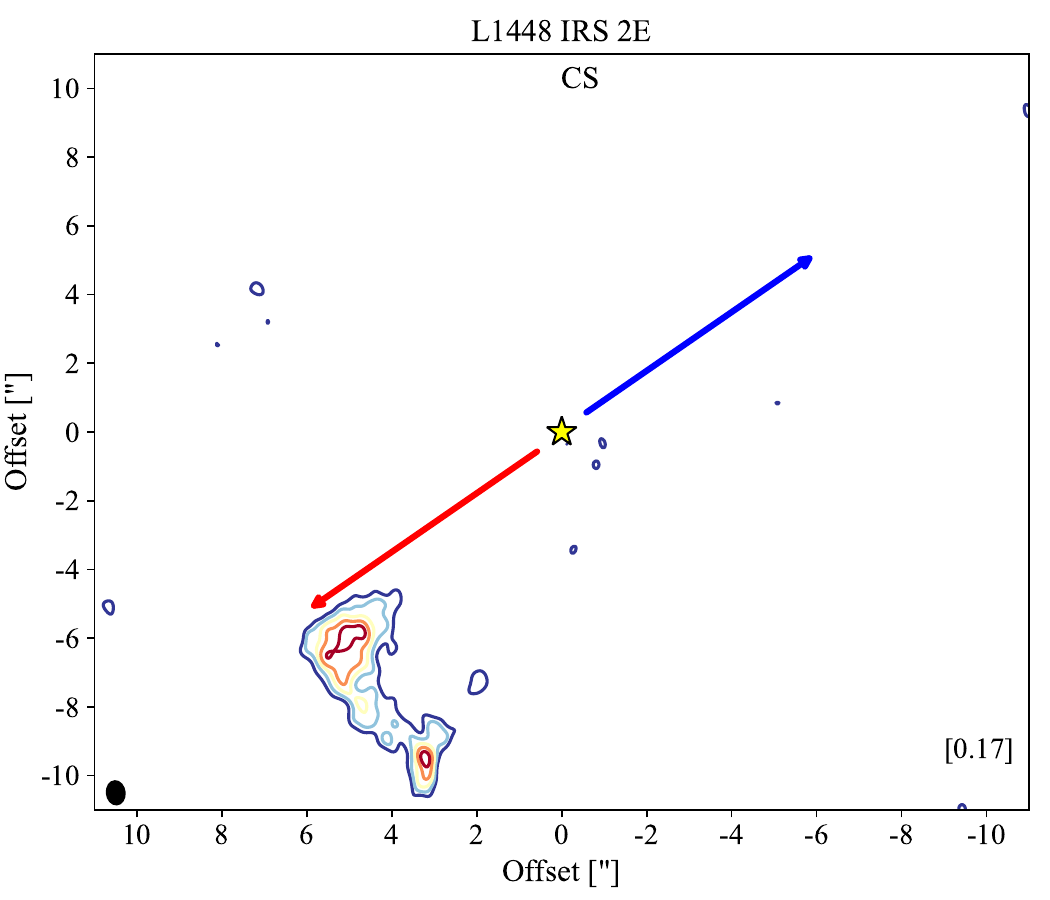}
        \includegraphics[width=.49\textwidth]{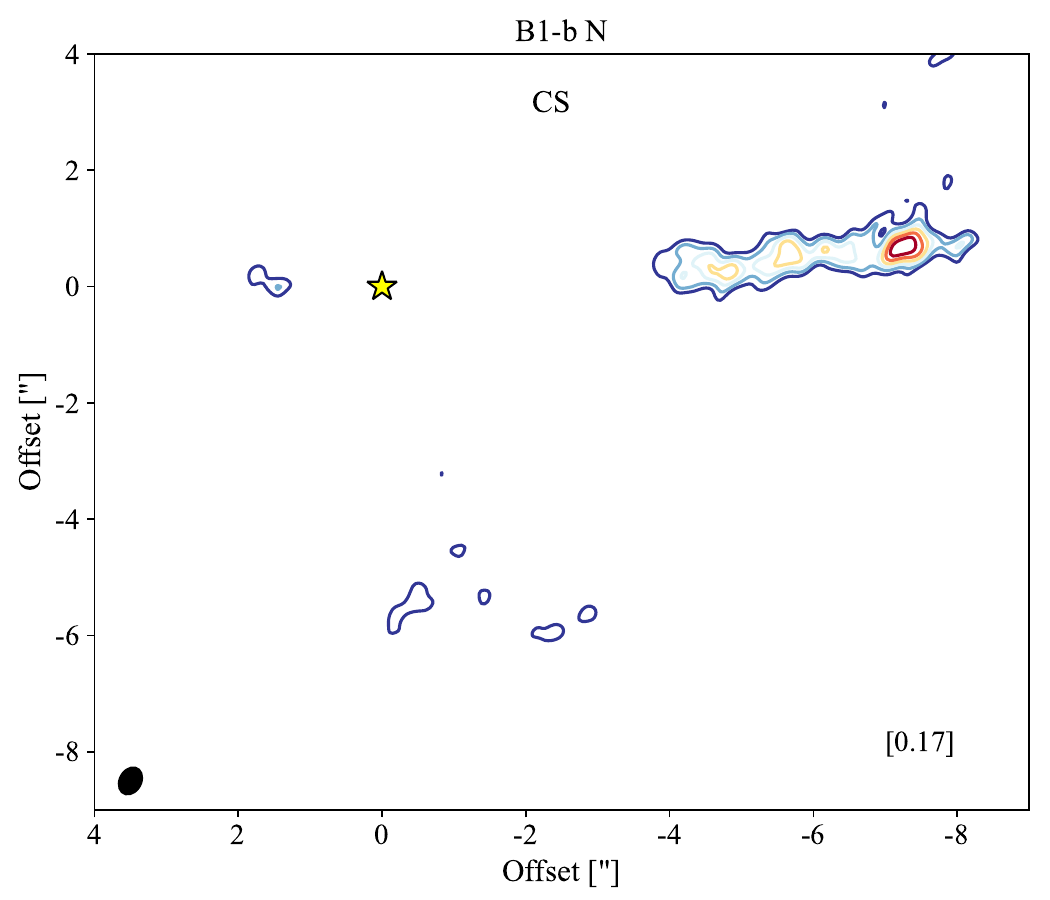}
        \caption[]{\label{fig:B1bN}
        Same as Fig.~\ref{fig:Mom0_1} but for large-scale emission of CS toward L1448 IRS 2E and B1-b N.
        }
\end{figure*}

Two Class 0 sources, L1448 IRS 2E and B1-b N, do not show molecular emission around the continuum peak position. However, extended CS emission is detected at large scales (Fig.~\ref{fig:B1bN}), possibly related with the outflow component.

%FIGURE 12

%FIGURE 13
\begin{figure*}%[H]
        \centering
        \includegraphics[width=.95\textwidth]{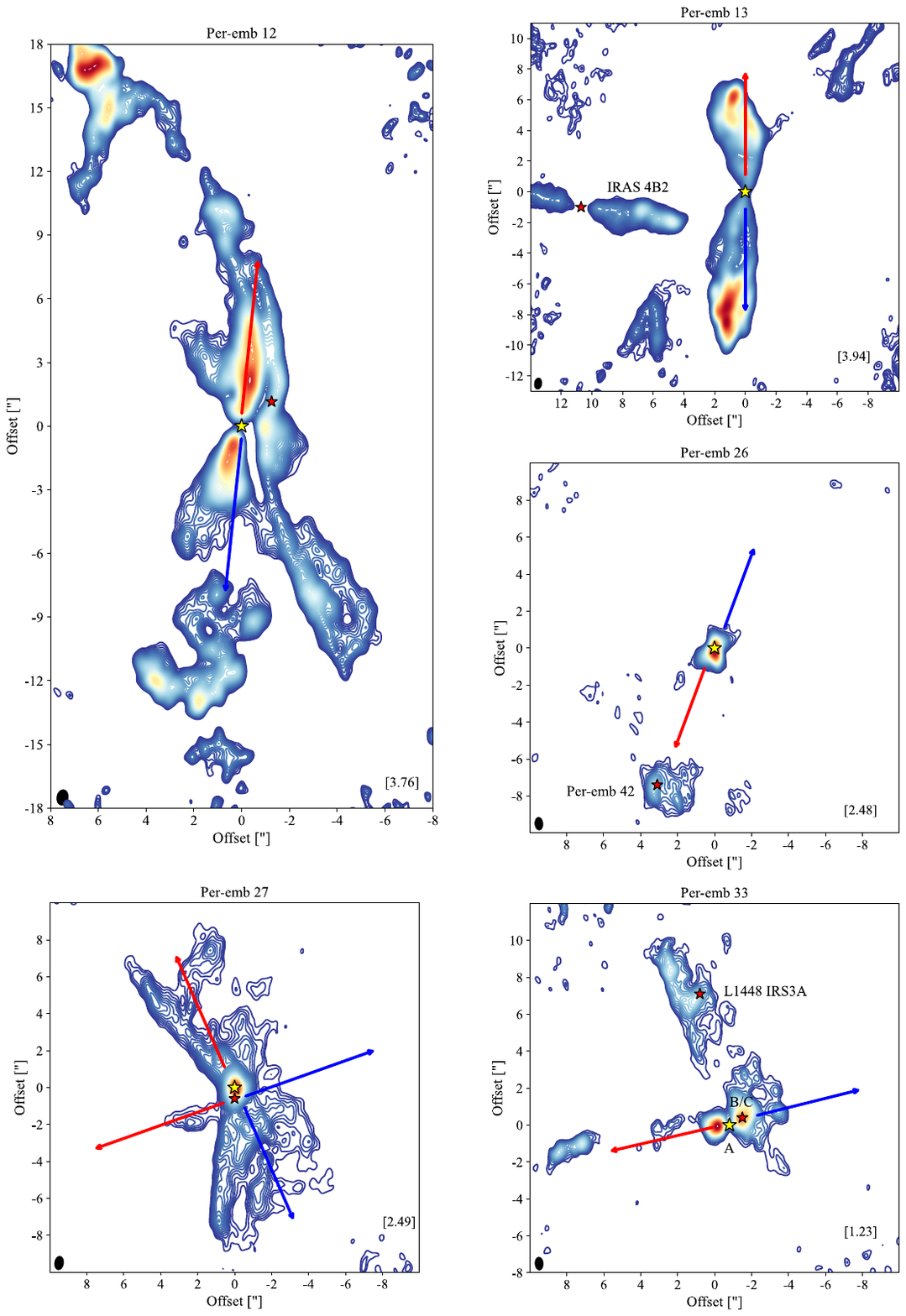}
        \caption[]{\label{fig:per12_large}
        Same as Fig.~\ref{fig:B1bN} for Per-emb 12, Per-emb 13 and IRAS 4B2, Per-emb 26 and Per-emb 42, Per-emb 27, and Per-emb 33 and L1448 IRS3A.
        }
\end{figure*}

Per-emb 12, Per-emb 13, and Per-emb 27 show strong and extended outflow structures that are shown in Fig.~\ref{fig:per12_large}. Per-emb 42 and L1448 IRAS3A lie very close to Per-emb 26 and Per-emb 33, respectively, and their close environments seem to be affected by the presence of these nearby sources (see Fig.~\ref{fig:per12_large}).

\section{Spectra}

\begin{figure*}[h]
        \centering
        \includegraphics[width=.97\textwidth]{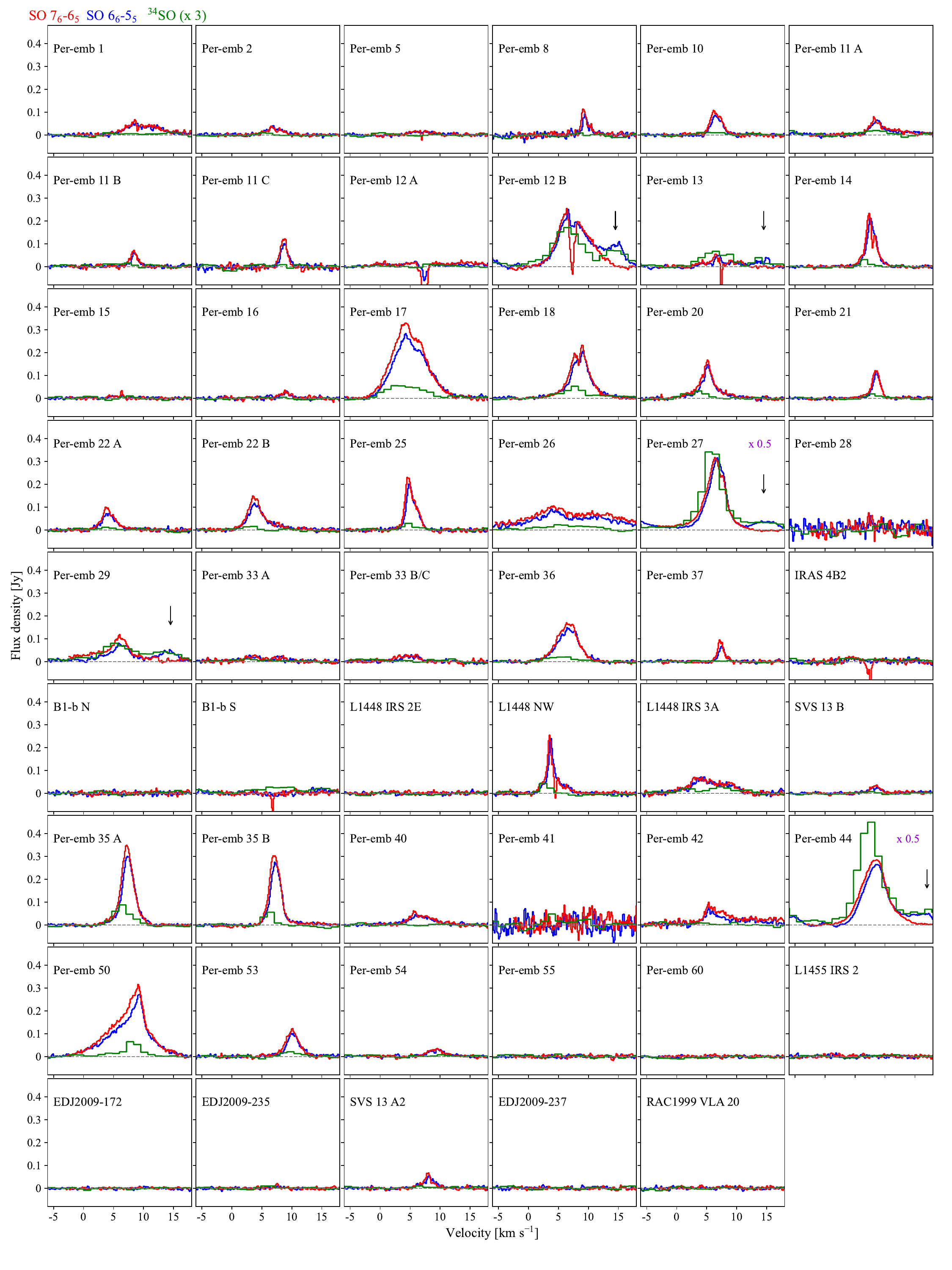}
        \vskip -20pt
        \caption[]{\label{fig:spectra_SO}
        Integrated spectra of SO 7$_{6}$--6$_{5}$ (red), SO 6$_{6}$--5$_{5}$ (blue), and $^{34}$SO 5$_{6}$--4$_{5}$ (green). The integration area for each source is listed in Table~\ref{table:integration} and the $^{34}$SO 5$_{6}$--4$_{5}$ spectra have been multiplied by three, for a better comparison. Black arrows in the panels that correspond to Per-emb 12 B, Per-emb 13, Per-emb 27, Per-emb 29, and Per-emb 44 indicate an offset emission, likely related with a transition from an unidentified COM.
        }
\end{figure*}

Spectra of SO 7$_{6}$--6$_{5}$, SO 6$_{6}$--5$_{5}$ and $^{34}$SO are presented in Fig.~\ref{fig:spectra_SO}, showing that both SO transitions exhibit a similar behavior. In addition, the bump in SO 6$_{6}$--5$_{5}$ (seen in a few sources) could be related to a different molecular transition (maybe to a COM) and not to the blueshifted or redshifted emission of SO.

%FIGURE 14

\section{Radiative transfer}

\begin{figure*}[h]
        \centering
        \includegraphics[width=.7\textwidth]{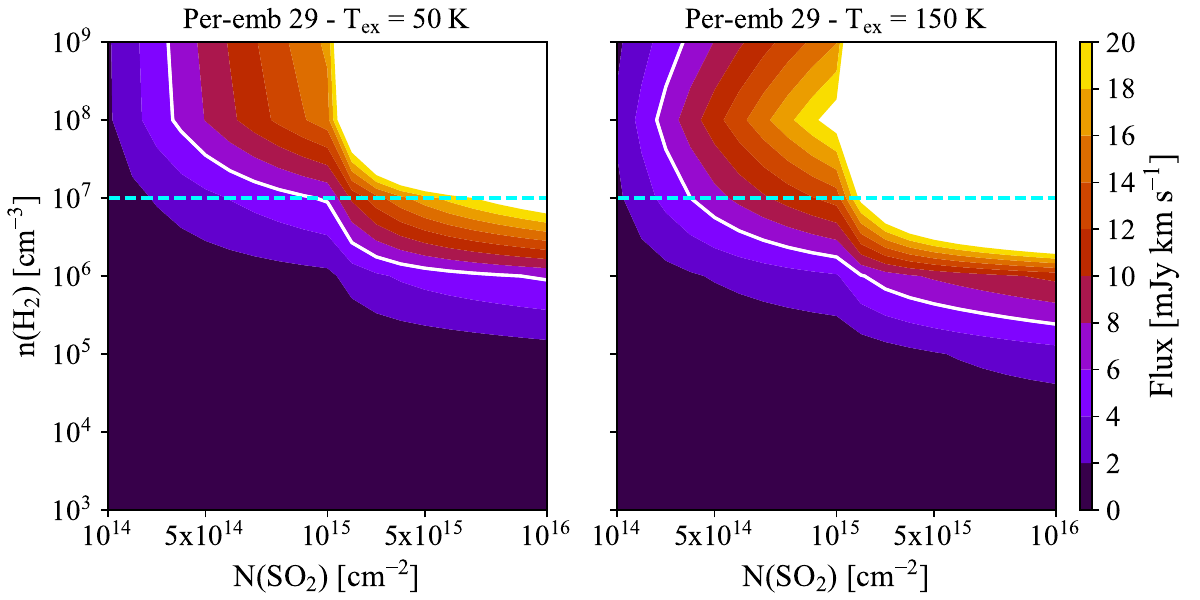}
        \caption[]{\label{fig:radex_n}
        RADEX models for Per-emb 29. Fluxes of the SO$_{2}$ 14$_{0,14}$--13$_{1,13}$ transition for two different kinetic temperatures: 50~K (\textit{left}) and 150~K  (\textit{right}). The white contour represents the flux observed for this source and the dashed cyan contour shows an H$_{2}$ number density of 10$^{7}$~cm$^{-3}$ (lower limit for \textit{n$_\mathrm{H}$} used in the calculations). In this example, the \textit{N$_\mathrm{SO_{2}}$} present a minimum value of 3~$\times$~10$^{14}$~cm$^{-2}$ (for \textit{T$_\mathrm{ex}$}~=~150) and a maximum value of 1~$\times$~10$^{15}$~cm$^{-2}$ (for \textit{T$_\mathrm{ex}$}~=~50), in a range between 10$^{7}$ and 10$^{9}$~cm$^{-3}$.
                }
\end{figure*}

Given that the SO$_{2}$ 14$_{0,14}$--13$_{1,13}$ transition is not thermalized below a H$_{2}$ number density of 10$^{9}$~cm$^{-3}$, RADEX was employed to estimate a range of molecular column densities. Figure~\ref{fig:radex_n} shows the example of Per-emb 29 for an excitation temperature of 50 and 150~K. A range of \textit{N$_\mathrm{SO_{2}}$} is estimated for \textit{n$_\mathrm{H}$} between 10$^{7}$ and 10$^{9}$~cm$^{-3}$, given the integrated flux (white contour), listed in Table~\ref{table:column_density}. 

\begin{figure*}[h]
        \centering
        \includegraphics[width=.99\textwidth]{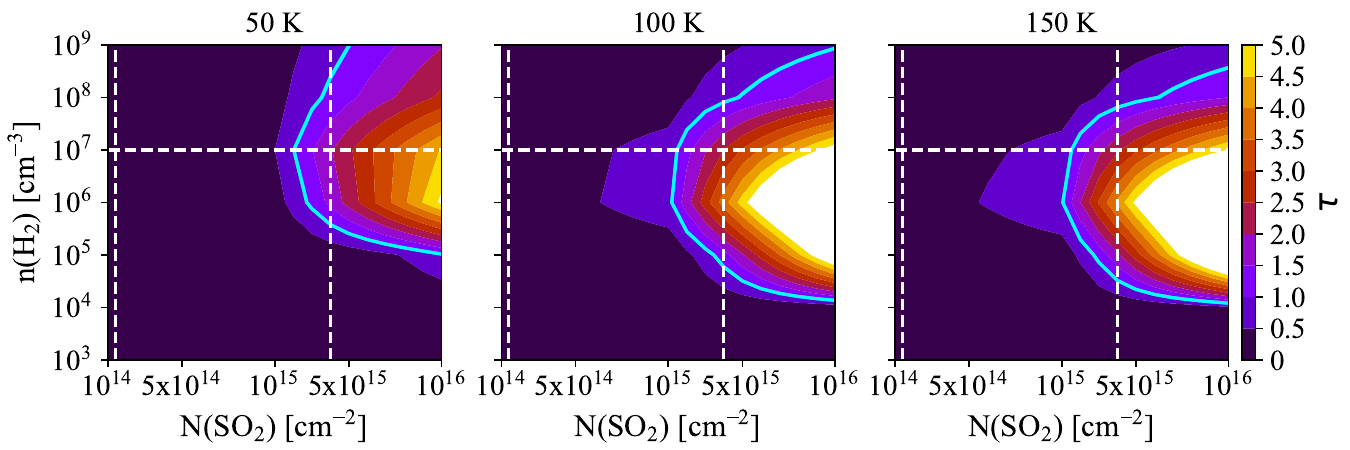}
        \caption[]{\label{fig:radex}
        RADEX models. Optical depth of the SO$_{2}$ 14$_{0,14}$--13$_{1,13}$ transition for three different excitation temperatures: 50,100, and 150~K. The cyan contour represents $\tau$~=~1, the two dashed white vertical lines indicate the range of values obtained for \textit{N$_\mathrm{SO_{2}}$} for most of the sources, and the dashed white horizontal line shows an H$_{2}$ number density of 10$^{7}$~cm$^{-3}$ (lower limit for \textit{n$_\mathrm{H}$} used in the calculations).
        }
\end{figure*}

RADEX models were also employed to assess to optical depth of the SO$_{2}$ 14$_{0,14}$--13$_{1,13}$ transition. Figure~\ref{fig:radex} shows the SO$_{2}$ molecular column density as a function of the H$_{2}$ number density, for three different values of \textit{T$_\mathrm{ex}$}: 50, 100, and 150~K. The range of calculated \textit{N$_\mathrm{SO_{2}}$} (see Table~\ref{table:column_density}) lies between 2~$\times$~10$^{14}$ and 5~$\times$~10$^{15}$~cm$^{-2}$; thus, the SO$_{2}$ emission could be taken as optically thin emission in most of the cases. Four sources, Per-emb 12, Per-emb 17, Per-emb 27, and Per-emb 44, are associated with very bright SO$_{2}$ emission and their estimated \textit{N$_\mathrm{SO_{2}}$} could reach values higher than 10$^{16}$~cm$^{-2}$. Therefore, the SO$_{2}$ emission toward these four sources is considered to be optically thick and only lower limits are presented for \textit{N$_\mathrm{SO_{2}}$} in Table~\ref{table:column_density}.

%FIGURE 16

%FIGURE 17

\section{[SO$_{2}$/H$_{2}$] from the literature}

There are only a handful of protostar where SO$_{2}$ emission has been detected at high angular resolution ($\leq$1$\farcs$0) in nearby star-forming regions ($\leq$300~pc): the Class 0 protostar IRAS 16293-2429, a sample of five Class I sources in Ophiuchus, different components of four Class I/II disks in Taurus, and the Class II disk IRS 48 \citep{Drozdovskaya2018, Artur2019a, Garufi2022, Booth2021}. All these sources are located at distances between 120 and 160~pc; therefore, special care has to be taken when comparing literature data with the PEACHES sample (Fig.~\ref{fig:Ntot}), located at $\sim$300~pc. 

The region of integration for the PEACHES sample is equivalent to the size of the beam ($\sim$0$\farcs$5), proving regions of $\sim$150~au in diameter around the central protostar. To trace similar regions in Taurus and Ophiuchus, a beam size (or integration area) of $\sim$1$\farcs$0 would be needed. Table~\ref{table:SO2_H2_lit} lists values taken from the literature to calculate the abundance of SO$_{2}$ (for an integration area of 1$\farcs$0~$\times$~1$\farcs$0). Given that in most of the cases only the SO$_{2}$ column density is available, \textit{N}(H$_{2}$) values were calculated using Eq.~\ref{eq:NH2} and the conversion:

% EQUATION 5
\begin{equation} 
    \frac{F_{\nu}^\mathrm{beam}}{\theta_\mathrm{HPBW}^{2}} = \frac{\pi}{4 \ \mathrm{ln(2)}} \frac{F_{\nu}}{\Omega}, \\
    \label{eq:flux}
\end{equation}

\noindent where \textit{F$_{\nu}$} in the integrated flux and $\Omega$ the source area. Dust opacities were taken from \citep{Ossenkopf1994}, where $\kappa_{\nu}$~=~0.0189~cm$^{2}$~g$^{-1}$ for $\lambda$~=~0.87~mm (Ophiuchus Class I sources), $\kappa_{\nu}$~=~0.00899~cm$^{2}$~g$^{-1}$ for $\lambda$~=~1.3~mm (Taurus Class I/II sources), and $\kappa_{\nu}$~=~0.0181~cm$^{2}$~g$^{-1}$ for $\lambda$~=~0.89~mm (IRS 48). \cite{Jorgensen2016} employed $\kappa_{\nu}$~=~0.0182~cm$^{2}$~g$^{-1}$ for $\lambda$~=~0.85~mm in IRAS 16293. 

The [SO$_{2}$/H$_{2}$] value for IRAS 16293 results in an upper limit, given that the continuum emission is optically thick \citep{Jorgensen2016}. \cite{Garufi2022} calculated SO$_{2}$ column densities by taken an integrated region equivalent to the beam size for the Taurus sources, which varies from 0$\farcs$27 to 0$\farcs$64. For these sources, we applied a factor (beam$_\mathrm{min}$~$\times$~beam$_\mathrm{maj}$/1$\farcs$0$^{2}$), where beam$_\mathrm{min}$ and beam$_\mathrm{maj}$ are the beam size in \cite{Garufi2022}, to be consistent with the integration area of 1$\farcs$0. This results in large error bars where the lower values correspond to abundances for an integration area of 1$\farcs$0 and the upper values were calculated using an integration area equivalent to the beam size.

This section highlights the difficulties in comparing data from different observations with diverse beam sizes, strategies, results, and different kind of information available in the literature. An unbiased survey is, therefore, crucial for a more precise comparison.

% TABLE 4
\begin{table*}[h!]
        \caption{Calculations for [SO$_{2}$/H$_{2}$] from literature values.}
        \label{table:SO2_H2_lit}
        \centering
        \begin{tabular}{l c c r r r l}
                \hline\hline
                Source             		& Distance	& \textit{N}(H$_{2}$) $^{a}$  	& \textit{T$_\mathrm{gas}$}   	& \textit{N}(SO$_{2}$)                  	& [SO$_{2}$/H$_{2}$]      & Ref.  	\\
                                         		& (pc)          	& ($\times$10$^{24}$~cm$^{-2}$) & (K)                                  	& ($\times$10$^{14}$~cm$^{-2}$) 	& ($\times$10$^{-10}$)    &          	\\
                \hline
                IRAS 16293   		& 120           	& >~12.0                           		& 125                                     	& 15.0                                          	& <~1.25                          & 1, 2  	\\
                \hline
                \multicolumn{7}{c}{Ophiuchus Class I} \\
                \hline
                GSS 30         		& 139           	& 0.20~$\pm$~0.02         		& 50 -- 150                               	& [67 -- 431]                                   	& [335 -- 2160]             	& 3      	\\
                Elias 29  			& 139           	& 0.24~$\pm$~0.02              	& 50 -- 150                               	& [150 -- 990]                                  	& [625 -- 4130]          	& 3          	\\
                IRS 43      			& 139           	& 0.26~$\pm$~0.05        		& 50 -- 150                               	& [49 -- 312]                                   	& [187 -- 1200]         		& 3      	\\
                IRS 44          		& 139           	& 0.23~$\pm$~0.03               	& 50 -- 150                               	& [239 -- 1540]                                 	& [1040 -- 6700]          	& 3      	\\
                IRS 67              		& 139           	& 0.92~$\pm$~0.10           	& 50 -- 150                               	& [26 -- 166]                                   	& [28 -- 180]                     	& 3         	\\
                \hline
                \multicolumn{7}{c}{Taurus Class I/II} \\
                \hline
                DG Tau (outer disk)	& 121           	& 8.1                                 		& 40 -- 300                     		& [0.15 -- 2.3]                                 	& [0.02 -- 0.28]            	& 4, 5	\\
                HL Tau (inner disk) 	& 147           	& 20.7                                  	& >~350                            		& >~1.6                                         	& >~0.08                          	& 4, 5  	\\
                HL Tau (outer disk)	& 147           	& 20.7                                	& 58                                		& [0.3 -- 10]                                   	& [0.02 -- 0.48]        		& 4, 5  	\\
                IRAS 04302          	& 161           	& 3.8                                   	& >~75                                    	& >~0.1                                         	& >~0.04                          	& 4, 5  	\\
                T Tau S             		& 144           	& 2.8                                        	& 40 -- 300                               	& [0.2 -- 37]                                   	& [0.08 -- 13.4]         		& 4, 5  	\\
                \hline
                IRS 48               		& 134           	& 1.1                                     	& 50 -- 150                               	& [12.7 -- 32.3]                               	& [12 -- 29]              		& 6, 7  	\\
                \hline\hline
        \end{tabular}
        \tablefoot{$^{(a)}$Assuming \textit{T$_\mathrm{dust}$}~=~30~K. References: [1] \cite{Jorgensen2016}, [2] \cite{Drozdovskaya2018}, [3] \cite{Artur2019a}, [4] \cite{Garufi2021}, [5] \cite{Garufi2022}, [6] \cite{Booth2021}, [7] \cite{vanderMarel2021}.
        
        }
\end{table*}

\end{appendix}

\end{document}